\documentclass[a4paper,11pt]{article}
\usepackage{jheppub} 
\usepackage{lineno}
\usepackage{amsmath}
\usepackage{physics}
\usepackage{multirow}
\usepackage{amssymb}
\usepackage{bbm}
\usepackage{bigints}
\usepackage{subfigure}
\usepackage{braket}
\usepackage{color}
\usepackage[normalem]{ulem}

\allowdisplaybreaks

\newcommand{\LG}{\text{LG}}
\newcommand{\LC}{\text{LC}}
\newcommand{\heli}[1]{\underline{#1}}
\newcommand{\dc}[1]{ {}^2 #1 }
\newcommand{\RNum}[1]{\uppercase\expandafter{\romannumeral #1\relax}}
\newcommand{\refvec}[1]{\vec{#1}_o}

\preprint{Preprint number: KEK-TH-2684}

\title{\boldmath Finite Volume Hamiltonian method for two-particle systems containing long-range potential on the lattice}

\author[a]{Kang Yu, }
\author[c]{Guang-Juan Wang, }
\author[a,b]{Jia-Jun Wu, }
\author[d]{Zhi Yang}

\affiliation[a]{%
School of Physical Sciences, University of Chinese Academy of Sciences, Beijing 100049, China}
\affiliation[b]{
outhern Center for Nuclear-Science Theory (SCNT), Institute of Modern Physics, Chinese Academy of Sciences, Huizhou 516000, China}
\affiliation[c]{
KEK Theory Center, Institute of Particle and Nuclear Studies (IPNS), High Energy Accelerator Research Organization (KEK), Tsukuba 305-0801, Japan
}
\affiliation[d]{
School of Physics, University of Electronic Science and Technology of China, Chengdu 610054, China
}

\emailAdd{ wujiajun@ucas.ac.cn}

\abstract{
We propose a systematic method to block-diagonalize the finite volume effective Hamiltonian for two-particle systems with arbitrary spin in both the rest and moving frame. 
The framework is convenient and efficient for addressing the left-hand cut issue arising from long-range potential, which are challenging in the framework of standard L\"uscher formula. Furthermore, the method provides a foundation for further extension to three-particle systems. 
We first benchmark our method by examining several toy models, demonstrating its consistency with standard L\"uscher formula in the absence of long-range potential. In the presence of long-range potential, we investigate and resolve the effects and issues of left-hand cut.
As a realistic application, we calculate the finite volume spectra of isoscalar $D\bar{D}^*$ system, where the well-known exotic state $\chi_{c1}(3872)$ is observed. The results are qualitatively consistent 
with the lattice QCD calculation, highlighting the reliability and potential application of our framework to the study of other exotic states in  hadron physics.

}

\begin{document}
\maketitle
\flushbottom
\newpage

\section{Introduction}

With the rapid advancement of computational resources, lattice QCD (LQCD) is now capable of providing increasingly precise calculations of the hadron physics based on the underlying QCD theory. 
The research paradigm of studying the scattering process of two hadrons with regular potential on the lattice is well-established. 
Firstly, energy levels are extracted by simulating Green functions of various types of interpolators based on LQCD Lagrangian. 
Provided that the continuum limit has been reached, these energies are then translated into scattering matrix utilizing the quantization condition formulated by L\"uscher in Ref.~\cite{Luscher:1990ux} or its extension applicable to the systems with spin and multiple channels in the rest or moving frame~\cite{Rummukainen:1995vs, Gockeler:2012yj, Hansen:2012tf, Kim:2005gf, Liu:2005kr, Li:2012bi}.
For more details, we refer to a review~\cite{Briceno:2017max} and references therein. 
To extract physical properties, the energy-dependence of scattering matrix is parameterized, for example, through effective range expansion (ERE) near the threshold~\cite{Bethe:1949yr, Blatt:1949zz, taylor2012scattering} or K-matrix formalism~\cite{taylor2012scattering, weinberg2013lectures}. By analyzing the resulting phase shift, poles corresponding to the bound states or resonances can be ultimately identified. 

However, since the discovery of doubly-charmed exotic state $T_{cc}$~\cite{LHCb:2021auc,LHCb:2021vvq,Lyu:2023xro,Padmanath:2022cvl}, whose mass is quite close to the $D^0D^{*+}$ threshold, it is realized that this paradigm is challenged when the left-hand cut arising from the one-pion-exchanged (OPE) long-range interaction approaches the threshold~\cite{Hansen:2024ffk, Du:2023hlu, Raposo:2023oru, Meng:2023bmz, Collins:2024sfi, Bubna:2024izx}. 
More specifically, OPE introduces a left-hand cut in the complex momentum plane starting at $\left(p_{\text{lhc}}^{\text{OPE}}\right)^2\approx -\frac{1}{4}m_\text{eff}^2$ where $m_{\text{eff}}^2 = m_\pi^2-(m_{D^*}-m_D)^2$.
At physical $m_\pi$, there is $m_{\text{eff}}^2<0$, indicating the need to recognize it as a three-particle $DD\pi$ system. 
Although $m_\pi$ can be larger than physical value in the lattice calculation, leading to $m_{\text{eff}}^2>0$, standard L\"uscher formula still encounters difficulties even above the left-hand cut. Specifically, the contribution from the terms proportional to $e^{-m_{\text{eff}} L}$, which are neglected during the derivation of L\"uscher formula, may be underestimated when $m_{\text{eff}}$ is sufficiently small. Moreover, as will be discussed, L\"uscher formula which is originally derived above the threshold can no longer be analytically continued to the regions below left-hand cut. 
To address these challenges, some modified L\"uscher formulas have been proposed in Refs.~\cite{Bubna:2024izx, Raposo:2023oru}, which suggest separating the short-range and long-range part. In Ref.~\cite{Hansen:2024ffk}, the effect of left-hand cut for $T_{cc}$ is incorporated by treating it as a $DD\pi$ three-particle system. In Ref.~\cite{Meng:2023bmz}, an EFT-based plane-waves expansion is used.
Furthermore, the convergence radius of ERE and the applicability of K-matrix parameterization are more limited due to the presence of left-hand cut. 
When the cut is close to the threshold, convergence radius can be quite small, necessitating modifications to ERE~\cite{vanHaeringen:1981pb,Du:2024snq}. These modifications allow the phase shift near the threshold to be parameterized more effectively when long-range potential matters.
In summary, addressing left-hand cut problem requires careful consideration of all these issues.

In fact, in addition to standard L\"uscher formula, there are some alternative and equivalent frameworks for extracting physical observables from lattice simulations, such as HALQCD method~\cite{Ishii:2012ssm} and Finite Volume Hamiltonian(FVH) method~\cite{Hall:2013qba, Wu:2014vma, Li:2021mob}. 
The HALQCD framework employs Nambu-Bethe Salpeter wave functions to extract an effective potential, which is subsequently used to solve Schr\"odinger equation. 
In the present work, we focus on FVH method, which was firstly introduced over a decade ago in Ref.~\cite{Hall:2013qba} and has been successfully applied to investigate various resonances including $\Lambda(1405)$~\cite{Hall:2014uca}, $N^*(1535)/N^*(1650)$~\cite{Liu:2015ktc, Abell:2023nex}, $N^*(1440)$~\cite{Liu:2016uzk, Wu:2017qve}, and $D_{s0}(2317)$~\cite{Yang:2021tvc}. 
Recently, a method called as EFT-based plane-wave expansions has been used in Refs.~\cite{Meng:2023bmz,Meng:2021uhz,Meng:2024kkp} to investigate $DD^*$ and $NN$ systems on the lattice. 
It is important to note that its workflow is fundamentally identical to that of FVH method. In these two frameworks, the effective Hamiltonian plays a central role.
In the finite volume, Hamiltonian is expressed as a matrix whose eigenvalues are related to the energy levels extracted from lattice simulations. \footnote{The Hamiltonian is assumed to be energy-independent by default.}
In the infinite volume, Hamiltonian serves as the kernel of Lippmann-Schwinger equation (LSE), which determines the scattering $T$-matrix. 
Notably, FVH method does not impose specific assumptions about the effective potential except for the exclusion of loop integrals, allowing it to address left-hand cut issue directly.

However, a technical issue arises due to the lattice symmetry. Instead of SO(3) symmetry, lattice simulations respect only the discrete symmetry of point group.  
The energy levels are extracted using interpolators belonging to specific irreducible representations (irreps) of lattice symmetry group. 
As a result, in order to analyze the lattice spectra, finite volume Hamiltonian should be projected onto the corresponding irreps. 
In Refs.~\cite{Li:2019qvh,Li:2021mob}, a partial wave method has been developed to facilitate such projections, which is convenient if potential is expressed via a partial-wave expansion with few terms.
However, this method breaks down for long-range potential, which may require a very large number of partial waves to give an accurate description.
In Refs.~\cite{Meng:2021uhz, Meng:2023vxy}, an alternative method based on the projection operator, which is valid even there is long-range potential, is employed. 
%
As will be discussed in the subsection~\ref{subsec:compare}, the procedure for the projection adopted there largely relies on numerical calculations and therefore the efficiency primarily depends on the quality of codes, which may be inconvenient for unsophisticated readers wishing to employ it. 
In this paper, we propose a systematic and sophisticated method to perform irrep decomposition of finite volume Hamiltonian. The method is applicable to two-particle systems with arbitrary spin, and enable us to explicitly compute the matrix elements of any given irrep block, offering great convenience. 
Furthermore, this method can be easily generalized to three-particle systems, a rapidly growing topic of interest, highlighting its potential applications in the future.

The paper is organized as follows: 
In Sec.~\ref{sec:QM on torus}, the FVH method is briefly introduced.
In Sec.~\ref{sec:pw:method}, the partial wave method for irrep decomposition given in Ref.~\cite{Li:2019qvh} is revisited and reformulated, with a few further advancements discussed. It is demonstrated that this method fails when long-range interaction is present.  
Sec.~\ref{sec:projection} constitutes the body part of the paper, where the projection operator method for irrep decomposition is developed in detail. 
Discussions are organized based on the spin and total momentum of a system. In Sec.~\ref{sec:toy model}, we apply the proposed method to a variety of toy models, enabling a comparison with standard L\"uscher formula and illustrating the effects of long-range interaction. 
In Sec.~\ref{sec:X3872}, as a more practical application, the proposed method is employed to study the exotic state $\chi_{c1}(3872)$ on the lattice, where long-range interaction may play an important role.

\section{Introduction to the Finite Volume Hamiltonian method}\label{sec:QM on torus}

First of all let us make a brief introduction to FVH method. 
For a given hadronic system, the effective Hamiltonian describing its dynamics can be constructed using either effective field theory or phenomenological model, incorporating necessary symmetries. 
The interacting potential in the coordinate space is denoted by $V(\vec{x}_1,\vec{x}_2,\cdots,\vec{x}_i\,;\{\sigma\})$, where $\{\sigma\}$ collectively denotes the discrete indices such as spin 
and isospin, and $\vec{x}_i$ denotes the coordinates of the $i$-th particle.

In the lattice calculation, periodic boundary conditions are often imposed in the configuration space of path integral. 
From the perspective of effective theory, this modifies the potential to an effective finite-volume potential $V^L$, given by~\cite{Luscher:1986pf,Luscher:1990ux}
\begin{align}
    V^L(\vec{x}_1,\,\cdots,\,\vec{x}_i\,;\{\sigma\}) = \left(\prod\limits_{j=1}^i \sum\limits_{\vec{n}_j\in \mathbb{Z}^3}\right) V\left(\vec{x}_1 + \vec{n}_1 L,\,\cdots,\,\vec{x}_i + \vec{n}_i L\,;\{\sigma\} \right), 
    \label{eq:VL in coordinate rep}
\end{align}
where $L$ is the size of finite volume. 
The periodicity of coordinate space discretizes the momentum space, permitting only the modes $\vec{p}_{\vec{n}}=\frac{2\pi}{L}\vec{n}$ with $\vec{n}\in\mathbb{Z}^3$. 
For the system consisting of two particles with definite total momentum $\vec{P}$ and spins $s_1$, $s_2$, the finite volume Hilbert space $\mathcal{H}_L$ is then given by\footnote{In the present paper we adopt the normalization convention as $ \langle \vec{k}^\prime | \vec{k} \rangle = \delta^3(\vec{k}^\prime-\vec{k})$ and $\langle \vec{n^\prime} | \vec{n} \rangle = \delta_{\vec{n}^\prime,\vec{n}}$ in the infinite and finite volume, respectively.} 
\begin{align}
    \mathcal{H}_L := \mathrm{span}\left\{ \ket{\vec{n},\lambda_1,\lambda_2}, \,\,\vec{n}\in \mathbb{Z}^3, \lambda_1=-s_1,\cdots,s_1,\lambda_2=-s_2,\cdots,s_2 \right\}.
\end{align}
Because of the space translation symmetry, potential only depends on the displacement of particles $\vec{r}=\vec{x}_1-\vec{x}_2$ and $\vec{r}^\prime=\vec{x}_3-\vec{x}_4$.\footnote{In principle, it also depends on $\Delta\vec{R}=\frac{1}{2}\left(\vec{x}_1+\vec{x}_2-\vec{x}_3-\vec{x}_4\right)$. However, its Fourier momentum corresponds to the total momentum, which is trivial. } In momentum space, the finite volume effective potential is given by
\begin{align}
    V^{L}\left(\vec{n}^{\,\prime},\vec{n}\right) 
    &= \frac{1}{L^3}\int_{L^3}\int_{L^3} d\vec{r} d\vec{r}^{\,\prime}  e^{i\frac{2\pi}{L}\left(\vec{n}^{\,\prime}\cdot\vec{r}^{\,\prime} - \vec{n}\cdot\vec{r}\right)} V^L\left(\vec{r}^{\,\prime},\vec{r}\right) 
    \\
    & = \frac{1}{L^3}\sum\limits_{\vec{m},\vec{m}^{\,\prime}\in\mathbb{Z}^3} \int_{L^3}\int_{L^3} d\vec{r} d\vec{r}^{\,\prime}  e^{i\frac{2\pi}{L}\left(\vec{n}^{\,\prime}\cdot\vec{r}^{\,\prime} - \vec{n}\cdot\vec{r}\right)} V\left(\vec{r}^{\,\prime}+\vec{m}^{\,\prime} L,\vec{r}+\vec{m}L\right) 
    \\
    & = \frac{1}{L^3} \int_{\mathbb{R}^3}\int_{\mathbb{R}^3} d\vec{r} d\vec{r}^{\,\prime}  e^{i\frac{2\pi}{L}\left(\vec{n}^{\,\prime}\cdot\vec{r}^{\,\prime} - \vec{n}\cdot\vec{r}\right)} V\left(\vec{r}^{\,\prime},\vec{r}\right) 
    \\
    &= \left(\frac{2\pi}{L}\right)^3 V\left(\vec{p}^\prime_{\vec{n}^\prime},\vec{p}_{\vec{n}}\right)\label{eq:vLL}
\end{align}
where the discrete indexes $\{\sigma\}$ are suppressed. 
%
In a word, up to a factor due to the different Fourier normalization in finite and infinite volume, the matrix elements of $\hat{V}^L$ and $\hat{V}$ in momentum space share exactly the same expression as well as parameters, as long as no loop integration contributes to $\hat{V}$\footnote{Integration anywhere in the infinite volume should be converted to summation when turning to the finite volume.} .
Note that the relation holds regardless of the range of interaction. 
%


Provided that the finite volume effective Hamiltonian describes the relevant system effectively enough, its eigenvalues correspond to the lattice energy levels. In the infinite volume, the partial wave scattering matrix is solved from partial wave LSE,
\begin{align}\label{eq:partial wave LSE}
    T^J_{\alpha\beta}(p,k;E) = V^J_{\alpha\beta}(p,k) +  \sum\limits_{\gamma} \int q^2 dq\, V^J_{\alpha\gamma}(p,q)\,G_\gamma(q;E+i\epsilon)\,T^J_{\gamma\beta}(q,k;E)
\end{align}
where $J$ is the total angular momentum, and $\alpha$, $\beta$, and $\gamma$ collectively denote the $z$-component of total angular momentum ($M$), the orbit angular momentum ($l$), total spin ($S$), and the channel. 
$T^J_{\alpha\beta}$ and $V^J_{\alpha\beta}$ are the partial wave amplitude of transition matrix $T$ and potential $V$ in the $JMlS$ basis~\cite{Chung:1971ri,weinberg2013lectures}. 
The propagator is given by $G_\alpha(q;E)=\left(E-\sqrt{q^2+m_{\alpha_1}^2}-\sqrt{q^2+m_{\alpha_2}^2}\right)^{-1}$, where $m_{\alpha_1}$ and $m_{\alpha_2}$ are masses of the particles in $\alpha$-channel. 
Depending on whether lattice energy levels or experimental scattering data are available to constrain the parameters of effective Hamiltonian, the calculations or predictions can be performed either from lattice to experiment or vice versa.

Because of the discretized momentum space, the lattice symmetry group $G$ is typically a point group, such as $O_h$, $C_{4v}$, $C_{2v}$ and $C_{3v}$ (or their double covers for fermionic systems) when the total momentum $\frac{\vec{P}L}{2\pi}=(0,0,0)$, $(0,0,1)$, $(0,1,1)$ and $(1,1,1)$, respectively. 
In order to compare with the lattice energy levels in a given irrep of $G$, the effective Hamiltonian in the finite volume should be projected onto the corresponding subspace. 
In the following section, we will briefly introduce the partial wave method, which performs the irrep decomposition by restricting $O(3)$ to $G$, and then provide a more detailed discussion on the projection operator method. This latter approach circumvents $O(3)$ group and focus on $G$ directly, making it more elegant and convenient when dealing with the long-range interaction.

Finally, it is important to highlight the model-dependence and model-independence of FVH method. 
%
%

Indeed, the effective Hamiltonian itself is model-dependent. It may rely on, for example, the approximation or the truncation of corresponding effective Lagrangian. However, as guaranteed by L\"uscher formula, the relationship between the finite volume energy levels and infinite volume scattering amplitude is model-independent. More specifically, suppose that there are two Hamiltonian constructed from two different EFT, and both of them can fit the lattice spectra accurately enough. Then, these two Hamiltonian will yield the same scattering amplitude, at least at the corresponding lattice energy~\cite{Wu:2014vma, Abell:2021awi}. \footnote{The scattering amplitude given by these two Hamiltonian can be very different at the energy outside the fitting region.}  In this regard the FVH method is model-independent.

\section{Partial wave method}\label{sec:pw:method}

In Refs.~\cite{Li:2019qvh,Li:2021mob}, the authors built a systematic method to extract the eigenvalues when $V$ is defined by a partial-wave expansion.
Here, we reformulate it in a slightly different manner and make some extensions. 
For simplicity, we firstly consider a system consisting of two spinless particles with $\vec{P}=0$. 
The result can be easily generalized to the system with nonzero spin by incorporating Clebsch-Gordon coefficient.
Assume that potential $V$ is given by
\begin{align}\label{eq:pw:expansion:infV}
    V(\vec{p},\vec{k}) =  \langle\vec{p}| \hat{V}|\vec{k}\rangle = \sum\limits_{l=0}^{l_{\text{cut}}} v_l(p,k)\sum\limits_{m=-l}^l  Y_{lm}(\hat{p}) Y^*_{lm}(\hat{k}),
\end{align}
where $Y_{lm}$ is spherical harmonics.  
The finite volume potential in the operator form, $\hat{V}^L= \sum\limits_{\vec{n},\,\vec{n}^\prime} V^L\left(\vec{n}^\prime,\vec{n}\right)\ket{\vec{n}^\prime}\bra{\vec{n}}$, reads
\begin{align}\label{eq:pw:expansion:finV}
    \hat{V}^L  = \left(\frac{2\pi}{L}\right)^3 \sum\limits_{N,N^\prime\in \mathbb{N}}
    \sum\limits_{l=0}^{l_{\text{cut
    }}}v_l\left(\frac{2\pi \sqrt{N^\prime}}{L},\frac{2\pi \sqrt{N}}{L}\right) \sum\limits_{m=-l}^l \ket{N^\prime,l,m}\bra{N,l,m},
\end{align}
where $\ket{N,l,m}$ is defined as
\begin{align}\label{eq:pwbasis:fin}
    \ket{N,l,m} = \begin{cases}
        \delta_{l0}\delta_{m0}\sqrt{\frac{1}{4\pi}}\ket{\vec{0}} \quad&\text{if}\, N=0 \\
        \sum\limits_{\vec{n}^2=N} Y_{lm}(\vec{n})\ket{\vec{n}} \quad&\text{if}\, N\neq0 
    \end{cases}.
\end{align}
Such states can be regarded as the finite volume analogs of the infinite volume partial wave state.  
Typically, $l_{\text{cut}}\leq4$ in practice. In such case, the interacting space $\mathcal{H}_{L,\text{int}}$ is actually the subspace given by,
\[
\mathcal{H}_{L,\text{int}} = \bigoplus \limits_{N\in\mathbb{N}}\left(\sum\limits_{l=0}^{l_{\text{cut}}}\mathcal{S}_{l}(N)\right) \subseteq \bigoplus \limits_{N\in\mathbb{N}}\mathcal{S}(N) \equiv \mathcal{H}_L,
\]
where 
\begin{align}
    \mathcal{S}_l(N)&:=\mathrm{span}\left\{\ket{N,l,m},\,\, m=-l,\cdots,l \right\},
    \\
    \mathcal{S}(N)&:=\mathrm{span}\{\ket{\vec{n},\,\vec{n}^2=N}\}.
\end{align}
Because the symmetry group $G$ is a subgroup of $O(3)$, for any $g\in G$ there is,
\begin{align}
    g\ket{N,l,m} = \sum\limits_{m^\prime} D^l_{m^\prime m}(g) \ket{N,l,m^\prime},
\end{align}
where $D^l$ is Wigner-D matrix, the representation matrix of $O(3)$. 
Thus, $\mathcal{S}_l(N)$ furnishes the restricted representation of $O(3)$ to $G$, if it is nonempty.
With coefficients $[C^l]_{\Gamma r \alpha,\,m}$, one can construct the states that furnish the $\Gamma$-irrep of $G$
\begin{align}\label{eq:pwmethod:basis1}
    \ket{N,l,\Gamma,r,\mu}=\sum\limits_{m}[C^l]_{\Gamma r \mu, \,m}\ket{N,l,m} .
\end{align}
Here, $r$ and $\mu$ denote the occurrence of $\Gamma$ in $D^l$ and the column index of $D^\Gamma$, respectively. 
Explicit values of $[C^l]$ for $l\leq 4$ can be found in Ref.~\cite{Bernard:2008ax}.  

%
Due to the breaking of $O(3)$ symmetry in the finite volume, different partial waves are mixed. 
To be specific, $\ket{N,l,m}$, and therefore $\ket{N,l,\Gamma,r,\mu}$, with different $l$-indices are not orthogonal in general and can even be linearly dependent. 
Consequently, for each $N$, it is necessary to orthonormalize the states with respect to $l$- and $r$- indices in order to obtain an orthonormal basis $\ket{N,\Gamma,F,\mu}, F=1,\cdots,F_{\text{max}}$. 
Here, $F_{\text{max}}$ denotes the occurrence of $\Gamma$ in the space $\sum\limits_{l}^{l_{\text{cut}}}\mathcal{S}_l(N)$. 

When $l_{\text{cut}}$ is sufficiently large such that $\sum\limits_{l=0}^{l_{\text{cut}}}\mathcal{S}_l(N) = \mathcal{S}(N)$, the workflow remains applicable in principle but becomes quite cumbersome.  
To optimize, we note that $\mathcal{S}(N)$ can be further decomposed into $G$-invariant subspace $\mathcal{S}(\vec{n})$, defined as
\begin{align}
    \mathcal{S}(\vec{n})=\mathrm{span}\{g\ket{\vec{n}},\,g\in G\},
\end{align}
representing the space spanned by the $G$-orbit of $\ket{\vec{n}}$. 
For example, $\mathcal{S}(9)=\mathcal{S}((0,0,3))\oplus\mathcal{S}((2,2,1))$. 
By appropriately selecting a set of reference momentum vectors $\refvec{n}$, $\mathcal{H}_L$ can be decomposed into $\mathcal{H}_L=\bigoplus\limits_{\refvec{n}}\mathcal{S}(\refvec{n})$.\footnote{For clarity, we denote $\refvec{n}$ as reference momentum hereafter.} 
For a system at rest, there are only seven distinct patterns of reference momentum vectors: $(0,0,0)$, $(0,0,a)$, $(0,a,a)$, $(a,a,a)$, $(0,a,b)$ and $(a,a,b)$, $(a,b,c)$ where $a,b,c$ represent the distinct nonzero integers. 
The structure of $\mathcal{S}(\refvec{n})$ is entirely determined by the pattern of $\refvec{n}$~\cite{Doring:2018xxx}. 
To proceed, we define states similar to Eq.~(\ref{eq:pwbasis:fin}) as follows:
\begin{align}\label{eq:pwbasis:fin2}
        \ket{\refvec{n},l,m} = \begin{cases}
        \delta_{l0}\delta_{m0}\sqrt{\frac{1}{4\pi}}\ket{\vec{0}} \quad&\text{if}\,\, |\refvec{n}|^2=0 \\
        \sum\limits_{\vec{n}^\prime\in\mathcal{S}(\refvec{n})} Y_{lm}(\vec{n}^\prime)\ket{\vec{n}^\prime} \quad&\text{if}\,\, |\refvec{n}|^2\neq0 
    \end{cases},
\end{align}
and $\mathcal{S}_l(\refvec{n})=\mathrm{span}\{\ket{\refvec{n},l,m},\,m=-l,\cdots,l\}$. The occurrence $F^{\refvec{n}}_{\text{max}}$ of $\Gamma$-irrep in a given $\mathcal{S}(\refvec{n})$ is given by, 
\begin{align}\label{eq:Fmax}
F^{\refvec{n}}_{\text{max}} = \frac{\dim(\Gamma)}{|G|} \sum\limits_{g\in G} \chi^{\Gamma*}(g) \chi_{\refvec{n}}(g),
\end{align}
where $|G|$ is the order of $G$. $\chi_{\vec{n}}(g)$ is the character of $D(G)$ with entries defined by
$
[D(g)]_{\vec{n}^\prime,\vec{n}^{\prime\prime}} := \delta_{\vec{n}^\prime ,\, g\vec{n}^{\prime\prime}}
$. 
Since the dimension of $\mathcal{S}(\refvec{n})$ is finite, one can always find $F_{\text{max}}$ distinct $(l;r)$, such that the states $\ket{\refvec{n},l,\Gamma,r,\mu}$(defined similarly to Eq.~(\ref{eq:pwmethod:basis1})) are linearly independent and span the subspace of $\mathcal{S}(\refvec{n})$ that furnishes $\Gamma$-irrep. 
As a consequence, when $l_{\text{cut}}$ is sufficiently large such that $\sum\limits_{l=0}^{l_{\text{cut}}}\mathcal{S}_l(\refvec{n})=\mathcal{S}(\refvec{n})$, it requires only identification of how the states $\ket{\refvec{n},l,\Gamma,r,\mu}$ with unchosen $(l;r)$ are expressed as the linear combination of the ones with chosen $(l;r)$ for the seven patterns of $\vec{n}$. However, such identification are still cumbersome. 

A choice of $(l;r)$ for the system consisting of two spinless particles with $\vec{P}=0$ is given in Ref.~\cite{Doring:2018xxx}. Here, we present the choice of $(l,J)$ for a system consisting of a (pseudo-)scalar particle and a (pseudo-)vector particle with $\vec{P}=0$ in Tab.\ref{tab:partial wave method}.\footnote{Here we always choose the occurrence $r=1$, i.e, for any irrep we only pick once even if it occur more than once for some $J$.} 
As an example, for $\refvec{n}=(0,0,1)$, $\ket{\refvec{n},l,J,T_1^+,\mu=1,2,3}$ with $(l,J)=(0,1), (2,1)$ form a complete basis for the subspace furnishing $T_1^+$ of $\mathcal{S}((0,0,1))$. 
The extension from the spinless system to the system with $S=1$ is not merely a naive addition of angular momentum. As an example, for $\refvec{n}=(0,0,a)$ or $(a,a,a)$, $\ket{\refvec{n},l,J,T_1^+,\mu}$ with $(l,J)=(2,1)$ and $(2,3)$ are actually linearly dependent, as we will see in a later section. 
These results can also be useful for the construction of two-mesons operator in the lattice simulation.

\begin{table}[tbp]
    \centering
    \begin{tabular}{|c|c|}
    \hline
    $(0,0,0)$  &  $\left\{(0,1)\right\}^{T_1^+}$ \\
    \hline
    {$(0,0,a)$} & $\left\{(1,0)\right\}^{A_1^-}$, $\left\{(1,2)\right\}^{E^-}$, $\left\{ (0,1), (2,1)\right\}^{T_1^+}$, 
    $\left\{(1,1)\right\}^{T_1^-}$, $\left\{(1,2)\right\}^{T_2^-}$, $\left\{(2,2)\right\}^{T_2^+}$
    \\
    \hline
    \multirow{2}{*}{$(0,a,a)$} &  $\left\{(1,0)\right\}^{A_1^-}$, $\left\{(3,3)\right\}^{A_2^-}$, $\left\{(2,3)\right\}^{A_2^+}$, $\left\{(2,2)\right\}^{E^+}$, $\left\{(1,2), (3,2)\right\}^{E^-}$,  \\
    & $\left\{(1,1), (3,3)\right\}^{T_1^-}$,  $\left\{(0,1),(2,1),(2,3)\right\}^{T_1^+}$ , $\left\{(1,2), (3,2)\right\}^{T_2^-}$, $\left\{(2,2),(2,3)\right\}^{T_2^+}$ \\
    \hline
    \multirow{2}{*}{$(a,a,a)$} & $\left\{(1,0)\right\}^{A_1^-}$, $\left\{(2,3)\right\}^{A_2^+}$, $\left\{(1,2)\right\}^{E^-}$, $\left\{(2,2)\right\}^{E^+}$,  $\left\{(1,1)\right\}^{T_1^-}$, \\
    & $\left\{(0,1),(2,1)\right\}^{T_1^+}$, $\left\{(1,2),(3,2)\right\}^{T_2^-}$, $\left\{(2,2)\right\}^{T_2^+}$ 
    \\ 
    \hline
    \multirow{4}{*}{$(0,a,b)$} & $ \left\{(1,0), (3,4)\right\}^{A_1^-}, \left\{(4,4)\right\}^{A_1^+}, \left\{(2,3)\right\}^{A_2^+}$, $\left\{(3,3), (5,6)\right\}^{A_2^-}$, $\left\{(2,2), (4,4)\right\}^{E^+}$,  \\
    & $\left\{ (1,2), (3,2), (3,4), (5,4)\right\}^{E^-}$, $\left\{(1,1), (3,3), (3,4), (5,4)\right\}^{T_1^-}$,  \\
    &  $\left\{ (0,1), (2,1), (2,3), (4,3), (4,4) \right\}^{T_1^+}$ , $\left\{ (1,2), (3,2), (3,3), (5,5) \right\}^{T_2^-}$,  \\
    &  $\left\{ (2,2), (2,3), (4,3), (4,5), (6,5)\right\}^{T_2^+}$ \\
    \hline 
    \multirow{4}{*}{$(a,a,b)$} & $\left\{(1,0),(3,4)\right\}^{A_1^-}$, $\left\{(4,4)\right\}^{A_1^+}$, $\left\{ (3,3)\right\}^{A_2^-}$, $\left\{(2,3), (4,3)\right\}^{A_2^+}$, \\
    & $\left\{(1,2),(3,2),(3,4)\right\}^{E^-}$, $\left\{(2,2),(4,4),(4,5)\right\}^{E^+}$, \\
    & $\left\{ (1,1), (3,3), (3,4), (5,4) \right\}^{T_1^-}, \left\{(0,1),(2,1),(2,3),(4,3),(4,4)\right\}^{T_1^+}$, \\
    & $\left\{(1,2),(3,2),(3,3),(3,4),(5,4)\right\}^{T_2^-}$, $\left\{(2,2),(2,3),(4,3),(4,4)\right\}^{T_2^+}$ \\
    \hline 
    \multirow{8}{*}{$(a,b,c)$} &  $\left\{(1,0),(3,4),(5,6)\right\}^{A_1^-}$, $\left\{ (4,4),(6,6),(8,9)\right\}^{A_1^+}$, \\
    &
    $\left\{ (3,3),(5,6),(7,6)\right\}^{A_2^-}$, $\left\{ (2,3),(4,3),(6,6)\right\}^{A_2^+}$, \\
    & $\left\{(1,2),(3,2),(3,4),(5,4),(5,6),(7,7) \right\}^{E^-}$, \\
    & $\left\{ (2,2),(4,4),(4,5),(6,6),(6,7),(8,7)\right\}^{E^+}$, \\ 
    & $\left\{ (1,1),(3,3),(3,4),(5,4),(5,5),(5,6),(7,6),(7,8),(9,8)\right\}^{T_1^-}$, \\
    & $\left\{ (0,1),(2,1),(2,3),(4,3),(4,4),(4,5),(6,5),(6,6),(8,8)\right\}^{T_1^+}$, \\
    & $\left\{ (1,2),(3,2),(3,3),(3,4),(5,4),(5,5),(5,6),(7,6),(7,7)\right\}^{T_2^-}$, \\
    & $\left\{ (2,2), (2,3), (4,3),(4,4), (4,5), (6,5),(6,6), (6,7),(8,7) \right\}^{T_2^+}$
    \\ \hline
    \end{tabular}
    \caption{The chosen $(l,J)$ to make $\ket{\vec{n},l,J,\Gamma,r=1,\mu}$ form a complete basis for the subspace that furnishes $\Gamma$-irrep in $\mathcal{S}(\vec{n})$.  The intrinsic parity is assumed to be positive.}
    \label{tab:partial wave method}
\end{table}

\section{Projection operator method}\label{sec:projection}

The partial wave method, while convenient and effective in certain cases, becomes tedious even after simplification when there are plenty of partial waves involved. 
To address it, we propose the projection operator method which is much more efficient and practical in many cases, especially when the long-range potentials are included. 
In subsections~\ref{subsec:rest,2spinless}-\ref{subsec:moving} organized by the total momentum and spins $s_1,s_2$, we systematically construct the orthonormal basis for each subspace that furnishes $\Gamma$-irrep of $G$. Accordingly, $V^L$ is block-diagonalized into multiple blocks, each corresponds to a certain $\Gamma$-irrep. 

As an example and preview, for the spinless system at rest, the entries of the block corresponding to $\Gamma$ are characterized by the reference momentum $\refvec{n}$ and the occurrence $r$ of $\Gamma$ in $\mathcal{S}(\refvec{n})$. These entries are expressed as:
\begin{align}
    V^\Gamma_{\refvec{n}^\prime r^\prime, \refvec{n}r} = \sum\limits_{g} \mathcal{O}^\Gamma_{\refvec{n}^\prime r^\prime, \refvec{n}r}(g) V^L\left(\refvec{n}^\prime,g\refvec{n}\right),
\end{align}
where $\mathcal{O}^\Gamma(g)$ is the coefficient independent of the dynamical $V^L$. A detailed derivation will be given in the following subsections. For the readers who want to skip the derivation and directly utilize the results, we refer to subsection~\ref{subsec:matrix element} and Appendix.~\ref{append:matrix elements}. 

Until the subsection~\ref{subsec:isospin}, it is assumed that two particles are not in the same isospin multiplets. The incorporation of isospin symmetry will be discussed in subsection~\ref{subsec:isospin}. From now on, the intrinsic parity of the two-particle state under the space inversion $\mathbb{P}$ is denoted by $\eta$. The order of any group or set $Q$ is denoted by $|Q|$.

\subsection{$|\vec{P}|=0\,,\,s_1=s_2=0$}\label{subsec:rest,2spinless}

We refer the case as $K\pi$-system.
This is the simplest but most fundamental case, from which the results can be extended to a more complex system. 
First, we introduce the well-known projection operator $P^\Gamma_{\mu\nu}$ defined as
\begin{align}\label{eq:def:of:projector}
    P^\Gamma_{\mu\nu} = \frac{\dim(\Gamma)}{\abs{G}}\sum\limits_{g\in G} \bar{D}^{\Gamma}_{\mu\nu}(g) g\, ,
\end{align}
where $\bar{D}$ denotes the complex conjugate of $D$. The convention for $D^\Gamma$ is clarified in Appendix~\ref{append:convention}. 
Given a $S(\refvec{n})$, we demonstrate that the complete basis of the subspace furnishing $\Gamma$ is exactly the maximal linearly independent subset(MaxLIS) of the set $\{P^\Gamma_{\mu\nu}\ket{\refvec{n}}, \mu,\nu=1,\cdots,\dim(\Gamma)\}$. 
To prove, note that for any $\ket{\vec{n}^\prime} \in \mathcal{S}(\refvec{n})$, there exists a $g_0\in G$ such that $\ket{\vec{n}^\prime} = g_0\ket{\refvec{n}}$. Consequently, 
\begin{align}
    P^\Gamma_{\mu\mu}\ket{\vec{n}^\prime} = \sum\limits_{\nu} D^{\Gamma}_{\mu\nu}(g_0) P^\Gamma_{\mu\nu}\ket{\refvec{n}},
\end{align}
implying that the component of $\ket{\vec{n}}$ corresponding to the $\mu$-th column of $\Gamma$-irrep is a linear combination of $P^\Gamma_{\mu\nu}\ket{\refvec{n}}$.\footnote{For any $G$-invariant space $\mathcal{V}$, it can be decomposed as the direct sum $\mathcal{V}=\bigoplus\limits_{\Gamma,\mu}\mathcal{V}^{\Gamma,\mu}$ where $\mathcal{V}^{\Gamma,\mu}$ is the subspace that furnishes the $\mu$-th column of $\Gamma$-irrep and its dimension is the occurrence of $\Gamma$ in $\mathcal{V}$. For any vector, $P^{\Gamma}_{\mu\mu}$ pick out its component belongs to $\mathcal{V}^{\Gamma,\mu}$.} 
Therefore, the goal is to find out the MaxLIS for any given $\refvec{n}$ and then orthonormalize them. In Ref.~\cite{Meng:2021uhz} this is achieved by checking the linear independence of several states numerically followed by Gram-Schmidt procedure.
In the current paper we adopt another systematic and sophisticated method. 
Briefly speaking, provided a set of states $\{\ket{v_i}\}$ and the inner product matrix $I_{ij}:=\langle v_i | v_j \rangle$, the orthonormal basis of this set can be constructed with the help of $I_{ij}$. Specifically, let $\vec{\beta}^i=(\beta^i_1,\beta^i_2,\cdots)$ denote the orthonormal eigenvectors of $I$ with non-zero eigenvalues $\kappa_i$, then $\left\{{\sqrt{\frac{1}{\kappa_i}}}\sum_j \beta^i_j\ket{v_j},i=1,\cdots\right\}$ serves as a orthonormal basis. 
Consequently, we introduce the inner product matrix $I^\Gamma(\refvec{n})$ up to a scaling factor, named as $I$-matrix, whose entries are given by 
\begin{align}\label{eq:Imat:2spinless}
    I^\Gamma_{\nu^\prime \nu}(\refvec{n}) &:= \frac{|G|}{\dim(\Gamma)}\bra{\refvec{n}} P^{\Gamma^\dagger}_{\mu\nu^\prime} P^\Gamma_{\mu\nu} \ket{\refvec{n}} =
    \frac{\dim(\Gamma)}{|G|}\sum\limits_{g,g^\prime
    \in G} D^{\Gamma}_{\mu\nu^\prime}(g^\prime)\bar{D}_{\mu\nu}^{\Gamma}(g) \bra{\refvec{n}} g^{\prime\dagger}g\ket{\refvec{n}}
    \notag \\
    & = \frac{\dim(\Gamma)}{|G|} \sum\limits_{g,h\in G} D^{\Gamma}_{\mu\nu^\prime}(gh^{-1})\bar{D}_{\mu\nu}^{\Gamma}(g) \bra{\refvec{n}} h\ket{\refvec{n}}
    = \sum\limits_{g\in G} \bar{D}^{\Gamma}_{\nu^\prime\nu}(g) \bra{\refvec{n}} g\ket{\refvec{n}}
    \notag \\
    &=  \sum\limits_{g\in \text{LG}(\refvec{n})} \bar{D}^{\Gamma_\eta}_{\nu^\prime\nu}(g) \,,
\end{align}
where the Schur orthogonality relations $\sum_{g\in G} D^\Gamma_{\mu^\prime\nu^\prime}(g)\bar{D}^{\Gamma}_{\mu\nu}(g)=|G|/\dim(\Gamma)\delta_{\mu^\prime\mu}\delta_{\nu^\prime\nu}$ has been applied. 
$\Gamma_\eta$, incorporating the intrinsic parity, denotes the irrep $A_1^\eta\times \Gamma$. (Recall that $\eta$ denotes the intrinsic parity. For two pseudoscalar meson it is $(-1)^2=1$.) For example, $E^+_{\eta=-1}=A_1^-\times E^+ = E^-$.
The subgroup LG($\refvec{n}$) denotes the little group regarding $\refvec{n}$, consisting of the element $g$ such that $g\refvec{n}=\refvec{n}$. 
The $\mu$-independence is guaranteed by the Wigner-Eckart theorem and manifest explicitly in the last equation. 

Furthermore, we demonstrate a pleasant property of $I^\Gamma(\refvec{n})$: it is idempotent up to a scaling factor. 
To prove this, we apply the rearrangement theorem,  
\begin{align}
    I^\Gamma(\refvec{n}) &=  \sum\limits_{g\in \text{LG}(\refvec{n})} \bar{D}^{\Gamma_\eta}(gh) = I^{\Gamma}(\refvec{n})\bar{D}^{\Gamma_\eta}(h) 
    \\
    &= \sum\limits_{g\in\LG(\refvec{n})} \bar{D}^{\Gamma_\eta}(hg) = \bar{D}^{\Gamma_\eta}(h) I^{\Gamma}(\refvec{n}) 
    ,\,\quad\forall h\in \text{LG}(\refvec{n}) ,
\end{align}
which yields $[I^\Gamma(\refvec{n})]^2 = |\text{LG}(\refvec{n})|\, I^\Gamma(\refvec{n})$ after summing over $h\in\LG(\refvec{n})$ on both sides. 
Consequently, the eigenvalues of $I^\Gamma(\refvec{n})$ can only be either $0$ or $|\text{LG}(\refvec{n})|$. 
Let $[c^{\Gamma,r}_{\refvec{n}}]$ denotes the $r$-th orthonormal eigenvectors with nonzero eigenvalues, then the orthonormalized MaxLIS $\ket{\refvec{n},\Gamma,r,\mu}$ are given by
\begin{align}\label{eq:irrep:basis:2spinless}
    \ket{\refvec{n},\Gamma,r,\mu} = N^\Gamma_{\refvec{n}} \sum\limits_{\nu} [c^{\Gamma,r}_{\refvec{n}}]_{\nu} P^\Gamma_{\mu\nu}\ket{\refvec{n}},
\end{align}
where the normalization factor $N_{\refvec{n}}^\Gamma = \sqrt{\frac{\abs{G}}{\abs{\text{LG}(\refvec{n})}\dim(\Gamma)}}$. 
It is apparent that the occurrence of $\Gamma$ in $\mathcal{S}(\refvec{n})$ is the algebraic multiplicity of the eigenvalue $|\LG(\refvec{n})|$.\footnote{Strictly speaking, it should be the geometric multiplicity. For the idempotent matrix, geometric multiplicity is identical to algebraic multiplicity as such matrix is diagonalizable.}
As a cross check, let us consider the case where the little group contains only the identity element, for example, $\refvec{n}=(1,2,3)$. 
For this case $I^\Gamma$ is proportional to identity matrix for any $\Gamma$ so the occurrence of $\Gamma$ is $\dim(\Gamma)$. 
On the other hand, because that for any $\vec{n}^\prime\in\mathcal{S}((1,2,3))$ there corresponds a distinct $g\in G$ such that $g\cdot(1,2,3)=\vec{n}^\prime$, $\mathcal{S}((1,2,3))$ is isomorphism to the group algebra of $O_h$. 
Therefore, $\mathcal{S}((1,2,3))$ actually furnishes the reducible regular representation, in which the occurrence of any irrep is indeed its dimension.

As mentioned earlier, there are only seven distinct patterns of reference momentum for a two-particle system. 
For two reference momenta $\refvec{n}$ in the same $G$-orbit, for example, $(1,0,0)$ and $(0,1,0)$, the corresponding two $I^\Gamma(\refvec{n})$ differ by a similarity transformation. Though the eigenvalues of Hamiltonian are exactly the same, some intermediate results provided later depend on the adopted convention of $\refvec{n}$.  
For clarity, therefore, we specify the pattern as $(0,0,0)$, $(0,0,a)$, $(0,a,a)$, $(a,a,a)$, $(0,a,b)$, $(a,a,b)$(or $(b,b,a)$) and $(a,b,c)$ with constraints $0<a<b<c$ from now on.
The eigenvectors $[c^{\Gamma,r}_{\refvec{n}}]$ of $I$-matrix in this convention are provided in Appendix~\ref{append:matrix elements}.

\subsection{$|\vec{P}|=0\,, s_1=s\in \mathbb{N}^+, s_2=0$}\label{subsec:rest,spin0spin1}

We refer to the system as $\rho\pi$-system. 
The subspace $\mathcal{S}(\refvec{n})$ is now defined as
\begin{align}
    \mathcal{S}(\refvec{n}) = \mathrm{span}\{g\ket{\refvec{n},\lambda},\,g\in G,\lambda=-s,\cdots,s\},
\end{align}
where $\lambda$ denotes the $z$-component of spin. 
Note that the group element will now not only change the momentum but also mix the polarization as 
\begin{align}
g\ket{\vec{n},\lambda}=\sum\limits_{\lambda^\prime}D^s_{\lambda^\prime\lambda}(g)\ket{g\vec{n},\lambda^\prime} \quad\forall g\in O .
\end{align} 
Inspired by the helicity formalism in the infinite volume~\cite{Chung:1971ri}, we introduce the helicity state for $\abs{\vec{n}}\neq 0$ as: 
\begin{align}
    \ket{\vec{n},\heli{\lambda}} := \sum\limits_{\lambda}  D^s_{\lambda \heli{\lambda}}(R_{\text{st}}(\vec{n}))\ket{\vec{n},\lambda},
\end{align}
where $R_{\text{st}}(\vec{n})$ is the standard rotation that aligns $z$-axis with $\vec{n}$, i.e, $R_{\text{st}}(\vec{n}) = e^{-i\phi_{\vec{n}} J_z} e^{-i\theta_{\vec{n}}J_y}$.
In this paper we adopt the convention $0\leq\theta_{\vec{n}}\leq\pi$ and $-\pi\leq\phi_{\vec{n}}<\pi$. 
For $\vec{n}\parallel \hat{e_z}$, we define $\phi_{\vec{n}} = 0$ if $n_z>0$ and $\phi_{\vec{n}} = -\pi$ if $n_z<0$ for consistency with the later discussion on space inversion. 
%
%
Hereafter, the states labeled by underlined index refer to the helicity states rather than polarization states. 

Let $Q^+$ and $Q^-$ denote the subset consisting of proper($\det=+1$) and improper($\det=-1$) elements of a given set $Q$, respectively. 
It can be proved that for any $g\in G^+$,
\begin{align}
     g\ket{\vec{n},\heli{\lambda}} &= \sum\limits_{\lambda} D^s_{\lambda \heli{\lambda}}(R_{\text{st}}(\vec{n})) g\ket{\vec{n},\lambda}
    \\&=
    \sum\limits_{\lambda,\lambda^\prime} D^s_{\lambda \lambda^\prime}(R_{\text{st}}(g\vec{n}))
    D^s_{\lambda^\prime \heli{\lambda}}\left(R^{-1}_{\text{st}}(g\vec{n})g R_{\text{st}}(\vec{n})\right) \ket{g\vec{n},\lambda}
    \\&=
    \sum\limits_{\lambda} D^s_{\lambda \heli{\lambda}}(R_{\text{st}}(g\vec{n})) e^{-i\heli{\lambda}\varphi_w(\vec{n},g)}  \ket{g\vec{n},\lambda}
    \\&=
    e^{-i\heli{\lambda}\varphi_w(\vec{n},g)}  \ket{g\vec{n},\heli{\lambda}}.
\end{align}
The Wigner angle $\varphi_w(\vec{n},g)$ is defined by the following equation\footnote{This is different from a more commonly known Wigner rotation related to the Lorentz boost as defined in Refs.~\cite{weinberg2013lectures, Jing:2024mag}}
\begin{align}
    D^s\left(e^{-i\varphi_w(\vec{n},g)J_z}\right) = D^s\left(R^{-1}_{\text{st}}(g\vec{n})g R_{\text{st}}(\vec{n})\right).
\end{align}
In a word, momentum is altered in the usual manner, while helicity is unchanged. 
For the improper element, it is sufficient to investigate the space inversion $\mathbb{P}$,
\begin{align}
    \mathbb{P}\ket{\vec{n},\heli{\lambda}} &=
    \eta \sum\limits_{\lambda} e^{-i\lambda\phi_{\vec{n}}} d^s_{\lambda \heli{\lambda}}(\theta_{\vec{n}})  \ket{-\vec{n},\lambda}
    \\
    &=\eta\sum\limits_{\lambda} e^{-i\lambda(\phi_{\vec{n}}\pm\pi)}e^{\mp i\pi s} d^s_{\lambda, -\heli{\lambda}}(\pi-\theta_{\vec{n}})  \ket{-\vec{n},\lambda}
    \\
    &= \eta e^{\mp i\pi s} \ket{-\vec{n},-\heli{\lambda}}
    \\
    &= \eta (-1)^s \ket{-\vec{n},-\heli{\lambda}},
\end{align}
where we have used the property of Wigner-d matrix $d^s_{\lambda\lambda^\prime}(\theta) = (-1)^{s+\lambda}d^s_{\lambda,-\lambda^\prime}(\pi-\theta)$. 
Both momentum and helicity flip under space inversion. 
For later convenience, the definition of Wigner angle is extended into $\varphi_w(g,\vec{n}) := \varphi_w(g\mathbb{P},\vec{n})$ for $g\in G^-$.

To proceed, we introduce the subspace $\mathcal{S}_{\heli{\lambda}}(\refvec{n}) := \mathrm{span}\{g\ket{\refvec{n},\heli{\lambda}},\,g\in G\}\subset\mathcal{S}(\refvec{n})$. 
Based on the preceding results, it is apparent that $ \mathcal{S}_{\heli{\lambda}}(\refvec{n}) \perp \mathcal{S}_{\heli{\lambda}^\prime}(\refvec{n})$ if $\abs{\heli{\lambda}} \neq \abs{\heli{\lambda}^\prime}$. 
Furthermore, if no improper element $g_0$ satisfies $g_0 \refvec{n} = \refvec{n}$ then $\mathcal{S}_{\heli{\lambda}}(\refvec{n}) \perp \mathcal{S}_{-\heli{\lambda}}(\refvec{n})$. Otherwise, $\mathcal{S}_{\heli{\lambda}}(\refvec{n}) = \mathcal{S}_{-\heli{\lambda}}(\refvec{n})$. 
To prove this, for any $g_0\in\LG(\refvec{n})^-$,
\begin{align}
    \mathcal{S}_{\heli{\lambda}}(\refvec{n}) = \mathrm{span}\{g\ket{\refvec{n},\heli{\lambda}}\,,g\in G\} = \mathrm{span}\{gg_0\ket{\refvec{n},-\heli{\lambda}}\,,g\in G\} = \mathcal{S}_{-\heli{\lambda}}(\refvec{n}),
\end{align}
where the rearrangement theorem is applied. 
If $g_1\ket{\refvec{n},\heli{\lambda}} \propto g_2\ket{\refvec{n},-\heli{\lambda}}$ then the desired $g_0$ is $g_1^{-1}g_2$. 
Therefore, $\mathcal{S}(\refvec{n})$ can be further decomposed into several $G$-invariant subspace as follows: 
\begin{align}
    \mathcal{S}(\refvec{n}) = 
    \begin{cases}
        \sum\limits_{\heli{\lambda}=0}^s \bigoplus \mathcal{S}_{\heli{\lambda}}(\refvec{n}) & \text{if}\,\, \LG(\refvec{n})^- \neq \O
        \\
        \sum\limits_{\heli{\lambda}=-s}^s \bigoplus \mathcal{S}_{\heli{\lambda}}(\refvec{n}) & \text{otherwise}
    \end{cases}.
\end{align}
This allows us to focus on analyzing $\mathcal{S}_{\heli{\lambda}}(\refvec{n})$ instead of the larger one $\mathcal{S}(\refvec{n})$. For a two-particle system at rest, all seven patterns of reference $\refvec{n}$, except for $(abc)$, have a non-empty $\LG(\refvec{n})^-$. 
Given a specific $\heli{\lambda}$, the structure of $\mathcal{S}_{\heli{\lambda}}(\refvec{n})$ is similar to the $\mathcal{S}(\refvec{n})$ for two spinless particles discussed in the subsection~\ref{subsec:rest,2spinless}. Consequently, the previous results can be directly applied here.

To proceed, we turn to $I$-matrix, $I^\Gamma_{\heli{\lambda}}(\refvec{n})$, defined as
\begin{align}\label{eq:Imat:spin0spin1}
    I^\Gamma_{\heli{\lambda}}(\refvec{n}) =  \sum\limits_{g\in\LG(\refvec{n})}\bar{D}^{\Gamma}(g)\bra{\refvec{n},\heli{\lambda}}g\ket{\refvec{n},\heli{\lambda}} =
    \begin{cases}
    C^\Gamma_{\refvec{n}}\sum\limits_{g\in \LG(\refvec{n})^+} \bar{D}^{\Gamma}(g)  & \text{if}\quad \heli{\lambda}=0 
    \\
    \sum\limits_{g\in \LG(\refvec{n})^+} e^{-i\heli{\lambda}\varphi_w(\refvec{n},g)} \bar{D}^{\Gamma}(g) & \text{if}\quad \heli{\lambda}\neq0 
    \end{cases},
\end{align}
where $C^\Gamma_{\refvec{n}}$ is identity matrix $\mathbbm{1}$ if $\LG(\refvec{n})^-$ is empty. Otherwise, $C^\Gamma_{\refvec{n}}=\mathbbm{1}+\eta(-1)^s \bar{D}^{\Gamma}(g_0)$ with arbitrary element $g_0\in\LG(\refvec{n})^-$. 
Besides, since any element $g\in\LG(\refvec{n})^+$ should be a rotation around $\refvec{n}$ with a certain angle $\omega$, there is $\varphi_w(\refvec{n},g) = \omega$. 

$I^\Gamma_{\heli{\lambda}}(\refvec{n})$ can also be proved to be idempotent up to a scaling factor. Inserting the closure relation $1=\sum_{\vec{n},\heli{\lambda}}\ket{\vec{n},\heli{\lambda}}\bra{\vec{n},\heli{\lambda}}$, we can derive the factorization
\begin{align*}
    \bra{\refvec{n},\heli{\lambda}} g h \ket{\refvec{n},\heli{\lambda}} = \bra{\refvec{n},\heli{\lambda}} g \ket{\refvec{n},\heli{\lambda}} \bra{\refvec{n},\heli{\lambda}} h \ket{\refvec{n},\heli{\lambda}},\quad \forall g,h \in \LG(\refvec{n};\heli{\lambda}) \,,
\end{align*}
where $\LG(\refvec{n};\heli{\lambda}) \equiv \LG(\refvec{n})^+ $ if $\heli{\lambda}\neq 0$ and $\LG(\refvec{n})$ if $\heli{\lambda}=0$. With the help of this factorization, the idempotence can be proved as similar as that in the subsection~\ref{subsec:rest,2spinless}. 
Accordingly, the nonzero eigenvalues of $I^{\Gamma}_{\heli{\lambda}}(\refvec{n})$ can only be $\abs{\LG(\refvec{n};\heli{\lambda})}$. 
Consequently, the orthonormal basis furnishing the $\Gamma$ irrep of $\mathcal{S}_{\heli{\lambda}}(\refvec{n})$ is given by 
\begin{align}\label{eq:irrep:basis:spin0spin1}
\ket{\refvec{n},\heli{\lambda},\Gamma,r,\mu} =  N^\Gamma_{\refvec{n},\heli{\lambda}} \sum\limits_{\nu} [c^{\Gamma,r}_{\refvec{n},\heli{\lambda}}]_{\nu}P^\Gamma_{\mu\nu}\ket{\refvec{n},\heli{\lambda}},\quad \abs{\refvec{n}}\neq0,
\end{align}
where the normalization factor $N^\Gamma_{\refvec{n},\heli{\lambda}}$ is $\sqrt{\frac{\abs{G}}{|\LG(\refvec{n};\heli{\lambda})| \dim\Gamma}}$. $[c^{\Gamma,r}_{\refvec{n},\heli{\lambda}}]$ continues to denote the orthonormal eigenvectors with nonzero eigenvalues of $I^\Gamma_{\heli{\lambda}}(\refvec{n})$. 

It is important to note that the helicity state is ill-defined when $\abs{\refvec{n}}=0$. In the case the irrep basis is constructed directly: 
\begin{align}
    \ket{\vec{0},\Gamma,r,\mu} = \sum\limits_{\lambda}[C^{J=s}]_{\Gamma r\mu,\,\lambda}\ket{\vec{0},\lambda},
\end{align}
where $[C^{J=s}]$ is the coefficient in Eq.~(\ref{eq:pwmethod:basis1}).

\subsection{$|\vec{P}|=0\,,\, s_1=s=\mathbb{N}+\frac{1}{2},s_2=0$}

We refer to the system as $N\pi$-system. 
The discussion on the structure of Hilbert space in the previous subsection~\ref{subsec:rest,spin0spin1} is also applicable to $N\pi$-system. 
However, a nontrivial issue is that the symmetry group $G$ is now the double cover of $O_h$, denoted as $\dc{O_h}$.\footnote{For any group $G$, we denote $\dc{G}$ as its double cover.}
For each $g\in O_h$ there exists a counterpart $\bar{g}\in\dc{O_h}$. \footnote{To be precise, in this subsection, $g\in O_h$ or something similar means its parameterization $(\vec{n},\omega)$ lies in the parameter space of $O_h$, say, $-\pi\leq \omega <\pi$. See Appendix.~\ref{append:convention}. Note that the set $\{ g\in \dc{O_h} |\,g\in O_h\}$ do not form a subgroup since they do not satisfy the closure relation as the multiplication rule has changed compared with that in $O_h$ group.}

For a fermionic system, there holds the relation $\bar{g} \ket{\vec{n},\heli{\lambda}} = - g \ket{\vec{n},\heli{\lambda}}$. 
Besides, the phase factor arising from the space inversion should be taken care of:
\begin{align}
    \mathbb{P}\ket{\vec{n},\heli{\lambda}} = \eta e^{\mp i\pi s}\ket{-\vec{n},-\heli{\lambda}},\label{eq:pm}
\end{align}
where ``$-$" sign corresponds to $-\pi \leq \phi_{\vec{n}}<0$ and``$+$" sign corresponds to $0\leq \phi_{\vec{n}}<\pi$.
In our convention of the reference momentum, the sign is always positive.
$I$-matrix now reads
\begin{align}\label{eq:Imat:spin0spin0.5}
    \dc{I^\Gamma_{\heli{\lambda}}(\refvec{n})} &= \sum\limits_{g \in \dc{O_h}} \bra{\refvec{n},\heli{\lambda}} g \ket{\refvec{n},\heli{\lambda}} \bar{D}^{\Gamma}(g)
    \\
    &= \sum\limits_{g \in \dc{O}} \bra{\refvec{n},\heli{\lambda}} g \ket{\refvec{n},\heli{\lambda}} \bar{D}^{\Gamma}(g)
    \\
    &= \sum\limits_{g \in {O}} \bra{\refvec{n},\heli{\lambda}} g \ket{\refvec{n},\heli{\lambda}} \left( \bar{D}^{\Gamma}(g) - \bar{D}^{\Gamma}(\bar{g}) \right),
\end{align}
where in the second equation $\dc{O_h}$ is replaced by $\dc{O}$ because $\heli{\lambda}\neq0$. 
As a cross check, notice that $\dc{I^\Gamma_{\heli{\lambda}}(\refvec{n})}\equiv0$ for $\Gamma= A_1^\pm,A_2^\pm,E^\pm,T_1^\pm,T_2^\pm$ since $D^\Gamma(g) = D^\Gamma(\bar{g})$. 
This is expected since these irreps only couple to integer spins. 
However, for $\Gamma=G_1^\pm,G_2^\pm,H^\pm$, there is $D^\Gamma(g)=-D^\Gamma(\bar{g})$. 
Explicitly, for these irreps there is 
\begin{align}\label{eq:Imat:spin0spin0.5:2}
\dc{I^\Gamma_{\heli{\lambda}}}(\refvec{n}) = 2 \sum\limits_{g \in {O}} \bra{\refvec{n},\heli{\lambda}} g \ket{\refvec{n},\heli{\lambda}}  \bar{D}^{\Gamma}(g) = 2 \sum\limits_{g \in H(\refvec{n})}  e^{-i\heli{\lambda}\varphi_w(\refvec{n},g)} \bar{D}^{\Gamma}(g),
\end{align}
where $H(\refvec{n})\subset O$ is the set consisting of the elements that remains $\refvec{n}$ unchanged.
%
The expression is identical to the Eq.~(\ref{eq:Imat:spin0spin1}), except for a factor $2$ arising from the double cover. 

The $I$-matrix is still idempotent, specifically, $(\dc{I})^2=|\LG(\refvec{n};\heli{\lambda})|(\dc{I})$. 
The expression for the irrep basis with $\abs{\refvec{n}}\neq 0$ are exactly the same as in Eq.~(\ref{eq:irrep:basis:spin0spin1}), with noting that $G$ and $\LG(\refvec{n})$ are $\dc{O}_h$ and its little group, respectively. 
Besides, Eq.~(\ref{eq:Imat:spin0spin0.5:2}) is independent of the parity of irrep, implying that $\dc{I}$ and its eigenvectors $[c^{\Gamma,r}_{\refvec{n},\heli{\lambda}}]$ are identical for $\Gamma^\pm$.

\subsection{$|\vec{P}|=0,\,s_1\neq0,\,s_2\neq0$}
The previous results can be easily extended into the system consisting of two particles both with nonzero spin. 
The helicity state is now defined as
\begin{align}
    \ket{\vec{n},\heli{\lambda_1},\heli{\lambda_2}} := \sum\limits_{\lambda_1,\lambda_2} D^{s_1}_{\lambda_1,\heli{\lambda_1}}(R_{\text{st}}(\vec{n})) D^{s_2}_{\lambda_2,-\heli{\lambda_2}}(R_{\text{st}}(\vec{n})) \ket{\vec{n},\lambda_1,\lambda_2}.
\end{align}
Here, the additional minus sign in $D^{s_2}_{\lambda_2,-\heli{\lambda_2}}$ arises because the momentum of the second particle is opposite to that of the first. 
For any $g\in O_h^+$ and space inversion $\mathbb{P}$, there are
\begin{align}
    g \ket{\vec{n},\heli{\lambda_1},\heli{\lambda_2}} &= e^{-i\left(\heli{\lambda_1}-\heli{\lambda_2}\right)\varphi_w\left(\vec{n},g\right)} \ket{g\vec{n},\heli{\lambda_1},\heli{\lambda_2}} , \quad\forall g\in O_h^+,
    \\
    \mathbb{P}\ket{\vec{n},\heli{\lambda_1},\heli{\lambda_2}} &= \eta e^{\mp i \pi (s_1+s_2)} \ket{-\vec{n},-\heli{\lambda_1},-\heli{\lambda_2}}.
\end{align}
where the sign convention is the same as that in Eq.~(\ref{eq:pm}). 

The space $\mathcal{S}(\refvec{n})=\mathrm{span}\{\ket{\refvec{n},\lambda_1,\lambda_2},\,\lambda_1=-s_1,\cdots,s_1,\lambda_2=-s_2,\cdots,s_2\}$ can be decomposed as
\begin{equation}
    \mathcal{S}(\refvec{n}) = 
    \begin{cases}
        \sum\limits_{\heli{\lambda_1}=-s_1}^{s_1}\sum\limits_{\heli{\lambda_2}=-s_2}^{s_2} \bigoplus \mathcal{S}_{\heli{\lambda_1},\heli{\lambda_2}}(\refvec{n}) & \text{otherwise}  \\
        \sum\limits_{\heli{\lambda_1}\geq0}^{s_1}\sum\limits_{\heli{\lambda_2}=-s_2}^{s_2} \bigoplus \mathcal{S}_{\heli{\lambda_1},\heli{\lambda_2}}(\refvec{n}) = \sum\limits_{\heli{\lambda_1}=-s_1}^{s_1}\sum\limits_{\heli{\lambda_2}\geq0}^{s_2} \bigoplus \mathcal{S}_{\heli{\lambda_1},\heli{\lambda_2}}(\refvec{n}) & \text{if}\,\, \LG(\refvec{n})^- \neq \O 
    \end{cases},
\end{equation}
where 
\begin{align}
    \mathcal{S}_{\heli{\lambda_1},\heli{\lambda_2}}(\refvec{n})=\mathrm{span}\left\{g\ket{\refvec{n},\heli{\lambda_1},\heli{\lambda_2}},\,g\in G\right\}.
\end{align}
$I$-matrix, $I^\Gamma_{\heli{\lambda_1},\heli{\lambda_2}}(\refvec{n})$, now reads 
\begin{align}\label{eq:Imattwospin}
   I^\Gamma_{\heli{\lambda_1},\heli{\lambda_2}}(\refvec{n}) &= \sum\limits_{g\in\LG(\refvec{n})}\bar{D}^{\Gamma}(g)\bra{\refvec{n},\heli{\lambda_1},\heli{\lambda_2}}g\ket{\refvec{n},\heli{\lambda_1},\heli{\lambda_2}} 
   \\
   &=
   \begin{cases}
    C^\Gamma_{\refvec{n}}\sum\limits_{g\in \LG(\refvec{n})^+} \bar{D}^{\Gamma}(g)  & \text{if}\quad \heli{\lambda_1} = \heli{\lambda_2} = 0 
    \\
    \sum\limits_{g\in \LG(\refvec{n})^+} e^{-i(\heli{\lambda_1} - \heli{\lambda_2})\varphi_w(\refvec{n},g) } \bar{D}^{\Gamma}(g) & \text{if}\quad \heli{\lambda_1} + \heli{\lambda_2} \,\,\,\text{is integer}
    \\
    2\sum\limits_{g\in H(\refvec{n})} e^{-i(\heli{\lambda_1} - \heli{\lambda_2})\varphi_w(\refvec{n},g) }  \bar{D}^{\Gamma}(g) & \text{if}\quad \heli{\lambda_1} + \heli{\lambda_2} \,\,\,\text{is half-integer}
    \end{cases},
\end{align}
where $C^\Gamma_{\refvec{n}}$ and $H(\refvec{n})$ are the same as in Eq.~(\ref{eq:Imat:spin0spin1}) and Eq.~(\ref{eq:Imat:spin0spin0.5:2}), respectively.

\subsection{$|\vec{P}|\neq0$}\label{subsec:moving}

For a moving system, the above results can be extended straightforwardly. %
The only primary difference is that the lattice symmetry group now becomes the subgroup of $O_h$ or its double cover. For example, for the total momenta $\frac{\vec{P}L}{2\pi} = \left(0,0,1\right),\,\left(0,1,1\right)$ and $\left(1,1,1\right)$, the respective symmetry groups are $C_{4v},\,C_{2v}$ and $C_{3v}$ (or their double cover for fermionic system).
As mentioned earlier, for a two-particle system at rest, there are seven patterns of reference momentum $\refvec{n}$, fully characterizing the structure of the whole Hilbert space. 
For a moving system, the number of patterns reduce. Below, we list the patterns for the three most common cases: $\frac{\vec{P}L}{2\pi}=(0,0,1),\,(0,1,1)$ and $(1,1,1)$. 

For $\frac{\vec{P}L}{2\pi}=\left(0,0,1\right)$, $G=C_{4v}$ consists of the elements that permute or flip the sign of the first two components of $\refvec{n}$. 
Therefore, there are four patterns as follows ($0<a<b$ and $*$ denotes arbitrary integer),
\begin{itemize}
    \item 
    $(0,0,*)_1$, i.e., the first two components are zero. 
    \item 
    $(0,a,*)_1$,i.e., the first component is zero while the second is not. 
    \item $(a,a,*)_1$, i.e., the first two components are nonzero and identical. 
    \item $(a,b,*)_1$, i.e., the first two components are nonzero and distinct.
\end{itemize}
For $\frac{\vec{P}L}{2\pi} = \left(0,1,1\right)$, $G=C_{2v}$ consists of the elements that permute the last two components or flip the sign of the first component of $\refvec{n}$. 
Therefore, there are four patterns as follows ($a\neq b$ and $*$ denotes the nonzero integer)
\begin{itemize}
    \item $(0,a,a)_2$, i.e., the first component is zero and the last two are identical.
    \item $(0,a,b)_2$, i.e., the first component is zero and the last two are distinct.
    \item 
    $(*,a,a)_2$, i.e., the first component is nonzero and the last two are identical.
    \item $(*,a,b)_2$, i.e., the first component is nonzero and the last two are distinct.
\end{itemize}
For $\frac{\vec{P}L}{2\pi} = \left(1,1,1\right)$, $G=C_{3v}$ consists of the elements that permute the three components of $\refvec{n}$. 
Therefore, there are three patterns as follows ($a\neq b \neq c$)
\begin{itemize}
    \item $(a,a,a)_3$, i.e., all three components are identical. 
    \item $(a,a,b)_3$, i.e., only the first two components are identical. 
    \item $(a,b,c)_3$, i.e., all three components are distinct.
\end{itemize}

\subsection{Matrix element of finite volume Hamiltonian}\label{subsec:matrix element}

Based on the irrep basis provided in the previous subsections, effective Hamiltonian is block-diagonalized into several blocks corresponding to different irreps. 
The matrix elements of effective potential in the $\Gamma$-block for $K\pi$-, $\rho\pi$-, $N\pi$- and their coupled-channel systems are present in this subsection. 
The entries are specified by the reference momentum $\refvec{n}$, the helicity index $\heli{\lambda}$, and the occurrence index $r$ of $\Gamma$ within the space $\mathcal{S}_{\heli{\lambda}}(\refvec{n})$. 
%

For $K\pi$-system, the matrix element reads,
\begin{align}\label{eq:Kpi mat entries}
    V^\Gamma_{\refvec{n}^\prime r^\prime,\refvec{n} r} 
     =& N_{\refvec{n}} N_{\refvec{n}^\prime} \sum\limits_{\nu,\nu^\prime} [c^{r^\prime}_{\refvec{n}^\prime}]^*_{\nu^\prime} [c^r_{\refvec{n}}]_{\nu} \bra{\refvec{n}^\prime} P^{\Gamma\dagger}_{\mu \nu^\prime} \hat{V}^L P^{\Gamma}_{\mu\nu}\ket{\refvec{n}}
    \notag \\
    =& \sqrt{\frac{1}{\abs{\LG(\refvec{n})} \abs{\LG(\refvec{n}^\prime)}}} \sum\limits_{g\in G} \left( [c^{r^\prime}_{\refvec{n}^\prime}]^\dagger\cdot \bar{D}^{\Gamma}(g)\cdot [c^r_{\refvec{n}}]  \right) \bra{\refvec{n}^\prime} \hat{V}^L g\ket{\refvec{n}}
    \notag \\
    =& \sqrt{\frac{1}{\abs{\LG(\refvec{n})}\abs{\LG(\refvec{n}^\prime)}}} \sum\limits_{\substack{g\in \frac{G}{\LG(\refvec{n})}\\ h \in \LG(\refvec{n})}} \left( [c^{r^\prime}_{\refvec{n}^\prime}]^\dagger\cdot \bar{D}^{\Gamma}(g) \cdot \bar{D}^{\Gamma}(h)\cdot [c^r_{\refvec{n}}] \right) \bra{\refvec{n}^\prime} \hat{V}^L gh\ket{\refvec{n}}
    \notag \\
    =&  \frac{N_{\refvec{n}^\prime}}{N_{\refvec{n}}} \sum\limits_{g\in \LC(\refvec{n})} \left( [c^{r^\prime}_{\refvec{n}^\prime}]^\dagger\cdot \bar{D}^{\Gamma}(g)\cdot [c^r_{\refvec{n}}]  \right) \bra{\refvec{n}^\prime}  \hat{V}^L g \ket{\refvec{n}}
    \notag \\
    =&   \frac{N_{\refvec{n}^\prime}}{N_{\refvec{n}}}\sum\limits_{g\in \LC(\refvec{n})} \left( [c^{r^\prime}_{\refvec{n}^\prime}]^\dagger\cdot \bar{D}^{\Gamma_\eta}(g)\cdot [c^r_{\refvec{n}}]  \right) V^L\left(\refvec{n}^\prime,g\refvec{n}\right),
\end{align}
where 
$N_{\refvec{n}}\equiv N^\Gamma_{\refvec{n}}$ and $[c^r_{\refvec{n}}]\equiv[c^{\Gamma,r}_{\refvec{n}}]$. $V^L$ is given in Eq.(\ref{eq:vLL}). $\LC(\refvec{n}):= G/\LG(\refvec{n})$ denotes the set of representative elements for the left coset of $\LG(\refvec{n})$ in $G$. 
In the second equation the rearrangement theorem and Schur orthogonality relation has been applied. 
In the third equation we employ the left-coset decomposition of $G$ regarding $\LG(\refvec{n})$.  
In the fourth equation we make use of the definition of $I$-matrix and the fact that $[c]$ is its eigenvector. 
In the last equation, the term in the parentheses is ``kinematical'' for it is independent of dynamical Hamiltonian and is solely determined by the lattice symmetry group and reference momentum. 
This kind of ``factorization" is universal. 

For $\rho\pi$- and $N\pi$-systems, the matrix elements for $\abs{\refvec{n}^\prime},\abs{\refvec{n}}\neq 0$ are given by 
\begin{align}
V^\Gamma_{\refvec{n}^\prime \heli{\lambda^\prime} r^\prime\,,\,\refvec{n} \heli{\lambda} r} &= \frac{N_{\refvec{n}^\prime,\heli{\lambda}^\prime}}{N_{\refvec{n},\heli{\lambda}}}\sum\limits_{g\in \LC(\refvec{n};\heli{\lambda})} \left( [c^{r^\prime}_{\refvec{n}^\prime,\heli{\lambda}^\prime}]^\dagger\cdot \bar{D}^{\Gamma}(g)\cdot [c^r_{\refvec{n},\heli{\lambda}}] \right) \bra{\refvec{n}^\prime,\heli{\lambda}^\prime} \hat{V}^L g \ket{\refvec{n},\heli{\lambda}} ,
\\
&= \frac{N_{\refvec{n}^\prime,\heli{\lambda}^\prime}}{N_{\refvec{n},\heli{\lambda}}}
\left[
\sum\limits_{g\in \LC(\refvec{n};\heli{\lambda})^+ } \left( [c^{r^\prime}_{\refvec{n}^\prime,\heli{\lambda}^\prime}]^\dagger\cdot \bar{D}^{\Gamma}(g)\cdot [c^r_{\refvec{n},\heli{\lambda}}]  \right) e^{-i\heli{\lambda}\varphi_w(g,\refvec{n})}V^L_{\heli{\lambda}^\prime \heli{\lambda}}\left(\refvec{n}^\prime,g\refvec{n}\right) 
\right. 
\notag \\
& \quad + \left.
\sum\limits_{g\in \LC(\refvec{n};\heli{\lambda})^- } \left( [c^{r^\prime}_{\refvec{n}^\prime,\heli{\lambda}^\prime}]^\dagger\cdot \bar{\tilde{D}}^{\Gamma}(g)\cdot [c^r_{\refvec{n},\heli{\lambda}}]  \right) e^{-i\heli{\lambda}\varphi_w(g,\refvec{n})}V^L_{\heli{\lambda}^\prime, -\heli{\lambda}}\left(\refvec{n}^\prime,g\refvec{n}\right) \right],
\end{align}
where $N_{\refvec{n},\heli{\lambda}}\equiv N^\Gamma_{\refvec{n},\heli{\lambda}}$ and $[c^r_{\refvec{n},r}]\equiv [c^{\Gamma,r}_{\refvec{n},r}]$ are defined in Eq.(\ref{eq:irrep:basis:spin0spin1}). $\tilde{D}^\Gamma= \eta e^{i\pi s} D^\Gamma$ with $s=\frac{1}{2}(N\pi)$ or $1(\rho\pi)$ incorporates the intrinsic parity and the statistical property of the system. 
If either of or both of $\abs{\refvec{n}}$ and $\abs{\refvec{n}^\prime}$ vanish, then 
\begin{align}
   V^\Gamma_{\vec{0} r^\prime,\refvec{n} \heli{\lambda} r}
   =& \frac{\dim(\Gamma)}{\abs{G}}N_{\refvec{n},\heli{\lambda}}\sum\limits_{g\in G} \sum\limits_{\lambda,\nu} [C^{J=s}]^*_{\Gamma r^\prime \mu,\lambda} \bra{\vec{0},\lambda} g \hat{V}^L  \ket{\refvec{n},\heli{\lambda}}\bar{D}_{\mu\nu}^{\Gamma}(g)[c^{r}_{\refvec{n},\heli{\lambda}}]_\nu,
   \notag \\
   =& \frac{\dim(\Gamma)}{\abs{G}}N_{\refvec{n},\heli{\lambda}}\sum\limits_{g\in G}  \sum\limits_{\lambda^\prime,\lambda,\nu} [C^{J=s}]^*_{\Gamma r^\prime \mu,\lambda} \bar{D}^{s}_{\lambda^\prime\lambda}\left(g^{-1}\right)  V^L_{\lambda^\prime\heli{\lambda}}(\vec{0},\refvec{n}) \bar{D}_{\mu\nu}^{\Gamma}(g)[c^{r}_{\refvec{n},\heli{\lambda}}]_\nu,
   \notag \\
   =& \frac{\dim(\Gamma)}{\abs{G}}N_{\refvec{n},\heli{\lambda}}\sum\limits_{\mu^\prime,\lambda^\prime,\nu} \left(\sum\limits_{g\in G}  D^{\Gamma}_{\mu\mu^\prime}\left(g\right) \bar{D}_{\mu\nu}^{\Gamma}(g) \right)[C^{J=s}]^*_{\Gamma r^\prime \mu^\prime,\lambda^\prime} V^L_{\lambda^\prime \heli{\lambda}}(\vec{0},\refvec{n}) [c^{r}_{\refvec{n},\heli{\lambda}}]_\nu,
   \notag \\
   =& N_{\refvec{n},\heli{\lambda}} \sum\limits_{\lambda,\mu} [C^{J=s}]^*_{\Gamma r^\prime \mu,\lambda} [c^{r}_{\refvec{n},\heli{\lambda}}]_\mu V^L_{\lambda\heli{\lambda}}(\vec{0},\refvec{n}),
   \\
   V^\Gamma_{\vec{0} r^\prime,\vec{0} r} =& \sum\limits_{\lambda^\prime,\lambda}  [C^{J=s}]^*_{\Gamma r^\prime \mu,\lambda^\prime}  V^L_{\lambda^\prime\lambda}(\vec{0},\vec{0})  [C^{J=s}]_{\Gamma r \mu,\lambda}.
\end{align}
For the $\rho\pi-K\pi$ coupled-channel system, the off-diagonal matrix elements are given by
\begin{align}
    V^\Gamma_{\refvec{n}^\prime\heli{\lambda}r^\prime,\,\refvec{n} r} &= \frac{N_{\refvec{n}^\prime,\heli{\lambda}}}{N_{\refvec{n}}}\sum\limits_{g\in \LC(\refvec{n})} \left( [c^{r^\prime}_{\refvec{n}^\prime,\heli{\lambda}}]^\dagger\cdot \bar{D}^{\Gamma_\eta}(g)\cdot [c^r_{\refvec{n}}]  \right)  V^L_{\heli{\lambda}}(\refvec{n}^\prime,\refvec{n}), 
    \\
V^\Gamma_{\vec{0} r^\prime,\,\refvec{n} r} &= N_{\refvec{n}}  \sum\limits_{\lambda,\mu} [C^{J=s}]^*_{\Gamma r^\prime \mu,\lambda} [c^{r}_{\refvec{n}}]_\mu  V^L_{\lambda}(\vec{0},\refvec{n}) .
\end{align}
All the helicity index $\heli{\lambda}$ above ranges from $-s$ to $s$ if the reference momentum of is of the pattern $(a,b,c),\,(a,b,*)_1,\,(*,a,b)_2$ and $(a,b,c)_3$. Otherwise, it only takes the positive values.  

Note that the expressions are not symmetric with respect to column and row indices. Since the Hamiltonian is hermitian, the transposed entries can be obtained directly. 
Matrix elements of more complex system are similar and thus not present. 
For readers who want to utilize our results, we provide the necessary ingredients for calculating the ``kinematical'' part in Appendix~\ref{append:matrix elements}.  

In most cases, it is most convenient to parameterize the potential in the rest frame. On the other hand, under Lorentz transformation $U(\Lambda)$, the states for $\rho\pi$-system, for example, behaves as~\cite{weinberg2005quantum}
\begin{align}
U(\Lambda)\,\ket{\vec{p},\lambda;\vec{P}} = 
\left(\frac{\mathcal{E}(\vec{p}_\Lambda;\vec{P}_\Lambda)}{\mathcal{E}(\vec{p};\vec{P})}\right)^\frac{1}{2}
\sum\limits_{\lambda^\prime}D^{s=1}_{\lambda^\prime \lambda}\left(W\left(\Lambda,\vec{p}\right)\right) \ket{\vec{p}_{\Lambda},\lambda^\prime;\vec{P}_{\Lambda}} \,, \label{eq:boost of state}
\end{align}
where $W(\Lambda,\vec{p})$ is the Wigner rotation induced by the Lorentz boost, differing from the rotation in the previous subsections. The explicit expression of $W(\Lambda,\vec{p})$ is given in Ref.~\cite{Jing:2024mag}. $\mathcal{E}(\vec{p};\vec{P}):=\omega_1(\vec{p})\omega_2(\vec{P}-\vec{p})$ with $\omega_i(\vec{p})=\left(\vec{p}^2+m_i^2\right)^\frac{1}{2}$ and $\vec{p}_\Lambda$ is the three-momenta component of $\Lambda p$ with $p^0=\omega(\vec{p})$. This prescription of Lorentz boost applying on the lattice is named as \textbf{LWLY}-scheme and justified in Ref.~\cite{Li:2024zld}. The potential in the moving frame and polarization representation, $V^{L,\vec{d}\neq0}_{\lambda^\prime\lambda}$, can be defined by ~\cite{Li:2021mob}
\begin{align}  V^{L,\vec{d}}_{\lambda^\prime\lambda}(\vec{n}^\prime,\vec{n}) := 
\mathcal{J}^\frac{1}{2}_{\vec{d}}(\vec{n})\mathcal{J}^\frac{1}{2}_{\vec{d}}(\vec{n}^\prime)
V^{L,\vec{0}}_{{\lambda}^{\prime}{\lambda}}(\vec{n}^{\prime*},\vec{n}^*) \,,
\end{align}
where $\mathcal{J}_{\vec{d}}(\vec{n}):=\frac{\mathcal{E}(\vec{p}_{\vec{n}^*};\vec{0})}{\mathcal{E}(\vec{p}_{\vec{n}};\vec{P}_{\vec{d}})}\frac{\omega_1(\vec{p}_{\vec{n}})+\omega_2(\vec{p}_{\vec{d}-\vec{n}})}{\omega_1(\vec{p}_{\vec{n}^*})+\omega_2(\vec{p}_{\vec{n}^*})}$.  
Note that the Wigner rotation in Eq.~(\ref{eq:boost of state}) is not necessary here since it is easy to verify that it is merely an unitary transformation so the eigenvalues are unaffected. $\vec{n}^*$ is the momentum in the rest frame, which is given by
\begin{align}
    \vec{n}^* = \vec{n} + \left[ \frac{\left(\gamma-1\right)\vec{n}\cdot\vec{d}}{\vec{d}^2} - \gamma \frac{\omega_1(\vec{p}_{\vec{n}})}{\omega_1(\vec{p}_{\vec{n}})+\omega_2(\vec{p}_{\vec{d}-\vec{n}})} 
 \right] \vec{d} \,,
\end{align}
with $\gamma=\frac{\omega_1(\vec{p}_{\vec{n}})+\omega_2(\vec{p}_{\vec{d}-\vec{n}})}{\sqrt{(\omega_1(\vec{p}_{\vec{n}})+\omega_2(\vec{p}_{\vec{d}-\vec{n}}))^2+\vec{P}_{\vec{d}}^2}}$. 

In addition, for the $\rho\pi$-system at rest, the polarization index appears only in the 
polarization vector $\epsilon(\vec{n};\lambda)$ in most cases, resulting in the factorization
\[V_{\lambda^\prime\lambda}(\vec{n}^\prime,\vec{n}) \sim f_{\mu\nu}(\vec{n}^\prime,\vec{n}) \epsilon^\mu(\vec{n}^\prime;\lambda^\prime)\epsilon^\nu(\vec{n};\lambda).\] 
Consequently, the potential in terms of helicity basis $\ket{\vec{n},\heli{\lambda}}$ can be directly derived by replacing $\epsilon(\vec{n},\lambda)$ with $\epsilon_H(\vec{n},\heli{\lambda}):= \sum\limits_{\lambda}D^s_{\lambda\heli{\lambda}}(R_{\text{st}}(\vec{n}))\epsilon(\vec{n},\lambda) $. 
The specific expression for $\epsilon$ and $\epsilon_H$ are given in the Appendix~\ref{append:polarization and helicity} for convenience.


\subsection{Incorporation of isospin symmetry}\label{subsec:isospin}

In this subsection, we will discuss whether incorporating isospin symmetry requires modifications to our previous results, especially when the system consists of two particles in a same isospin multiplet.\footnote{If two particles are in different isospin multiplets, no modification is needed.} 
The state with definite isospin $I$ and total momentum $\vec{P}=\frac{2\pi}{L}\vec{d}$ is given by
\begin{align}
    \ket{\vec{n};\vec{d}}^{I,I_z} &= 
    \sum\limits_{i,j} C^{I I_z}_{I_1 i; I_1 j} \ket{\pi^i(\vec{n})\pi^j(\vec{d}-\vec{n})},
\end{align}
where $\pi^i$ denotes the charged state in the isospin-$I_1$ multiplets and $C^{jm}_{j_1m_1;j_2m_2}$ is the Clebsch-Gordon coefficient of $SU(2)$ group. The state satisfies the symmetry relation:
\begin{align}
    \ket{\vec{n};\vec{d}}^{I,I_z} &=  (-1)^{2s}\Delta^I_{I_1} \ket{\vec{d}-\vec{n};\vec{d}}^{I,I_z} \label{eq:symmetry of isospin state}, 
\end{align}
where $s=s_1=s_2$ is the spin of particle so $(-1)^{2s}=1$ for bosonic multiplets and $-1$ for fermionic multiplets. $\Delta^I_{I_1}=\pm1$ is defined by the symmetry property $C^{I,I_z}_{I_1i;I_1j}=\Delta^I_{I_1}C^{I,I_z}_{I_1j;I_1i}$. The state is normalized as
\begin{align}\label{eq:isospin state normalization}
    {}^{I,I_z}\langle \vec{n}^\prime;\vec{d} | \vec{n};\vec{d} \rangle^{I,I_z} = \delta_{\vec{n},\vec{n}^\prime} + (-1)^{2s} \Delta^I_{I_1} \delta_{\vec{n},\vec{d}-\vec{n}^\prime} \,. 
\end{align}
If $(-1)^{2s}\Delta^I_{I_1}=-1$ and $\frac{\vec{d}}{2}\in\mathbb{Z}^3$, then Eq.~(\ref{eq:isospin state normalization}) vanishes for $\vec{n}=\vec{n}^\prime=\frac{\vec{d}}{2}$ and the corresponding state should be removed from the Hilbert space.\footnote{Don't worry if this will result in a ``hole'' in the corresponding $G$-orbit since the orbit for such case is always one-dimensional.} For example, for the physical $\pi\pi$ system with isospin $I=1$ and $\vec{d}=(0,0,2)$, the state with $\vec{n}=(0,0,1)$ vanishes and should be removed. Such exclusion is implemented by default in the subsequent discussions.

%

The $I$-matrix introduced in the previous subsections may need modifications when $\vec{n}$ and $\vec{d}-\vec{n}$ are in the same $G$-orbit. For $|\vec{d}|=0$, this is inevitable. Eq.~(\ref{eq:symmetry of isospin state}) implies that $\ket{\vec{n};\vec{d}=\vec{0}}^{I,I_z}$ is the eigenstate of space inversion $\mathbb{P}$. 
Consequently, the $I$-matrix defined in Eq.~(\ref{eq:Imat:2spinless}), for example, should be revised to 
\begin{align}
    I^\Gamma_{\refvec{n}} = \sum\limits_{g\in G}\bar{D}^{\Gamma}(g)\, {}^{I,I_z}\bra{\refvec{n}} g \ket{\refvec{n}}^{I,I_z} &=  \left(1+\Delta^{I}_{I_1} \mathcal{P}(\Gamma)\right)\sum\limits_{g\in \LG(\refvec{n})} \bar{D}^{\Gamma}(g) ,\,\, &\text{for } \abs{\refvec{n}} \neq 0, \label{eq:isospin:rest}
    \\
    I^\Gamma_{\refvec{n}} & = 2\delta_{\Gamma,A_1^+} |G| ,  & \text{for} \abs{\refvec{n}} = 0,
\end{align}
where $\mathcal{P}(\Gamma)$ is the parity of $\Gamma$-irrep\footnote{The intrinsic parity is always positive for two particles in the same isospin multiplets.} and $\ket{\vec{n}}^{I,I_z}\equiv \ket{\vec{n};\vec{d}=\vec{0}}^{I,I_z}$. The second equation in Eq.~(\ref{eq:isospin:rest}) is because that both of the elements belong to $\LG(\refvec{n})$ as well as $\mathbb{P}\cdot\LG(\refvec{n})$ contribute now. 
Therefore, the incorporation of isopin symmetry eliminates some certain irreps. For the remaining irreps, the basis defined in Eqs.~(\ref{eq:irrep:basis:2spinless}) and (\ref{eq:irrep:basis:spin0spin1}) need an additional normalization factor $\sqrt{\frac{1}{2}}$. The matrix element in Eq.~(\ref{eq:Kpi mat entries}) also needs an addition $\sqrt{\frac{1}{2}}$.
%

For $|\vec{d}|\neq 0$, the necessary and sufficient condition for $\refvec{n}$ and $\vec{d}-\refvec{n}$ are in the same $G$-orbit is $\refvec{n}\cdot\vec{d}=\frac{|\vec{d}|^2}{2}$. 
For $|\vec{d}|^2\leq3$ this can only happen when $\vec{d}=(0,1,1)$ and $\refvec{n}$ is of the pattern $(0,a,b)_2$ and $(*,a,b)_2$ with $a+b=1$. Therefore, when $0<|\vec{d}|^2\leq3$, we pick the $\refvec{n}$ such that $\refvec{n}\cdot \vec{d}>\frac{|\vec{d}|^2}{2}$ and exclude those $\refvec{n}\cdot \vec{d}<\frac{|\vec{d}|^2}{2}$ to avoid double counting. For $\vec{d}=(0,0,1)$ and $(1,1,1)$, the results in the previous subsections can then be directly applied. However, for $\vec{d}=(0,1,1)$, if $\refvec{n}$ is of pattern $(0,a,b)_2$ or $(*,a,b)_2$ with $a+b=1$, $I$-matrix for these $\refvec{n}$ should be revised as,
\begin{align*}
    I^\Gamma_{\refvec{n};\vec{d}} &= \sum\limits_{g\in G \text{ for }\vec{d}}\bar{D}^{\Gamma}(g)\, {}^{I,I_z}\bra{\refvec{n};\vec{d}} g \ket{\refvec{n};\vec{d}}^{I,I_z} = 
   \left(1 + \Delta^I_{I_1} \bar{D}^{\Gamma}(g_{41})\right) \sum\limits_{g\in\LG(\refvec{n}) \text{ for }\vec{d}} \bar{D}^{\Gamma}(g)
\end{align*} 
where $g_{41}$ is the element in $C_{2v}$ that interchanges the last two components of reference momentum, see Appendix~\ref{append:convention}.
As in the rest frame, some irreps are eliminated while the survived irreps need an additional normalization factor $\sqrt{\frac{1}{2}}$. 
For example, in $\mathcal{S}((0,1,0))$, the $B_2$ irrep vanishes when $I=0,2$ while $A_1$ irrep vanishes when $I=1$. 
This is consistent with the results in Refs.~\cite{Dudek:2012gj,Dudek:2012xn}.\footnote{Their $B_1$ is the $B_2$ here.}


\subsection{Discussion with other approaches}\label{subsec:compare}

At the end of this section we compare our approach with other methods. 
As mentioned earlier, the partial wave method in Refs.~\cite{Li:2019qvh,Li:2021mob} is particularly convenient when potential $V$ is defined as a partial wave expansion involving only a few terms. 
In such cases, the basis constructed by the projection operator method, however, may be redundant. 
Specifically, the projection operator method is independent of the specific form of $V$ and the irrep decomposition is made for the whole Hilbert space $\mathcal{H}_L$.  
Consequently, if the interacting subspace $\mathcal{H}_{L,\text{int}}$ is smaller than $\mathcal{H}_L$, the eigenvalues of the Hamiltonian under the basis given in the previous subsections will include non-interacting energies, which should be excluded. 
In contrast, the partial wave method becomes increasingly cumbersome when the expansion involves sufficiently many partial waves. In such cases, the projection operator method is much more advantageous. 
Incidentally, for a three-particle system, the partial wave expansion is not convenient even in the infinite volume\cite{Elster:1998qv}, not to speak of in the finite volume.

In Ref.~\cite{Doring:2018xxx}, the authors construct an orthonormal basis named ``cubic harmonics'' $X^{\Gamma}_l$, the counterpart of spherical harmonics in the finite volume, to perform irrep decomposition of the whole $\mathcal{H}_L$ for a spinless system. 
Their cubic harmonics are actually equivalent to the state $\ket{\refvec{n},l,\Gamma,r,\mu}$ in Sec.~\ref{sec:pw:method}.
As an extension, in Tab.~\ref{tab:partial wave method}, we provide cubic harmonics for the system consisting of a (pseudo)scalar meson and a (axial)vector meson. 

The projection operator $P^\Gamma$ given in Eq.~(\ref{eq:def:of:projector}) is also employed in Ref.~\cite{Meng:2021uhz} for irrep decomposition. 
In that work, the authors firstly vary the states to be projected, to obtain a sufficient number of linear independent states numerically (the occurrence of a given $\Gamma$ can be known in advance from Eq.~(\ref{eq:Fmax})), and then perform orthonormalization by, for example, Gram-Schmidt procedure. The incorporation of isospin symmetry is not discussed there. 
%
%
In contrast, in this paper we construct the orthonormal basis directly and explicitly with the help of $I$-matrix. We are therefore able to provide explicit expressions of the matrix elements of finite volume Hamiltonian,  which differs from Ref.~\cite{Meng:2021uhz} and offers great convenience to readers who wish to apply our results. The further extension to a three-particle system is also straightforward in this framework. Besides, the incorporation of isospin symmetry, which plays a crucial role in reducing the dimension of the Hamiltonian matrix for a three-particle system, can be seamlessly integrated.

\section{Toy Models}\label{sec:toy model}

In this section we apply our method to several toy models and compare the results with standard L\"uscher formula. 
We focus on a $\rho\pi$-system at rest with a effective potential of the following form:
\begin{align}\label{eq:toymodel:V:general}
    V_{\lambda^\prime \lambda}(\vec{p},\vec{k}) = \frac{1}{(2\pi)^3}\left(\vec{\epsilon}^*_{\lambda^\prime}(\vec{0})\cdot\vec{\epsilon}_{\lambda}(\vec{0})\right) V(\vec{p},\vec{k}) = \frac{1}{(2\pi)^3} \delta_{\lambda^\prime \lambda} V(\vec{p},\vec{k}).
\end{align}
Its partial wave expansion is given by 
\begin{align}
    V^{J}_{l^\prime l}(p,k) &= \sum\limits_{\lambda^\prime,\lambda} \sqrt{\frac{(2l+1)4\pi}{2J+1}}\, C_{l^\prime,\lambda-\lambda^\prime;1\lambda^\prime}^{J\lambda}\,C^{J\lambda}_{l0;1\lambda} 
    \int d\hat{\vec{p}}\,  Y^*_{l^\prime,\lambda-\lambda^\prime}(\hat{\vec{p}})\,  V_{\lambda^\prime \lambda}(\vec{p},k\hat{e}_z),
    \\
    & =  \frac{2\pi \sqrt{(2l+1)(2l^\prime+1)}}{ 2J+1}\left(\sum\limits_{\lambda}C_{l^\prime0;1\lambda}^{J\lambda}C_{l0;1\lambda}^{J\lambda}\right) \int \frac{d\cos\theta}{(2\pi)^3} P_l(\cos\theta)V(\vec{p},k\hat{e_z}).
\end{align}
The integration variable is the relative angle between $\vec{p}$ and $\vec{k}$. 
It can be directly verified that $V^J_{l^\prime l}$ is diagonal with respect to $l$ and $l^\prime$ due to the summation in parentheses.
Thus, the partial wave LSE given in Eq.~(\ref{eq:partial wave LSE}) reduces to a single integral equation for any $J$ and $l$ as follows,
\begin{align}\label{eq:partial wave LSE:toymodel}
    T^J_{l}(p,k;E) = V^J_{l}(p,k) +   \int q^2 dq\, V^J_{l}(p,q)\,G(q;E+i\epsilon)\,T^J_{l}(q,k;E).
\end{align}
This indicates that different partial waves for a given $J$ are not dynamically coupled in the infinite volume for this toy model, which is not general.  
%
%
%
For different purposes, we investigate the following four toy models:
\begin{itemize}
    \item Model \RNum{1}: 
    $ V(\vec{p},\vec{k}) = c_S \left(\frac{\Lambda^2}{\vec{k}^2+\Lambda^2}\right)^2 \left(\frac{\Lambda^2}{\vec{p}^2+\Lambda^2}\right)^2 $
    \item Model \RNum{2}: 
    $ 
    V(\vec{p},\vec{k}) = c_D \left(\frac{1}{3}\left(\vec{p}\cdot\vec{k}\right)^2 - \vec{p}^2\,\vec{k}^2\right) \frac{1}{\Lambda^4}  \left(\frac{\Lambda^2}{\vec{k}^2+\Lambda^2}\right)^4 \left(\frac{\Lambda^2}{\vec{p}^2+\Lambda^2}\right)^4
    $
    \item Model \RNum{3}: 
    $
    V(\vec{p},\vec{k}) = c_S \left(\frac{\Lambda^2}{\vec{k}^2+\Lambda^2}\right)^2 \left(\frac{\Lambda^2}{\vec{p}^2+\Lambda^2}\right)^2 + \frac{1}{2} \bigintsss   d\cos\theta \frac{-g \, m_{\text{eff}}^2}{(\vec{p}+\vec{k})^2 + m_{\text{eff}}^2}
    $
    \item Model \RNum{4}: 
    $
    V(\vec{p},\vec{k}) = c_S \left(\frac{\Lambda^2}{\vec{k}^2+\Lambda^2}\right)^2 \left(\frac{\Lambda^2}{\vec{p}^2+\Lambda^2}\right)^2 +  \frac{-g \, m_{\text{eff}}^2}{(\vec{p}+\vec{k})^2 + m_{\text{eff}}^2}
    $
\end{itemize}

In Model \RNum{1} and Model \RNum{2} which feature pure $S$- and $D$-wave short-range potential respectively, we verify the equivalence between FVH method and standard L\"uscher formula in the absence of long-range interaction, and verify the correctness of the formulas we provide. 
The potential in Model \RNum{4}, which is much more non-trivial, consists of a $S$- wave short-range interaction and a long-range Yukawa-type interaction. 
In Model \RNum{3}, we isolate only the $S$ wave component of long-range interaction to disentangle the effects from left-hand cut and high partial waves.
The explicit expression for the $S$-wave component of the long-range interaction is: \footnote{Strictly speaking, $m_{\text{eff}}^2$ should be replaced by $m_{\text{eff}}^2-i\epsilon$ to clarify the Riemann sheets associated to left-hand cut.  
}
\begin{align}\label{eq:Swave:longrange}
    V^S_{\text{long-range}}(\vec{p},\vec{k}) = \frac{1}{(2\pi)^3}\frac{-g m_{\text{eff}}^2}{4pk}\left[ \ln\left(\left(p+k\right)^2+m_{\text{eff}}^2\right) - \ln\left((p-k)^2+m_{\text{eff}}^2\right)
    \right].
\end{align}
with $p=|\vec{p}|$ and $k=|\vec{k}|$.  
For all models, masses of vector particle and pseudoscalar particle are set to $m_1=500$ MeV and $m_2=400$ MeV, respectively. The effective exchanged mass is set to $m_{\text{eff}}=100$ MeV, resulting in a left-hand cut starting at $\left(p_{\text{lhc}}^{\text{OPE}}\right)^2 = -(50\,\text{MeV})^2$ in the complex momentum plane or $E_{\text{lhc}}^{\text{OPE}} - (m_1+m_2) \approx- 5.6$ MeV in the complex energy plane. The regulator mass $\Lambda$ is fixed at $1$ GeV. Throughout this section, we focus on the $T_1^+$-irrep, which is the most nontrivial one for the present models. The size of the finite volume size $L$ is set to $m_{\text{eff}} L = 2$. 

\subsection{Model \RNum{1} and Model \RNum{2}}\label{subsec:toymodel1and2}

\begin{figure}
    \centering
    \subfigure[Model \RNum{1}: repulsive potential: $c_S>0$]{
    \centering
    \includegraphics[width=0.47\linewidth]{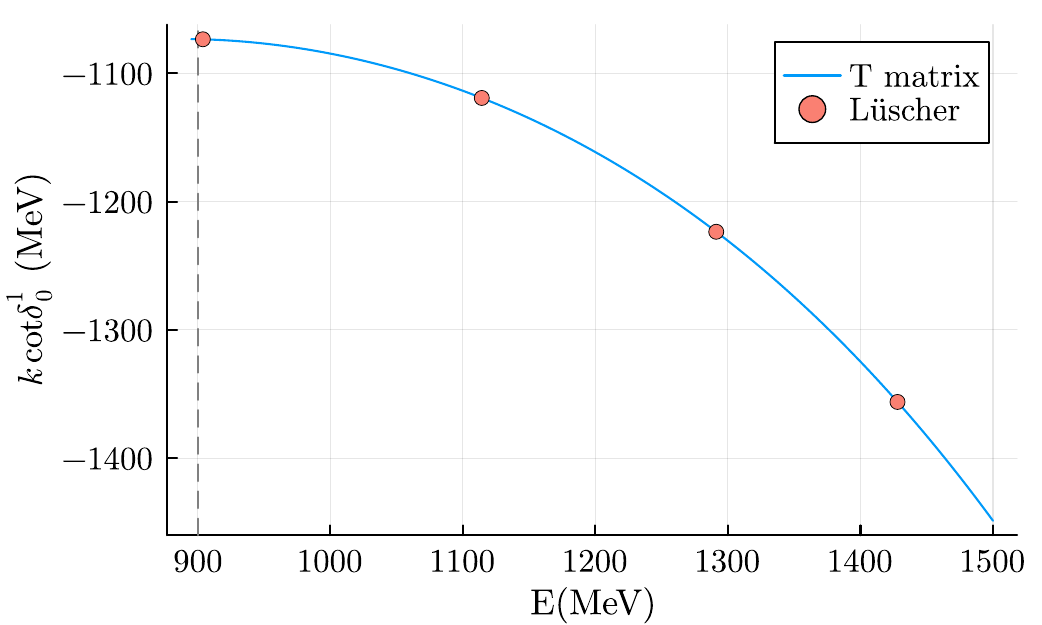}
    }
    \subfigure[Model \RNum{1}: attractive potential: $c_S<0$]{
    \centering
    \includegraphics[width=0.47\linewidth]{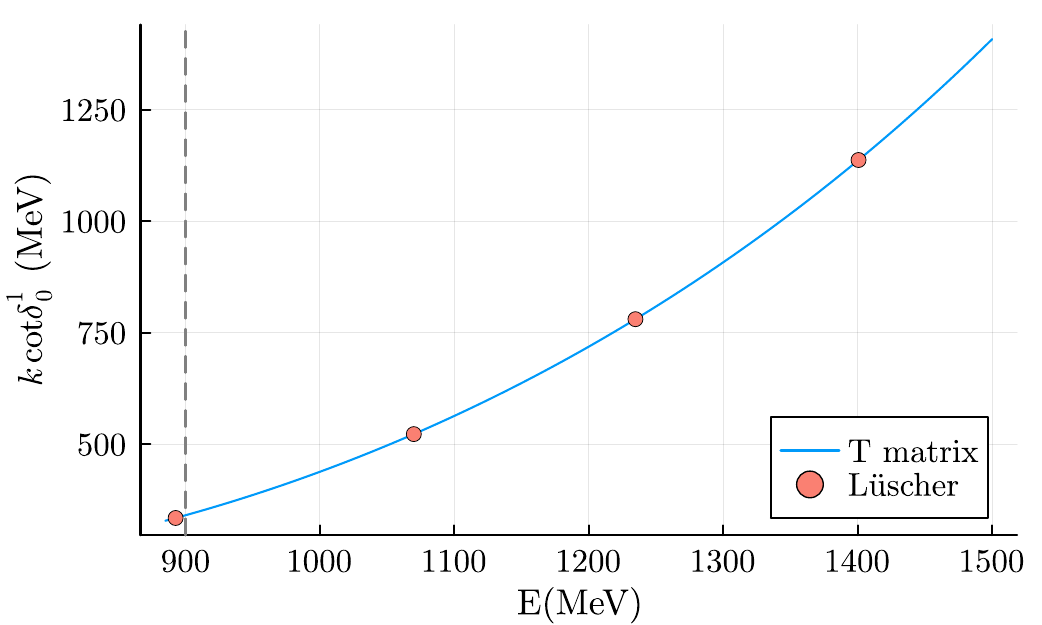}
    }
    \subfigure[Model \RNum{2}: repulsive potential: $c_D>0$]{
    \centering
    \includegraphics[width=0.47\linewidth]{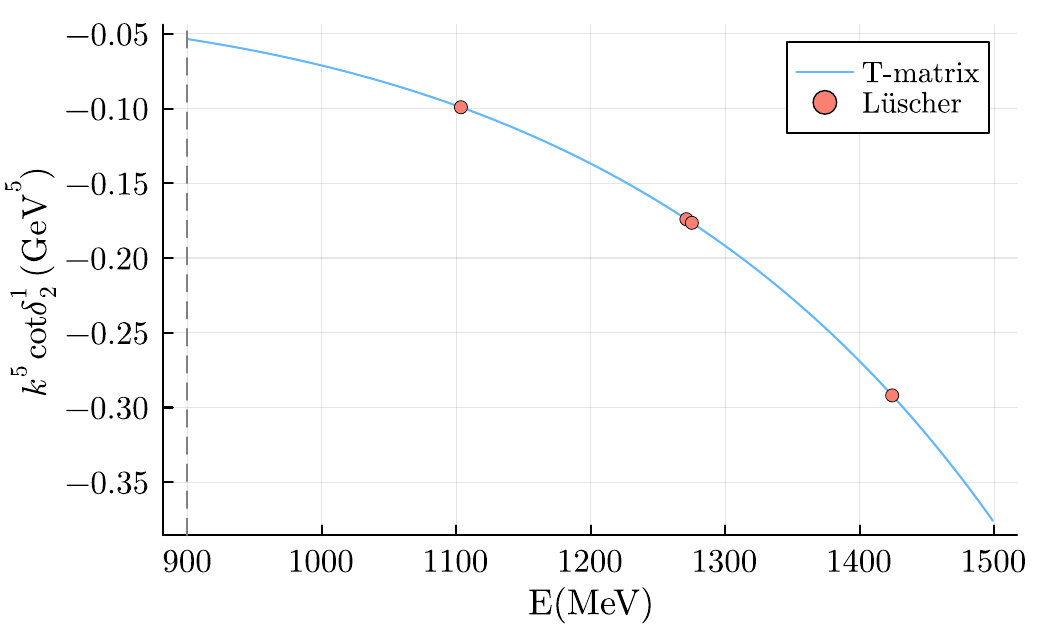}
    }
    \subfigure[Model \RNum{2}: attractive potential: $c_D<0$]{
    \centering
    \includegraphics[width=0.47\linewidth]{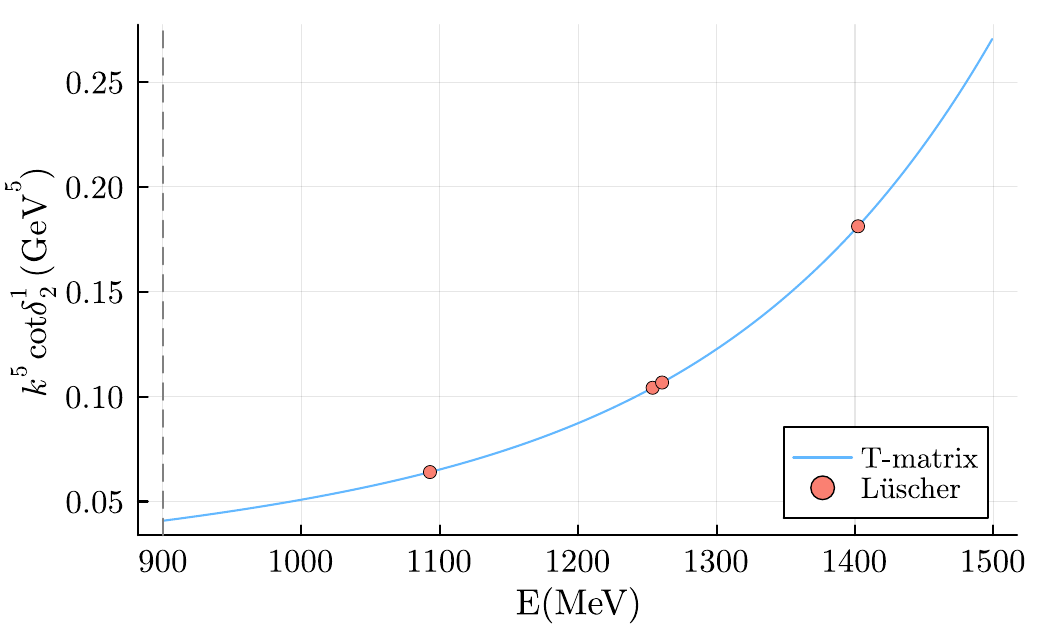}
    }
    \caption{Model \RNum{1} and \RNum{2}: $k_{\text{on}}\cot\delta^1_0$ and $k^5_{\text{on}}\cot\delta^1_2$ determined by standard L\"uscher formula(solid circle) and partial wave LSE(solid line). The dashed line denotes the threshold $m_1+m_2$. The employed parameters are $|c_S|=4\times10^{-5}/\text{MeV}^2$ and $|c_D|=3\times10^{-3}/\text{MeV}^2$. 
    }
   \label{fig:Model1and2}
\end{figure}

Investigating Model \RNum{1} and \RNum{2} allow us to verify the equivalence between FVH method and standard L\"uscher formula. The standard L\"uscher formula reads
\begin{align}\label{eq:swave:luscher}
    1 + i \,\rho\,T_{00}^{J=1} \left( 1 + i \mathcal{M}_{01,01} \right) = 0 \,,
\end{align}
for Model \RNum{1} and 
\begin{align}\label{eq:dwave:lushcer}
        \det\begin{pmatrix}
        1 + i\,\rho\, T^{J=1}_{22} \left(1+i\mathcal{M}_{21,21}\right)   &   -\,\rho\, T^{J=1}_{22}\mathcal{M}_{21,23} \\
        -\,\rho T^{J=3}_{22} \mathcal{M}_{23,21} & 1 + i\,\rho\, T^{J=3}_{22} \left(1+i\mathcal{M}_{23,23}\right) 
    \end{pmatrix} {=} 0 \,,
\end{align}
for Model \RNum{2}, where $\rho = - \pi k_{\text{on}} \frac{\omega_1(k_{\text{on}})\omega_2(k_{\text{on}})}{E_{\text{cm}}}$\footnote{The definition of $\rho$-factor is different from that in Eq.(7.1) of the Ref.~\cite{Woss:2018irj} due to different conventions of T-matrix.}
, $k_{\text{on}} = \frac{1}{2E_{\text{cm}}}\lambda^{\frac{1}{2}}\left(E_{cm},m_1,m_2\right)$ with triangle function $\lambda(a,b,c)=(a^2-(b-c)^2)(a^2-(b+c)^2)$. $\mathcal{M}_{l^\prime J^\prime,\,l J}$ depends on $E$ and encodes the finite volume effect. It is clear that Eq.~(\ref{eq:swave:luscher}) provides a one-to-one relationship between $E$ and $T^{J=1}_{00}$ while the Eq.~(\ref{eq:dwave:lushcer}) involves $E$, $T^{J=1}_{22}$ and $T^{J=3}_{22}$. Detailed expressions of the relevant $\mathcal{M}$ are given in Appendix~\ref{append:Luscher function}. In this toy model the on-shell T-matrix $T^J_{ll}(k_{\text{on}},k_{\text{on}};E)$ can be parameterized as
\begin{align}
    T^{J}_{ll} = \rho^{-1} \exp(i\delta^J_l)\sin\delta^J_l \,.
\end{align}
Given $c_S$ or $c_D$, on the other hand, the finite volume energies $\{E_n\}$ correspond to the eigenvalues of finite volume Hamiltonian, and the infinite volume scattering matrix $[T^{J}_{ll}]_{\text{Ham}}$ can be derived from Eq.~(\ref{eq:partial wave LSE:toymodel}). Our workflow to verify the equivalence of two frameworks is as follows: 
\begin{itemize}
    \item For Model \RNum{1}, we take $\{E_n\}$ as input to obtain $[T^{J=1}_{00}]_{\text{L\"uscher}}$ using L\"uscher formula in Eq.~(\ref{eq:swave:luscher}) and compare with $[T^{J=1}_{00}]_{\text{Ham}}$.
    \item  For Model \RNum{2}, we take $\{E_n\}$ as well as $[T^{J=3}_{22}]_{\text{Ham}}$ as input to obtain $[T^{J=1}_{22}]_{\text{L\"uscher}}$ using L\"uscher formula in Eq.~(\ref{eq:dwave:lushcer}) and compare with $[T^{J=1}_{22}]_{\text{Ham}}$. 
\end{itemize}
In Fig.~\ref{fig:Model1and2}, we present the results for both models. It is evident that $[T^{J}_{ll}]_{\text{L\"uscher}}$ agrees pretty well with $[T^{J}_{ll}]_{\text{Ham}}$, confirming the equivalence of two frameworks in the absence of long-range interaction. Besides, this also validates the formulas provided before.

Besides, it is noticed that for Model \RNum{2} there are two energy levels near the non-interacting energy $E_{\text{free}}(|\vec{n}|^2=2)$, but only one near $E_{\text{free}}(|\vec{n}|^2=1)$ and $E_{\text{free}}(|\vec{n}|^2=3)$, as shown in Fig.~\ref{fig:Model1and2}. We demonstrate that this behavior is attributed to $lS$ coupling and, as mentioned in Sec.~\ref{sec:pw:method}, the partial wave components $(l,J)=(2,1)$ and $(2,3)$ contribute to $T_1^+$ irrep independently for $|\vec{n}|^2=2$ while linearly dependently for $|\vec{n}|^2=1,3$.


%

\subsection{Model \RNum{3}}
\begin{figure}
    \centering
    \subfigure[repulsive potential: $c_S>0,\,g<0$ ]{
    \includegraphics[width=0.47\linewidth]{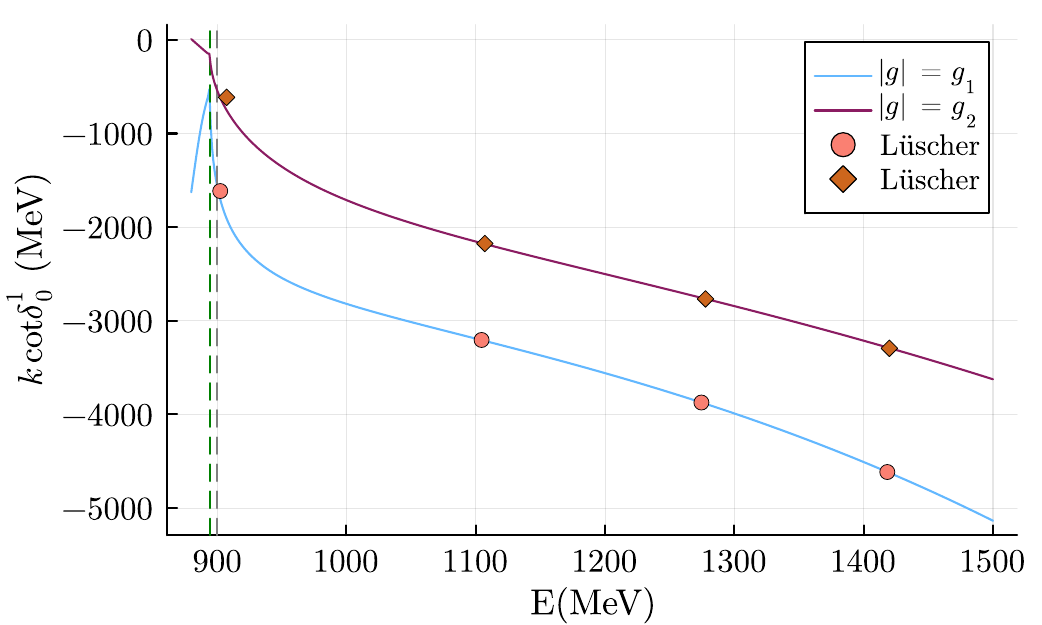}
    }
     \subfigure[attractive potential: $c_S<0,\,g>0$]{
    \includegraphics[width=0.47\linewidth]{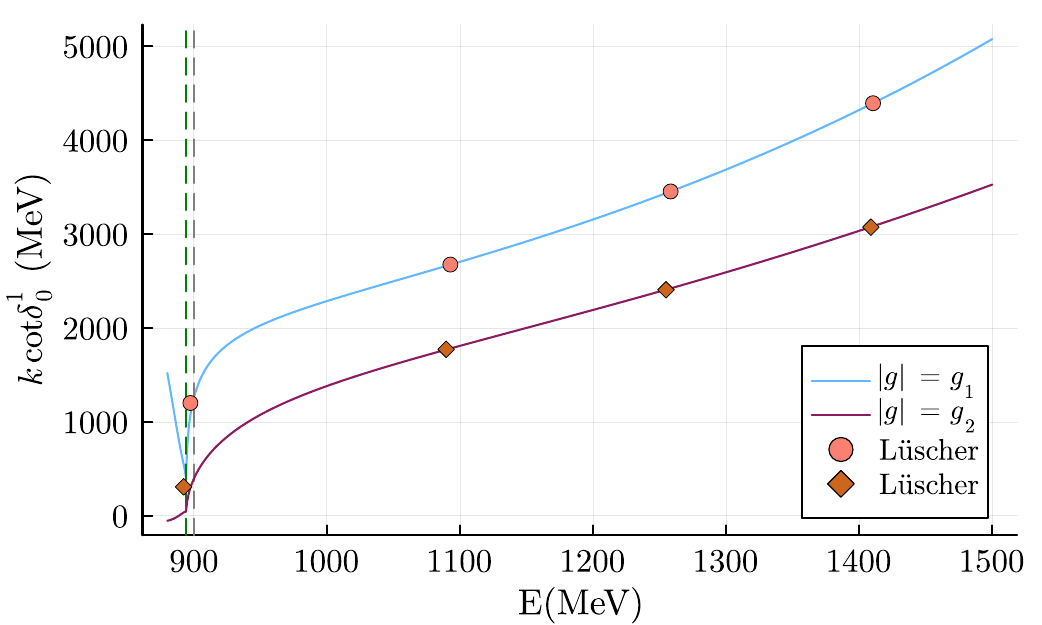}
    }
     \subfigure[detail near the threshold for (a)]{
    \includegraphics[width=0.47\linewidth]{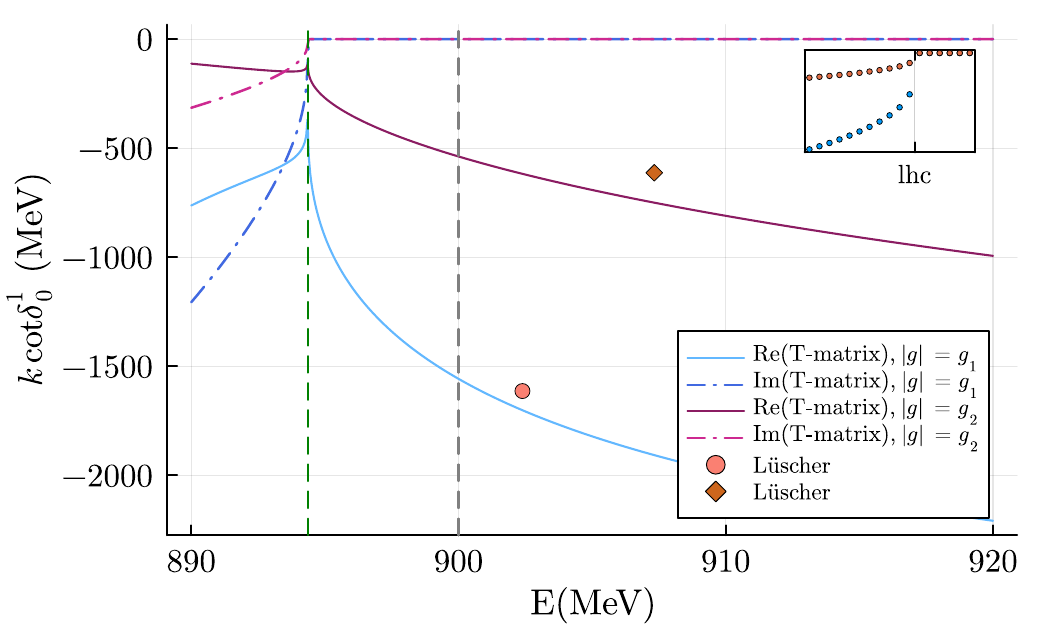}
    }
    \subfigure[detail near the threshold for (b)]{
    \includegraphics[width=0.47\linewidth]{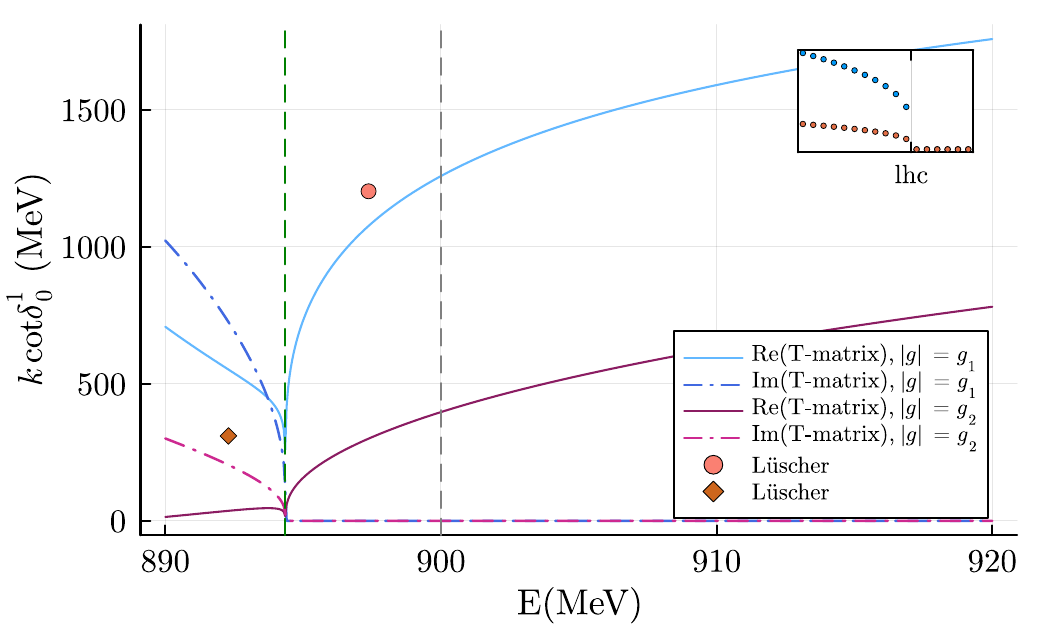}
    }
    \caption{ Model \RNum{3}: $k_{\text{on}}\cot\delta^1_0$ determined by standard L\"uscher formula(solid circle and diamond) and partial wave LSE(solid line). The gray and green dashed lines denote the threshold and endpoint of left-hand cut, respectively. Subfigures (c) and (d) provide detailed views of regions (a) and (b) near the threshold, offering a clearer perspective of the deviations is these regions. Below the left-hand cut,  $k_{\text{on}}\cot\delta^1_0$ determined by partial wave LSE acquires a non-zero imaginary component, depicted by the dot-dashed line. In subfigures (c) and (d) we zoom in the imaginary component near the cut in the upper right corner. The employed parameters are $(c_S,g_1,g_2)=(1,1,5)\times10^{-5}/\text{MeV}^2$. }
    \label{fig:Model3}
\end{figure}

We now introduce the $S$-wave component of the long-range interaction into Model \RNum{1} to investigate the effect arising from the left-hand cut. We continue to take $\{E_n\}$ as input to obtain $[T^{J=1}_{00}]_{\text{L\"uscher}}$ via Eq.~(\ref{eq:swave:luscher}) and compare it to $[T^{J=1}_{00}]_{\text{Ham}}$ derived from Hamiltonian formalism. The results for various values of $c_S$ and $g$ are presented in the Fig.~\ref{fig:Model3}. To clarify the effect from left-hand cut, the sign of $c_S$ and $g$ are chosen such that both of short- and long-range part are attractive or repulsive.\footnote{If one is attractive while the other is repulsive, the interplay of these two parts may lead to a zero of $T$-matrix or a pole of phase shift near the left-hand cut~\cite{Du:2023hlu,Meng:2023bmz}, which is not the intrinsic effect from left-hand cut.} We now make some discussions on the observations.

Firstly, at the the energy levels above the lowest one, $[T^{J=1}_{00}]_{\text{L\"uscher}}$ remains in agreement with the exact $[T^{J=1}_{00}]_{\text{Ham}}$, regardless of whether the interaction is attractive or repulsive. This suggests that the energy levels far from left-hand cut are scarcely affected, and standard L\"uscher formula remains valid in this regime. 

Secondly, for the lowest energy level, $[T^{J=1}_{00}]_{\text{L\"uscher}}$ apparently diverges from the exact $[T^{J=1}_{00}]_{\text{Ham}}$, regardless of whether it is below or above the threshold or left-hand cut. The degree of deviation is relevant to the coupling strength $g$. Therefore, the lowest energy level, which is close to left-hand cut, experience the inescapable effect. The greater the role of long-range interaction play, the more pronounced the deviation from standard L\"uscher formula.

Thirdly, for attractive interactions, the lowest finite volume energy level can lie below the left-hand cut if coupling is sufficiently strong. Under such conditions, exact $k_{\text{on}}\cot\delta^1_0$ acquires a non-zero imaginary part. To be more specific, note that $k_{\text{on}}\cot\delta^1_0 = i k_{\text{on}} + k_{\text{on}}(\rho T^{J=1}_{00})^{-1}$, where $T(E)=V+VG(E+i\epsilon)T(E)$. The unitarity of $S$-matrix ensures that $ik_{\text{on}}$ is always canceled by the corresponding factor in $T$-matrix arising from the propagator $G(E+i\epsilon)$ or unitary cut. Consequently, $k_{\text{on}}\cot\delta^1_0$ remains real above the threshold if the potential is real (or hermitian). However, below left-hand cut, due to the negative argument in log function, $V^{J=1}_{00}(k_{\text{on}},k_{\text{on}})$ introduces an additional non-vanishing imaginary part which is not canceled and ultimately results in complex $k_{\text{on}}\cot\delta^1_0$. 
In contrast, the $k_{\text{on}}\cot\delta^1_0$ derived from standard L\"uscher formula is always real when $q^2=\left(\frac{k_{\text{on}}L}{2\pi}\right)^2\in\mathbb{R}$. Therefore, at the energy below left-hand cut, standard L\"uscher formula suffers a much more severe problem, as it violates the analytical structure of the scattering matrix. 

At this point, we would like to make a comment on the application of the Unitarized Chiral Perturbation Theory(UChPT) in the finite volume proposed in Ref.~\cite{Doring:2011vk}. Within the framework of UChPT in the infinite volume, it is concluded that, under a certain renormalization procedure, only the on-shell point of a potential contributes to LSE. Consequently, LSE degenerates into a algebraic equation. When turning to the finite volume, it is concluded that the finite volume energy levels for a single partial wave and single channel system are determined by the equation:
\begin{align}\label{eq:UChPT:fin}
    V^{-1}(E) - \tilde{G}(E) = 0 ,
\end{align}
where $\tilde{G}(E)\sim\frac{1}{L^3}\sum\limits_{\vec{n}}\frac{\omega_1+\omega_2}{\omega_1\omega_2}\frac{1}{E^2-(\omega_1+\omega_2)^2}$. However, we argue that the Eq.~(\ref{eq:UChPT:fin}) cannot be applicable in the presence of long-range interaction, at least below the left-hand cut, since $\tilde{G}(E)$ is real but $V^{-1}(E)$ becomes complex so Eq.~(\ref{eq:UChPT:fin}) will never yield a real energy solution in that regime.

\begin{figure}
    \centering
    \subfigure[repulsive potential: $c_S>0\,,g<0$]{
    \includegraphics[width=0.47\linewidth]{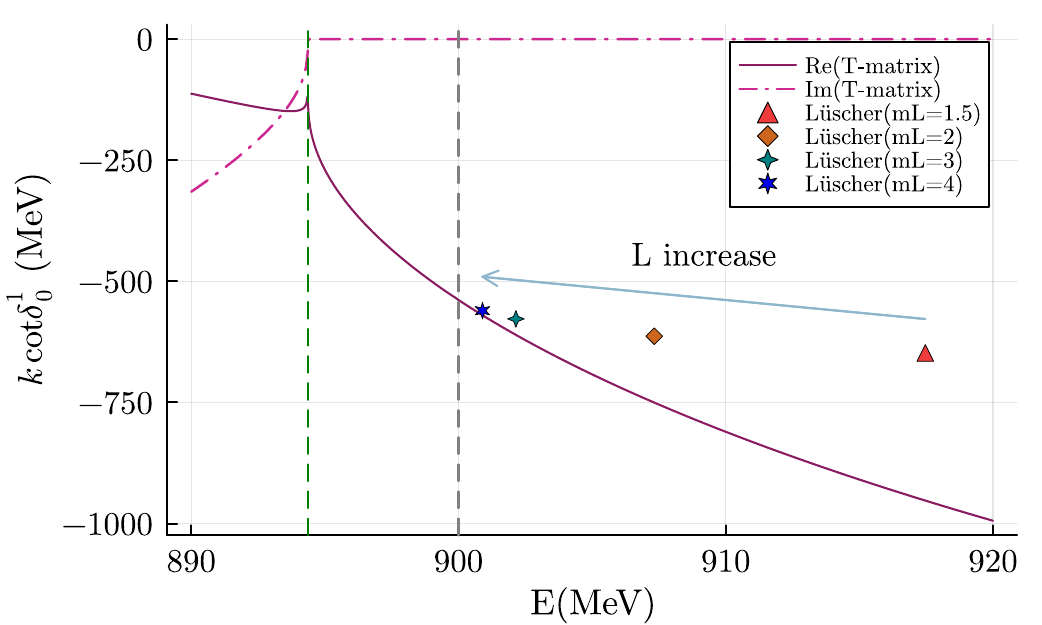}
    }
    \subfigure[attractive potential: $c_S<0\,,g>0$]{
    \includegraphics[width=0.47\linewidth]{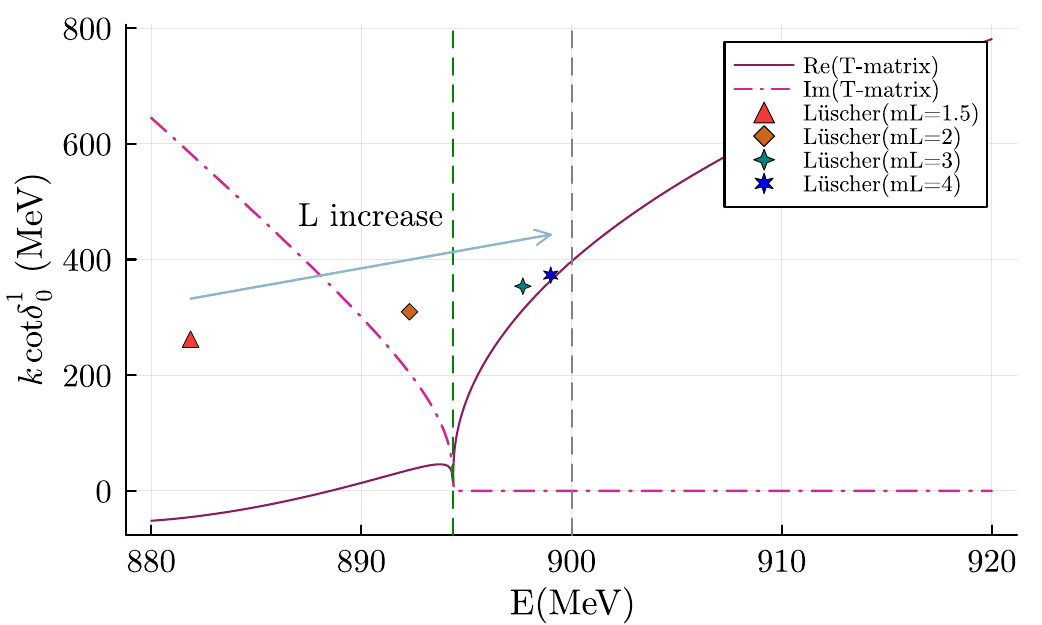}
    }
    \caption{Model \RNum{3}: Same as Figs.~\ref{fig:Model3}(c,d) with $|g|=g_2$. The solid makers denote the $k_{\text{on}}\cot\delta^1_0$ determined by standard L\"uscher formula at several $L$, specifically $m_{\text{eff}}L=1.5,\,2,\,3,\,4$. }
    \label{fig:Model3:largeL}
\end{figure}

At last, in Fig.~\ref{fig:Model3:largeL} we also present $[T^{J=1}_{00}]_{\text{L\"uscher}}$ for the first energy level at several other values of $L$, specifically for $m_{\text{eff}}L=1.5,3,4$. It is evident that $[T^{J=1}_{00}]_{\text{L\"uscher}}$ converges to the exact $[T^{J=1}_{00}]_{\text{Ham}}$ as $L$ increases. 
The effect of left-hand cut can be significant when $m_{\text{eff}}L$ is sufficiently small. In the presence of non-negligible long-range interaction, L\"uscher formula remains applicable only when $L$ is sufficiently large such that $m_{\text{eff}}L\gtrsim4$ and the lowest energy level is above left-hand cut. Unfortunately, this is impractical for the realistic systems where $m_{\text{eff}}$ are quite small. 

To gain further insight into these issues, we now briefly review the derivation of standard ($S$-wave only) L\"uscher formula. We focus initially on the above-threshold regime and subsequently perform analytical continuation. The finite volume energy corresponds to the poles of $T^L(E)$ whose on-shell matrix elements are defined as~\cite{Li:2024zld} 
\begin{align}\label{eq:fin:T}
    T^L(k_{\text{on}},k_{\text{on}};E) = V(k_{\text{on}},k_{\text{on}}) + \sum\limits_{\vec{q}} V(k_{\text{on}},q) G^L(q;E+i\epsilon) T^L(q,k_{\text{on}};E) \,, 
\end{align}
where $q=|\vec{q}|$ and $G^L(q;E)=\left(\frac{2\pi}{L}\right)^3 G(q;E)$. The Neumann series $T^L=\sum\limits_{n=0}^\infty V(G^L V)^n$ serves as a formal solution of Eq.~(\ref{eq:fin:T}). A key step in the derivation involves applying of Poisson formula:
\begin{align}\label{eq:keystep:Luscher}
    & \sum\limits_{\vec{q}} f(q) G^L(q;E+i\epsilon) =
     \sum\limits_{\vec{q}}  \left(f(q)-f(k_{\text{on}})\right)G^L(q;E+i\epsilon)  + f(k_{\text{on}})\sum\limits_{\vec{q}} G^L(q;E+i\epsilon) 
    \notag \\ & =
     \int d\vec{q}\, f(q) G(q;E+i\epsilon) + f(k_{\text{on}}) \left[\left(\frac{2\pi}{L}\right)^3\sum\limits_{\vec{q}}-\int d\vec{q}\right] G(q;E+i\epsilon) + \mathcal{O}(e^{-\kappa L}) \,\, ,
\end{align}
where $f(q)$ is assumed to be analytical in a strip including real $q$-axis. The first term relates to the infinite volume contribution and the summation-minus-integration term encodes the finite volume effect which will ultimately lead to the L\"uscher Zeta function. 
The exponential term, which is typically dropped, is given by the integral as 
\begin{align}\label{eq:integral:for:exponential}
   \mathcal{O}(e^{-\kappa L})  \sim  \sum\limits_{\vec{m}\neq \vec{0},\vec{m}\in \mathbb{Z}^3} \, \int d\vec{q} \,  \frac{f(q)-f(k_{\text{on}})}{E-\omega_1(q)-\omega_2(q)+i\epsilon}\, e^{i\vec{m}\cdot\vec{q} L}
\end{align}
where $\kappa$ is a certain energy scale. For a realistic hadronic system with only short-range interaction, $\kappa \gtrsim m_\pi$.  In order to illustrate the issue arising from long-range interaction, We now switch off the short-range part and apply the Poisson formula to the second term of Neumann series, i.e., the second term on the right-hand side of Eq.~(\ref{eq:fin:T}) with $T^L(q,k_{\text{on}};E)$ replaced by $V(q,k_{\text{on}})$. Then, $f(q;E)=\left[V^S_{\text{long-range}}(q,k_{\text{on}})\right]^2$ (Note that $f(q;E)$ also depends on $E$ implicitly), see Eq.~(\ref{eq:Swave:longrange}). Due to the logarithm function, $f(q;E)$ features four branch cuts at $\pm k_{\text{on}}\pm iz$ with $z\geq m_{\text{eff}}$ on the complex $q$-plane. With the help of Cauchy theorem the integral in Eq.~(\ref{eq:integral:for:exponential}) can be converted into the integration along these branch cuts, which is expected to yield a result of order $\mathcal{O}\left(A(E) e^{-m_{\text{eff}}L}\right)$, where $A(E)$ is a potentially $E$-dependent coefficient. As a consequence, the contribution from such exponential term may be underestimated if $m_{\text{eff}}$ is small enough, which disproves the neglect of it. Moreover, based on the previous observations, the coefficient $A(E)$ is expected to vanish as $E$ moves away from the left-hand cut, justifying the neglect of such term in that regime. In a word, in the above-threshold regime, the whole story is about the potential underestimated exponential term that is typically dropped in the derivation of L\"uscher formula. 

We now turn to the below-threshold regime. As shown in the subsection ~\ref{subsec:toymodel1and2}, the derivation can be safely analytically continued to the below-threshold regime in the absence of long-range interaction. However, this continuation is invalid in the presence of left-hand cut. Note that since $f(q;E)$ relates to the half on-shell elements of $V^S_{\text{long-range}}$, the most left-hand side of Eq.~(\ref{eq:keystep:Luscher}) is a analytical function of $E$ below the threshold along the real axis. In contrast, dropping the exponential term, the summation-minus-integration term in the right-hand side of the equation involves $f(k_{\text{on}}(E);E)$, which introduces a left-hand cut on the real $E$-axis. Therefore, Eq.~(\ref{eq:keystep:Luscher}) cannot be valid below the left-hand cut. (The continuation into the regime above the left-hand cut but below the threshold remains safe.) Obviously, it is the separation of on-shell point that ruins the continuation in the presence of long-range interaction~\cite{Hansen:2024ffk}, which is a necessary step when applying the Poisson formula above the threshold in the derivation.

%
%

\subsection{Model \RNum{4}}
We now turn to the complete model, in which energy levels also receive contributions from higher partial waves. As in the previous subsections, we firstly calculate the energy levels $\{E_n\}$ and the scattering matrix elements  $[T^J_{ll}]_{\text{Ham}}$ using the Hamiltonian method.
Due to the inclusion of high partial waves, additional energy levels emerge compared to Model \RNum{3}. These new levels are initially degenerate with non-interacting energies. As an distinct advantage of the Hamiltonian method, we are able to quantitatively investigate the contributions of different partial waves to $E_n$. Specifically, this can be achieved using the identity $\bra{E_n}\hat{V}_L\ket{E_n} = E_n - \bra{E_n} H_0 \ket{E_n}$ where $\ket{E_n}$ denotes the eigenvector corresponding to $E_n$. On the other hand, $\hat{V}_L$ can also be expressed as a partial-wave expansion as follows\footnote{In this toy model, the expansion is diagonal with respect to $l$. Such property does not generally hold in more realistic problems.}, 
\begin{align}\label{eq:explore high pw to energy}
    \hat{V}_L = \left(\frac{2\pi}{L}\right)^3 \sum\limits_{N^\prime,N} \sum\limits_{l,J} V^{J}_{l l}\left(
    \sum\limits_{M=-J}^J \ket{N^\prime,l,J,M}\bra{N,l,J,M} \right) , 
\end{align}
where
\begin{align}
    \ket{N,l,J,M} = \begin{cases}
        \delta_{l0}\delta_{J1}\sqrt{\frac{1}{4\pi}}\ket{\vec{0},M} \quad&\text{if}\, N=0 \\
        \sum\limits_{m,\lambda}C^{JM}_{lm,1\lambda} \sum\limits_{\vec{n}\,,\vec{n}^2=N} Y_{lm}(\vec{n})\ket{\vec{n},\lambda} \quad&\text{if}\, N\neq0 
    \end{cases} ,
\end{align}
which is similar to that in Eq.~(\ref{eq:pwbasis:fin}). 
Therefore, the contribution from $(l,J)$-component to $\bra{E_n}\hat{V}_L\ket{E_n}$, denoted by $[f^J_{ll}]_n$, is given by
\begin{align}\label{eq:fjll}
    [f^J_{ll}]_n = \left(\frac{2\pi}{L}\right)^3 \sum\limits_{N^\prime,N} V^J_{l,l}\left(\frac{2\pi\sqrt{N^\prime}}{L},\frac{2\pi\sqrt{N}}{L}\right) \sum\limits_{M=-J}^J \langle E_n | N^\prime,l,J,M \rangle \langle N,l,J,M | E_n \rangle .
\end{align}
The $f^J_{ll}$ at few lowest levels are examined. As examples, we present the $f^J_{ll}$ for an attractive interaction in Tabs.~\ref{tab:toymodel:pw:contribution} and ~\ref{tab:toymodel:pw:contribution2}. 

\begin{figure}[htbp]
    \centering
    \subfigure[repulsive potential: $c_S>0,\,g<0$]{
    \centering
    \includegraphics[width=0.47\linewidth]{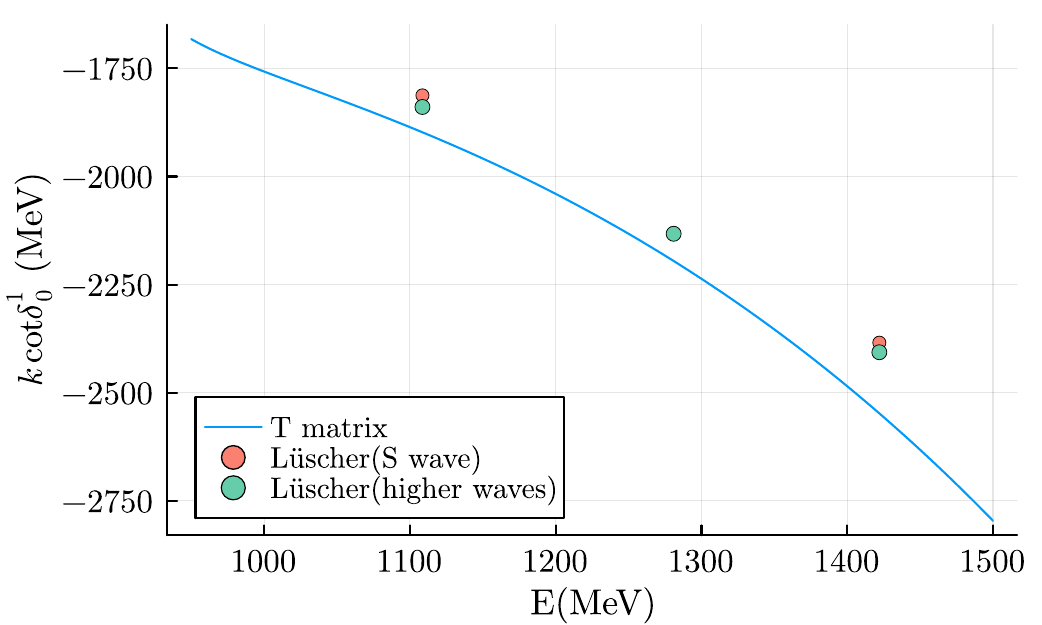}
    }
    \subfigure[attractive potential: $c_S<0,\,g>0$]{
    \centering
    \includegraphics[width=0.47\linewidth]{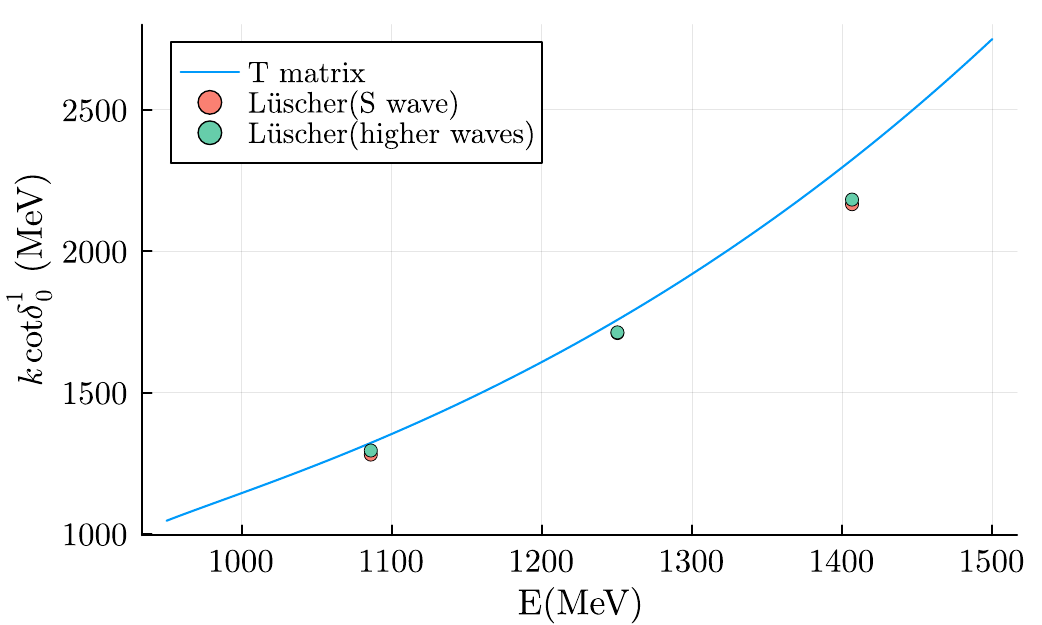}
    }
    \subfigure[repulsive potential: $c_S>0,\,g<0$]{
    \centering
    \includegraphics[width=0.47\linewidth]{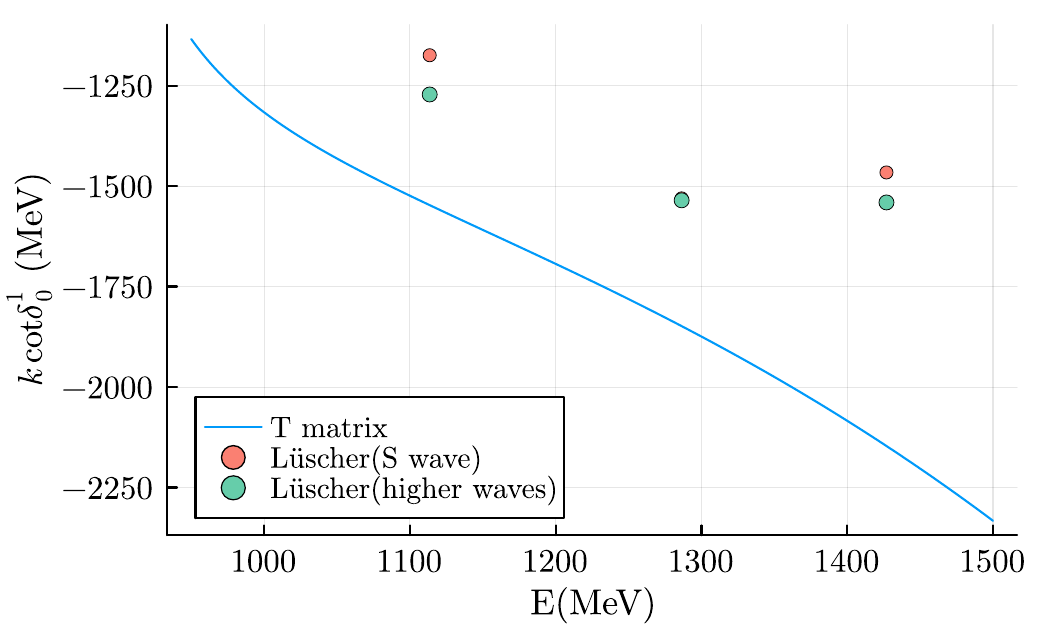}
    }
    \subfigure[attractive potential: $c_S<0,\,g>0$]{
    \centering
    \includegraphics[width=0.47\linewidth]{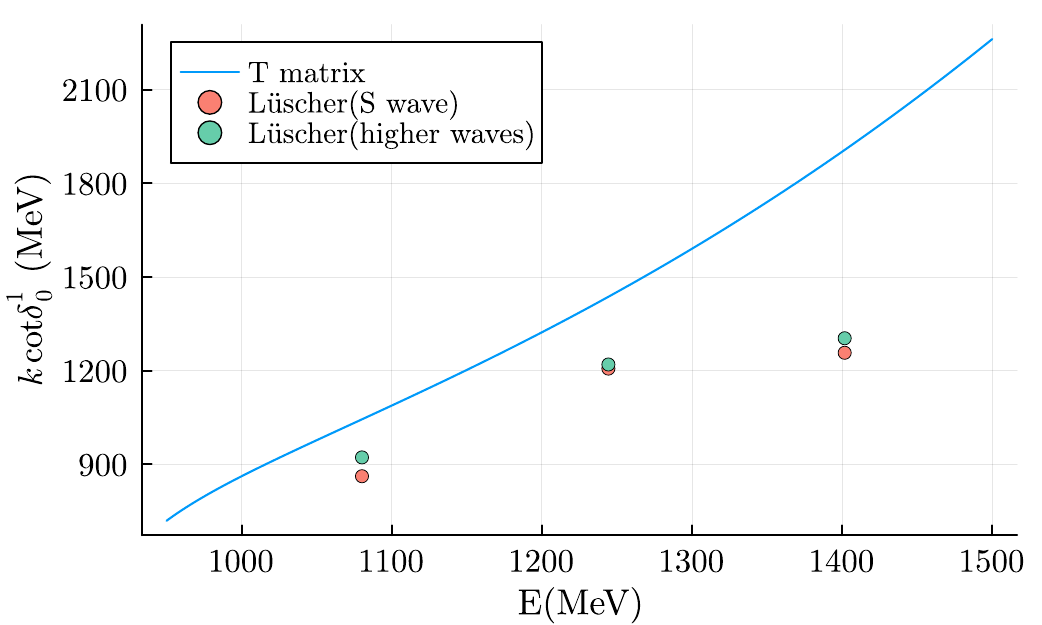}
    }
    \caption{Model \RNum{4}: $k_{\text{on}}\cot\delta^1_0$ determined by standard L\"uscher formula(solid circles) and partial wave LSE(solid line). The orange and green solid circles, which corresponds to the energy levels mainly contributed  by the $S$-wave components of potential, are determined by Eq.(\ref{eq:swave:luscher}) and Eq.(\ref{eq:complete:luscher}), respectively. The employed parameters are $|c_S|=2\times10^{-5}/\text{MeV}^2$ and $|g|=0.5\times10^{-5}/\text{MeV}^2$ for the top row or $|g|=4\times10^{-5}/\text{MeV}^2$ for the bottom row. }
    \label{fig:model4}
\end{figure}

On the other hand, because of the inclusion of high partial waves, the complete standard L\"uscher formula is now given by
\begin{align}\label{eq:complete:luscher}
\det\left[ \delta_{ll^{\prime\prime}}\delta_{JJ^{\prime\prime}}\delta_{rr^{\prime\prime}} + i\rho \sum\limits_{l^\prime,J^\prime,r^\prime} \delta_{JJ^\prime}\delta_{rr^\prime} T^J_{l^\prime,l} \left(\delta_{l^\prime l^{\prime\prime}}\delta_{J^\prime J^{\prime\prime}} \delta_{r^\prime r^{\prime\prime}} + i\mathcal{M}_{l^\prime J^\prime r^\prime,l^{\prime\prime}J^{\prime\prime}r^{\prime\prime}}  \right)   \right] = 0 ,
\end{align}
where $r$ is the occurrence of $\Gamma$-irrep in $D^J$.\footnote{Here, $r\equiv1$ for $T_1^+$ and $J\leq4$.} For the first three energy levels that are 1) far from left-hand cut and 2) mainly contributed by $f^{J=1}_{00}$, we derive the $[T^{J=1}_{00}]_{\text{L\"uscher,S-wave}}$ using Eq.~(\ref{eq:swave:luscher}) and $[T^{J=1}_{00}]_{\text{L\"uscher,higher waves}}$ using Eq.~(\ref{eq:complete:luscher}). The latter incorporates a truncation of $l\leq4,J\leq4$ and takes $[T^{J=1,3}_{22}]_{\text{Ham}},[T^{J=3,4}_{44}]_{\text{Ham}}$ as input. The relevant $\mathcal{M}$-function are provided in Appendix~\ref{append:Luscher function}. By comparing two results, we are able to estimate the contributions of some high partial wave components from another perspective. The results are presented in Fig.~\ref{fig:model4}.  

\begin{table}[htbp]
    \centering
    \begin{tabular}{|c|c|c|c|c|c|c|c|c|}
    $n$ & 1 & $2^*$ & 3 & $4^*$ & 5 & 6 & $7^*$ & 8  \\ \hline
     $\Delta E_{\text{free}}$ &  0.0 & \multicolumn{2}{|c|}{199.1} & \multicolumn{3}{|c|}{366.7} & \multicolumn{2}{|c|}{514.3} \\ \hline
    $\Delta E_n$ &  -3.6 & 186.0 & 198.5 &  350.3  & 366.0 & 366.1 & 506.6 & 513.7  \\ \hline
    $\bra{E_n}\hat{V}_L\ket{E_n}$ & -4.2 & -15.3 &   -0.6 & -17.2 & -0.6 & -0.6 & -7.6 & -0.6  \\
    $f^1_{00}$ &  -4.1 & -14.5 & 0.0 & -16.2  & 0.0 & 0.0 & -6.9 & 0.0   \\
    $f^1_{22}$ &   0.0 & 0.0 & -0.1  & 0.0 & -0.1  & -0.04 & 0.0 & -0.08   \\
    $f^3_{22}$ &   0.0 &  0.0  & -0.2  & 0.0  & -0.1 & -0.06 & 0.0 & -0.06 \\
    $f^3_{44}$ &   -0.5 E-5  & -0.06 & -0.02 & -0.01 & -0.01 & -0.04 & -0.02 & -0.01 \\
    $f^4_{44}$ &   -0.7 E-5  & -0.07 & -0.04 &  -0.01 & -0.01 & -0.1 & -0.03 & -0.02\\
    $f^5_{44}$ &   -0.8 E-5  & -0.09 & -0.03 &  -0.01 & -0.06 & -0.1 & -0.04 & -0.08\\
    $\sum f$ & -4.1  & -14.7 &  -0.39 &  -16.2 & -0.28 & -0.34 & -7.0 & -0.25 \\
    \end{tabular}
    \caption{Model \RNum{4}: Contributions to energy levels from different partial wave components for the attractive potential characterized by  $(c_S,g)=(-2,0.5)\times10^{-5}/\text{MeV}^2$. $\Delta E_{\text{free}}(\Delta E_n)$ denotes the difference between $E_{\text{free}}(E_n)$ and threshold $m_1+m_2$. The quantity $\sum f$ is the sum $\sum_{l\leq4,J\leq5}f^J_{ll}$. During the evaluation of $f^J_{ll}$, we truncate the momentum with $N\leq 10$ in Eq.~(\ref{eq:fjll}), since the higher momentum components of $\ket{E_n}$ are supposed to negligible at low $E_n$. Energy levels marked with an asterisk ($*$) correspond to the solid circles in Fig.~\ref{fig:model4}.  All dimensional quantities in the table are presented in the units of MeV.}
    \label{tab:toymodel:pw:contribution}
\end{table}

\begin{table}[htbp]
    \centering
    \begin{tabular}{|c|c|c|c|c|c|c|c|c|}
    $n$ & 1 & $2^*$ & 3 & $4^*$ & 5 & 6 & $7^*$ & 8  \\ \hline
     $\Delta E_{\text{free}}$ &  0.0 & \multicolumn{2}{|c|}{199.1} & \multicolumn{3}{|c|}{366.7} & \multicolumn{2}{|c|}{514.3} \\ \hline
    $\Delta E_n$ & -8.1 & 180.2 & 194.5 & 344.3 &  361.9 & 362.0 & 501.7 & 509.5 \\  \hline
    $\bra{E_n}\hat{V}_L\ket{E_n}$ & -8.9 & -21.9 & -4.6 & -23.3 & -4.8 & -4.7 & -12.5 & -4.9 \\
    $f^1_{00}$ & -8.8 & -18.1 & 0.0 & -19.3 & 0.0 & 0.0 & -8.2 & 0.0    \\
    $f^1_{22}$ & 0.0 &  0.0 & -0.9 & 0.0 & -0.94 & -0.31 & 0.0 & -0.67 \\
    $f^3_{22}$ & 0.0 & 0.0 & -1.3 & 0.0 & -0.63 & -0.47 & 0.0  & -0.44 \\
    $f^3_{44}$ & -0.5E-3 & -0.5 & -0.1 & -0.1 & -0.04 & -0.34 & -0.2 & -0.06  \\
    $f^4_{44}$ &  -0.7E-3 & -0.6 & -0.3 & -0.1 & -0.10 & -0.86 & -0.2 & -0.14 \\
    $f^5_{44}$ &   -0.9E-3 & -0.7 & -0.2 & -0.1 & -0.44 &  -0.64 & -0.3 & -0.62 \\
    $\sum f$ & -8.8 &  -19.9 &  -2.8 & -19.6 & -2.2 & -2.6 & -8.9 & -1.9 \\
    \end{tabular}
    \caption{ Model \RNum{4}: Same as Tab.~\ref{tab:toymodel:pw:contribution}, but with stronger coupling parameters $(c_S,g)=(-2,4)\times10^{-5}/\text{MeV}^2$.}
    \label{tab:toymodel:pw:contribution2}
\end{table}


Several discussions can be made accordingly. 
Firstly, from Tabs.~\ref{tab:toymodel:pw:contribution} and \ref{tab:toymodel:pw:contribution2}, it is found that either $f_{00}^{J=1}=0,f^{J=1,3}_{22}\neq0$ or vice versa, indicating that $E_n$ would not be simultaneously affected by $V^{J=1}_{00}$ and $V^{J=1,3}_{22}$. 
This result is straightforward to understand when Eq.~(\ref{eq:complete:luscher}) is truncated with $l\leq2$ since in our toy model there is $T^J_{l^\prime l}\propto\delta_{l^\prime l}$ (which does not hold in general cases) and $\mathcal{M}_{2J,01}=\mathcal{M}_{01,2J}=0$ so Eq.~(\ref{eq:complete:luscher}) is factorized into two decoupled equations, namely, Eq.~(\ref{eq:swave:luscher}) and Eq.~(\ref{eq:dwave:lushcer}). We expect such factorization to hold even when no truncation is made. 

\begin{figure}[htbp]
    \centering
    \includegraphics[scale=0.4]{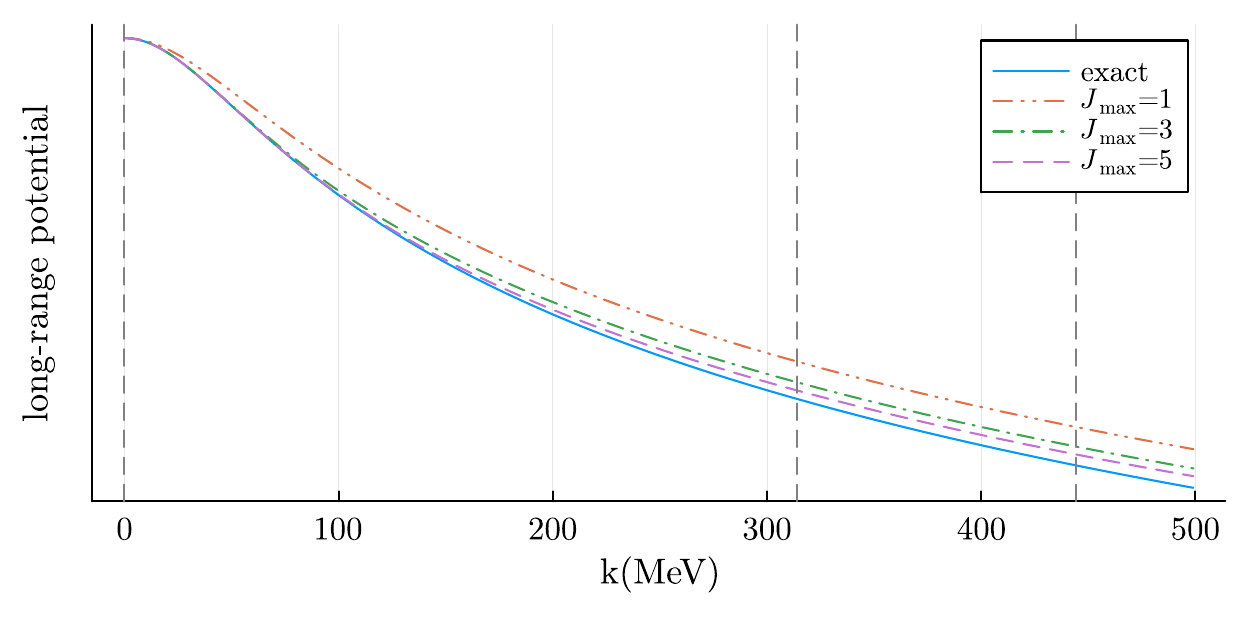}
    \caption{Model \RNum{4}: partial wave expansion of potential $|V_{00}(k\hat{e}_z,k\hat{e_z})|$ with truncation $J\leq J_{max}$ and $c_S=0$. The $y$-axis is set as log scale. The vertical gray dashed lines from left to right denote the lowest three lattice momentum $k=0,\frac{2\pi}{L}$ and $\frac{2\sqrt{2}\pi}{L}$. }
    \label{fig:pwexpand:of:potential}
\end{figure}

Secondly, as shown in Fig.~\ref{fig:model4}, it is evident that $[T^{J=1}_{00}]_{\text{L\"uscher,S-wave}}$ exhibits larger deviations from the $[T^{J=1}_{00}]_{\text{Ham}}$ as $g$ and $E_n$ increases. 
This observation aligns with the results in Tabs.~\ref{tab:toymodel:pw:contribution} and ~\ref{tab:toymodel:pw:contribution2}, where $f^{J}_{ll\geq2}$ becomes increasingly significant for larger $g$ and $E_n$. 
For a deeper insight, we switch off the short-range part and present the partial wave expansion of $V_{00}(k\hat{e_z},k\hat{e_z})$ with truncation $J\leq J_{max}$ in Fig.~\ref{fig:pwexpand:of:potential}. 
Near the lowest lattice momentum mode $k=0$, contributions from the high partial waves are highly suppressed and partial wave expansion converges rapidly, as expected. 
However, even near the second lowest mode, $k=\frac{2\pi}{L}\approx300$ MeV, the contributions from higher partial waves become significant and the expansion converges much more slowly, as also mentioned in Refs.~\cite{Epelbaum:2004fk, Meng:2021uhz}. 
The convergence may be even worse in the finite volume, especially for the energy levels not dominantly contributed by $S$-wave component. 
As an example, for the fifth and sixth energy levels in Tab.~\ref{tab:toymodel:pw:contribution2}, $\sum_{l\leq4,J\leq5} f^J_{ll}$ accounts for only half of the total contributions.

Thirdly, in Fig.~\ref{fig:model4}, $[T^{J=1}_{00}]_{\text{L\"uscher, higher waves}}$ undergoes a noticeable shift compared to $[T^{J=1}_{00}]_{\text{L\"uscher, S wave}}$ at the first and the third energy level, while it remains nearly unchanged at the second energy. 
This can also be explained from Tabs.~\ref{tab:toymodel:pw:contribution} and ~\ref{tab:toymodel:pw:contribution2}. 
It is found that $f^{J=3,4}_{44}$ at the second energy are quite small, rendering the negligible corrections from $(l,J)=(4,3),(4,4)$ to L\"uscher formula. 
To achieve a noticeable shift at this energy, the even higher partial waves $l\geq6$ needs to be included.  
In conclusion, higher partial waves can have sizable contributions, particularly when long-range interaction is sufficiently strong.

\subsection{Conclusion}

While some calculations in this section are specific to the current toy model, most of the conclusions are universal. 
In summary, FVH method is equivalent to standard L\"uscher formula up to a negligible exponential suppressed correction when long-range interaction is absent. 
However, when long-range interaction becomes significant and left-hand cut approaches the threshold, standard L\"uscher formula is in trouble in the energy region near the cut and requires modification. 
This issue becomes particularly severe if a lattice energy locates below the cut, as standard L\"uscher formula violates the analytical structure of scattering matrix in this regime.
Moreover, if long-range interaction is potentially significant, careful consideration of the contributions from high partial waves is suggested when applying L\"uscher formula to analyze high energy levels. Neglecting these contributions can lead to inaccurate results in this case.

In practice, FVH method provides a convenient and efficient framework to bridge the finite volume energy levels and infinite volume scattering observables when long-range potential is significant. The long-range part, which is generally constrained by experimental data, can be fixed (with potential assumptions such as the pion-mass dependence of coupling constant) during fitting the lattice energy levels. The short-range part can be parameterized with a few parameters, whose model-independence is guaranteed by (modified) L\"uscher formula.~\cite{Bubna:2024izx, Raposo:2023oru}


\section{A more realistic application: $\chi_{c1}(3872)$}\label{sec:X3872}

As a more realistic application, we calculate the finite volume energy levels of $I=0$ sector of $D\bar{D}^* + c.c.$\footnote{Hereafter, $c.c.$ is omitted for simplicity.} in the $T_1^+$ irrep, where the signal of the well-known exotic state $\chi_{c1}(3872)$, also known as $X(3872)$, can be observed. 
The discovery of $\chi_{c1}(3872)$ at 2003 starts a new era for hadron physics~\cite{Belle:2003nnu}. 
Since its observation, numerous phenomenological studies have been conducted to explore the nature of $\chi_{c1}(3872)$.
Possible interpretations include a compact tetraquark or, more convincingly, a molecular bound state of $D\bar{D}^*$, mixed with $c\bar{c}$ core known as $\chi_{c1}(2P)$. 
We refer to Refs.~\cite{Guo:2017jvc,Brambilla:2019esw,Chen:2016qju,Esposito:2016noz,Chen:2024eaq} for comprehensive summaries of the relevant phenomenological studies. 

The signal of $\chi_{c1}(3872)$ has also been observed in lattice studies. Refs.~\cite{Li:2024pfg,Prelovsek:2013cra,Lee:2014uta} report the identification of a candidate for $\chi_{c1}(3872)$ on the lattice using $D\bar{D}^*$ and $c\bar{c}$ interpolators.
Furthermore, the coupling of $\chi_{c1}(3872)$ to the $c\bar{c}$ core is suggested to be strong in Ref.~\cite{Padmanath:2015era,Bali:2011rd}. 
More recently, the $D\bar{D}^*(I=0)$ scattering at four different $m_\pi$ has been studied. A bound state corresponding to $\chi_{c1}(3872)$ was reported at the highest $m_\pi$ in the standard L\"uscher formalism~\cite{Li:2024pfg}.

In this section, we calculate the energy levels based on the model built in Ref.~\cite{Wang:2023ovj}, which successfully reproduces the binding energy of $\chi_{c1}(3872)$ as well as the line shape of $T_{cc}$ and $Z_c(3900)$~\cite{Yu:2024sqv}. 
As mentioned earlier, $\chi_{c1}(3872)$ can be interpreted as a mixture of $c\bar{c}$ core and a $D\bar{D}^*$ molecular component. %
The interaction between $\ket{D\bar{D}^*}$ is given by the following potential:\footnote{For simplicity, the polarization or helicity indices are suppressed in the notation.}
\begin{align}
    V^{D\bar{D}^*}(\vec{p},\vec{k}) = \frac{1}{2(2\pi)^3}\left( -3 \bar{V}^u_\pi -  3 \bar{V}^u_\rho - \bar{V}^u_\omega + 3 \bar{V}^t_\rho + \bar{V}^t_\omega \right)  \left(\frac{\Lambda^2}{\Lambda^2 + \vec{k}^2}\right)^2  \left(\frac{\Lambda^2}{\Lambda^2 + \vec{p}^2}\right)^2,
\end{align}
where coefficients are the isospin factor. 
Note that the isospin breaking effect is not accounted for in this model. This is consistent with the current lattice simulation, which assumes $m_u = m_d$. 
Specifically, $\bar{V}^{t,u}_{\pi,\rho,\omega} = \frac{1}{2}\left( V^{t,u}_{\pi,\rho,\omega}(q_{t,u}) + V^{t,u}_{\pi,\rho,\omega}(\tilde{q}_{t,u})\right)$ and
\begin{align}
    V^u_\pi(q) &= -\left( \frac{g}{f_\pi} \right)^2 \frac{\left(q\cdot\epsilon\right)\left(q\cdot\epsilon^*\right)}{q^2-m_\pi^2} 
    \\
    V^u_{\rho,\omega}(q) &= -2 \left(g_v \lambda\right)^2 \frac{\left(q\cdot \epsilon\right)\left(q\cdot \epsilon^*\right)-q^2\left(\epsilon^*\cdot \epsilon\right)}{q^2 - m_{\rho,\omega}^2}
    \\
    V^t_{\rho,\omega}(q) &= \frac{1}{2} \left(g_v\beta\right)^2 \frac{\epsilon\cdot\epsilon^*}{q^2 - m_{\rho,\omega}^2}
\end{align}
where 
\begin{align}
    q_u &= \left(\omega_{D^*}(\vec{k}) - \omega_{D}(\vec{p}),\,\vec{p} + \vec{k} \right) \quad,\quad 
    \tilde{q}_u = \left(\omega_{D^*}(\vec{p}) - \omega_{D}(\vec{k}),\,\vec{p} + \vec{k} \right)
    \\
    q_t &= \left(\omega_{D^*}(\vec{k}) - \omega_{D^*}(\vec{p}),\,\vec{p} + \vec{k} \right) \quad , \quad \tilde{q}_t = \left(\omega_{D}(\vec{p}) - \omega_{D}(\vec{k}),\,\vec{p} + \vec{k} \right)
\end{align}
where $\epsilon$ denotes the polarization or helicity vector given in Appendix~\ref{append:polarization and helicity}. 
The parameters used are: $g=0.57$, $f_\pi=132.0$ MeV, $ g_v = 5.8$, $\beta = 0.687$, $\lambda=0.683$/GeV and $\Lambda=1.0$ GeV. 
All particle masses, except for $m_\pi$, are taken as physical values. 
We set $m_\pi=150$ MeV, which is slightly higher than the physical value to avoid the three-body decay $D^*\to D \pi$. 
Otherwise, the system, strictly speaking, should be treated as a three-particle system. 
Based on the Quark-Pair-Creation(QPC) model, the pure $S$-wave interaction between the $\ket{c\bar{c}}$ and $\ket{D\bar{D}^*}$ is given by 
\begin{align}
     V^{c\bar{c}}(\vec{k}_{D\bar{D}^*}) = \gamma I_{D\bar{D}^*,c\bar{c}}(|\vec{k}_{D\bar{D}^*}|)
\end{align}
where $\gamma=4.69$ represents the amplitude of producing the light quark pair and $I_{D\bar{D}^*,c\bar{c}}$ is the overlap of the meson wave functions. 
The mass of $c\bar{c}$ core is set as $m_{\chi_{c1}(2P)}=3.95$ GeV. 
Further details of the QPC model can be found in Ref.~\cite{Micu:1968mk}.

\begin{figure}
    \centering
    \subfigure[full model]{
    \includegraphics[width=0.47\linewidth]{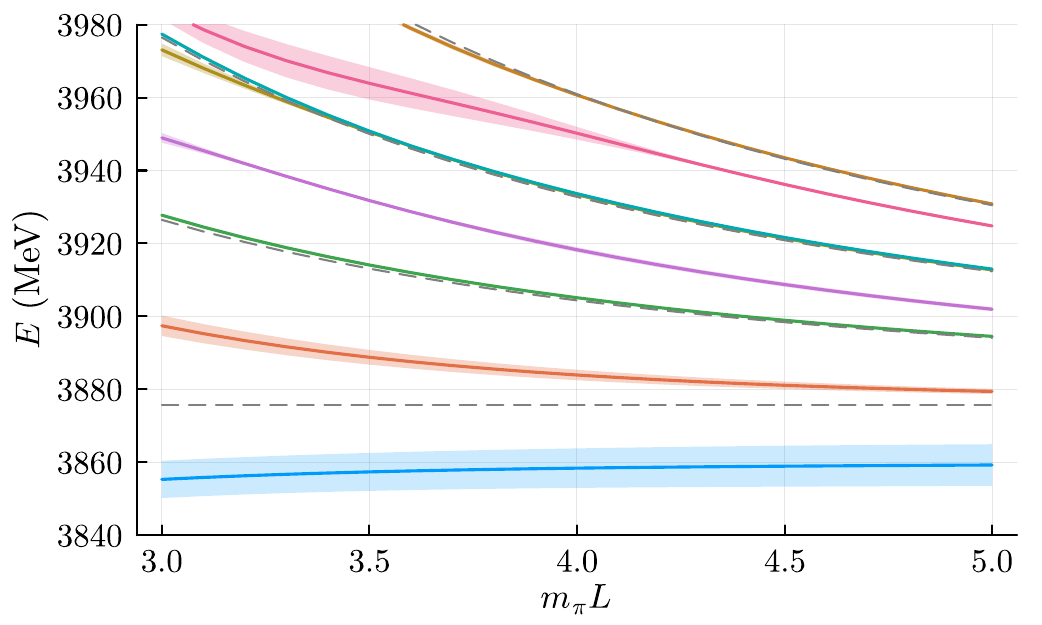}
    }
    \subfigure[model excluding $V^{c\bar{c}}$]{
    \includegraphics[width=0.47\linewidth]{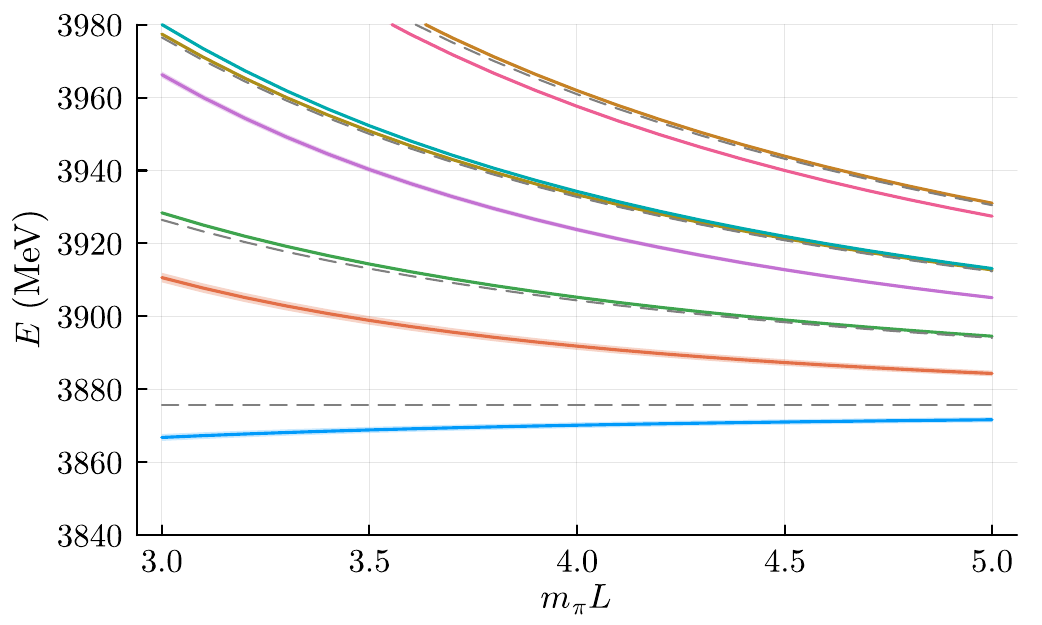}
    }
    \caption{Finite volume energy levels of $\chi_{c1}(3872)$ in $T_1^+$-irrep. The dashed lines denote the non-interacting energy of $D\bar{D}^*$. The shaded bands denote the uncertainty.}
    \label{fig:spectrum:X3872}
\end{figure}

The energy levels in $T_1^+$-irrep calculated using full model and the model excluding $V^{c\bar{c}}$ are presented in the Fig.~\ref{fig:spectrum:X3872}. 
The shaded bands denote the uncertainties estimated by bootstrapping parameter errors in the model.
The uncertainty of $\gamma$ is not explicitly provided in the literature, but is expected to be large. Here, we adopt  $\gamma=4.69$  and manually assign a relative error of $10\%$. 
In both cases, the energy levels are qualitatively consistent with the lattice results reported in %
Refs.~\cite{Li:2024pfg,Prelovsek:2013cra,Lee:2014uta}. 
Specifically speaking, there is a level below the $D\bar{D}^*$ threshold, a level situated between the non-interacting energies $D\bar{D}^*(\abs{\vec{n}}^2=1)$ and $D\bar{D}^*(\abs{\vec{n}}^2=2)$, and a level very close to and slightly above the non-interacting energy $D\bar{D}^*(\abs{\vec{n}}^2=2)$. 
As summarized in the Tab.~\ref{tab:pw:X3872}, the contributions from different partial waves to the energy levels $f^J_{l^\prime,l}$ are analyzed. 
The first and the second energy levels are primarily contributed by the $S$-wave component. The third energy level is predominantly influenced by $V^{J=1,3}_{22}$, reflecting $D$-wave dominance.
Note that although the third energy can be extracted by a $S$-wave interpolator due to the partial wave mixing~\cite{Li:2024pfg}, the $S$-wave L\"ushcer formula cannot be applied to get $S$-wave phase shift at this level.
Furthermore, a energy situated between the non-interacting energies $D\bar{D}^*(\abs{\vec{n}}^2=2) $ and $D\bar{D}^*(\abs{\vec{n}}^2=3)$ is anticipated if more interpolators are employed. 

\begin{table}[]
    \centering
    \begin{tabular}{|c|c|c|c|c|c|c|c|c|c|}
    $m_\pi L$ &  \multicolumn{3}{c|}{3} & \multicolumn{3}{c|}{4} & \multicolumn{3}{c|}{5} \\ \hline
    $n$  &  1 & 2 & 3 &  1 & 2 & 3 &  1 & 2 & 3 \\ \hline
    $\bra{E_n}\hat{V}_L\ket{E_n}$ & -50.80 & -23.03 & 0.74 & -55.24 & -12.98 & 0.57 & -58.06 & -6.46 & 0.33 \\
    $f^1_{00}$($V^{c\bar{c}}$) & -23.88 & -12.83 & -0.20 & -25.87 & -6.88 & -0.10 & -26.76 & -3.32 & -0.04 \\
    $f^1_{00}(V^{D\bar{D}^*})$ &  -12.57 & -2.44 & -0.01 & -13.43 & -1.92 & -0.02 & -13.38 & -1.00 & -0.01 \\
    $f^{1}_{02} = f^1_{20}$ & -3.07 & -0.74 & 0.13 & -1.97 & -0.30 & 0.08 & -1.16 & -0.08 & 0.03\\
    $f^1_{22}$ & 0.01 & 0.00 & 0.73 & 0.02 & 0.00 & 0.40 & 0.01 & 0.00 & 0.23 \\
    $f^3_{22}$ & 0.02 & 0.00 & 0.47 & 0.00 & 0.00 & 0.32 & 0.00 & 0.00 & 0.19 \\
    \end{tabular}
    \caption{Contributions from several lowest partial wave components to the finite volume energy levels of the full model for $\chi_{c1}(3872)$ at $m_\pi L=3,4,5$. $f^1_{00}(V^{c\bar{c}})$ and $f^1_{00}(V^{D\bar{D}^*})$ denote the diagonal $S$-wave contribution from $V^{c\bar{c}}$ and $V^{D\bar{D}^*}$, respectively.}
    \label{tab:pw:X3872}
\end{table}

\begin{figure}
    \centering
    \subfigure[full model]{
    \includegraphics[width=0.47\linewidth]{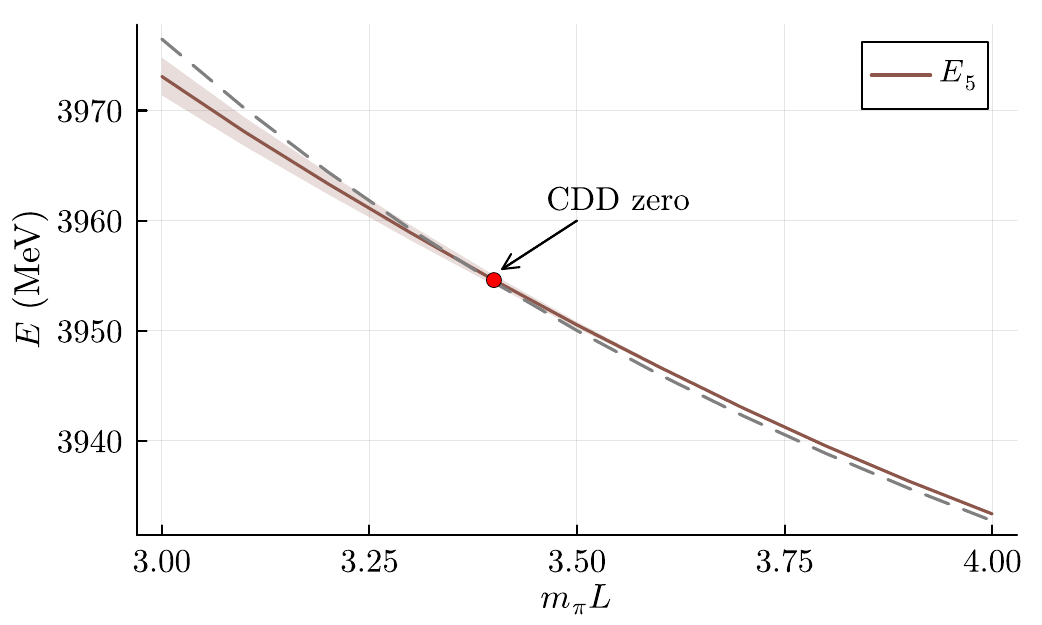}
    }
    \subfigure[model excluding $V^{c\bar{c}}$]{
    \includegraphics[width=0.47\linewidth]{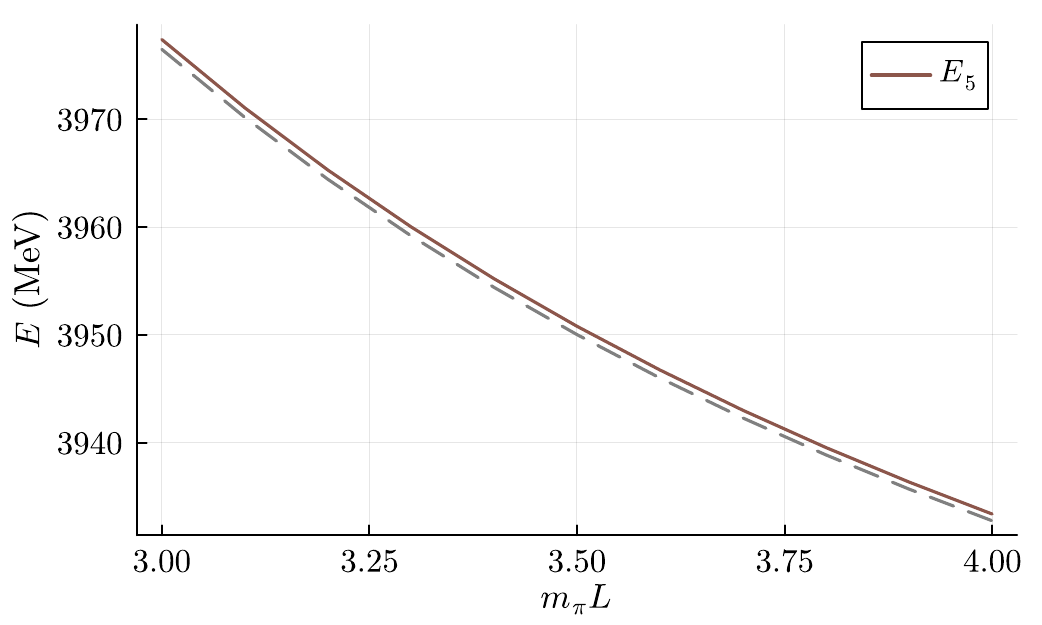}
    }
    \caption{Detail of the fifth energy level in Fig.\ref{fig:spectrum:X3872}. The red circle in the left subfigure indicates the intersection with the non-interacting energy, suggesting the presence of a CDD zero due to the competition between $V^{c\bar{c}}$ and $V^{D\bar{D}^*}$. No intersection is found in the right subfigure.}
    \label{fig:CDD pole:X3872}
\end{figure}

By comparing the qualitative behavior of energy levels as a function of $L$ in the subfigures of Fig.~\ref{fig:spectrum:X3872}, we can discern the difference arising from the $c\bar{c}$ core. 
As depicted by the pink line, the presence of the $c\bar{c}$ core causes an energy level near the $m_{\chi_{c1}(2P)}$ to align closely with a lower non-interacting energy level at small $L$ and transition toward a higher non-interacting level as $L$ increases. 
Furthermore, it is possible for some energy levels to intersect with the non-interacting as shown in Fig. \ref{fig:CDD pole:X3872}. 
The intersection corresponds to Castillejo-Dalitz-Dyson(CDD) zeros arising from the competition between $V^{c\bar{c}}$ and $V^{D\bar{D}^*}$ interactions~\cite{Castillejo:1955ed,Li:2022aru,Yang:2022vdb}.
The observation of these two behaviors will imply the existence of $c\bar{c}$ core. 
However, we acknowledge the significant technical challenges involved on the lattice, as these observations require lattice data of very high precision across multiple values of $L$. 

\section{Summary}

In this paper, we introduce a systematic and sophisticated method for performing irrep decomposition of finite volume Hamiltonian for a two-particle system with general potential and arbitrary spins. This framework is applicable to the system in both of the rest and moving frames. 
Based on the method, we provide the explicit expressions for matrix elements of finite volume effective Hamiltonian, offering practical convenience for readers who wish to apply FVH method to address the left-hand cut issue on the lattice. The method inherently incorporates the isospin symmetry and can be straightforwardly generalized into a three-particle system in the foreseeable future. 

Utilizing this method, we investigate several toy models to compare the FVH method with standard L\"uscher formula. 
The two frameworks agree pretty well when long-range interaction is absent. 
However, in the presence of long-range interaction, standard L\"uscher formula needs modification near left-hand cut, especially for the energies situated below the cut. 
The influence of the cut decreases as the size of finite volume increases. 
Furthermore,  higher energy levels can receive non-negligible contributions from higher partial waves if long-range interaction play a significant role. 
Therefore, to obtain reliable results, careful examination of the truncation of partial waves is suggested when applying standard L\"uscher formula to these energy levels. 

In addition, we calculate finite volume spectra of $\chi_{c1}(3872)$ using FVH method with the phenomenological model given in Refs.~\cite{Wang:2023ovj,Yu:2024sqv} where the binding energy and line shapes of $T_{cc}$ and $Z_c(3900)$ are successfully reproduced. 
Our spectrum is qualitatively consistent with the lattice spectrum provided in Refs.~\cite{Li:2024pfg,Prelovsek:2013cra,Lee:2014uta}. 
We also explore the contributions from different partial waves to energy levels and examined the impact of $c\bar{c}$ core. In a word, this approach can effectively addresses challenging problems in lattice QCD and holds potential for applications to the study of exotic states in hadron physics.

\begin{acknowledgments}

We thank useful discussions and valuable comments from Akaki Rusetsky, Anthony Thomas, Derek Leinweber, Yu Lu, Yan Li, Lu Meng, and Bingsong Zou. 
This work is partly supported  by
the National Natural Science Foundation of China under Grant Nos. 12175239 and 12221005 (J.J.W), 
and by the KAKENHI under Grant Nos. 23K03427 and 24K17055 (G.J.W), and by the National Natural Science Foundation of China (NSFC) under Grants Nos. 12275046 (Z.Y.), 
and by the Chinese Academy of Sciences under Grant No. YSBR-101 (J.J.W).

\end{acknowledgments}

\appendix

\section{Conventions of symmetry group} \label{append:convention}

In this appendix we clarify our conventions of the symmetry groups and their irreducible representations.
Though the finite volume effective Hamiltonian are equivalent up to a similarity transformation and its eigenvalues are exactly identical, the specific expression of $I$-matrix and its eigenvectors differ from the conventions adopted. 
We follow the conventions in Refs.~\cite{Bernard:2008ax, Doring:2018xxx, Morningstar:2013bda} and provide a brief summary here. 

For $O_h$, we follow the conventions in Refs.~\cite{Bernard:2008ax, Doring:2018xxx}. 
The 48 elements of $O_h$ can be divide into two subsets: $g_i$ with $i=1, \cdots\,, 24$ making up $O$ group, and another 24 elements generated by multiplying parity inversion, $g_i=\mathbb{P} g_{i-24}$ for $25\leq i\leq48$.
As outlined in Tab.~2 of the Ref.~\cite{Bernard:2008ax}, $g_i$ with $1\leq i\leq 24$ are uniquely parameterized by a angle $\omega$ with $-\pi\leq\omega<\pi$ and a unit vector $\vec{m}$ with $m_z\geq0$. 
The $O_h$ group has ten bosonic irrep $A_1^\pm,A_2^\pm,E^\pm,T_1^\pm,T_2^\pm$ with respecitve dimensions $1,1,2,3,3$. 
The specific matrix expressions for these irreps are provided in Tab.~1 of Ref.~\cite{Doring:2018xxx}, where the matrix elements are all real.  

The order of $\dc{O}_h$ is as twice as that of $O_h$. 
For each element $g\in O_h$ specified by $(\vec{m},\omega)$ there is a counterpart element $\bar{g}\in\dc{O}_h$ specified by $(\vec{m},\omega+2\pi)$. 
The irreps of $O_h$ are also the irreps of $\dc{O}_h$. 
Besides, there are six additional fermionic irreps, $G_1^\pm,G_2^\pm,H^\pm$. 
For bosonic irrep,  $D^\Gamma(\bar{g})=D^\Gamma(g)$, while for fermonic irrep, $D^\Gamma(\bar{g})=-D^\Gamma(g)$. 
The specific matrix expressions are provided in Ref.~\cite{Bernard:2008ax}.
Briefly speaking, the matrices of $G_1$ and $H$ corresponds to the Wigner-D matices $D^{J=\frac{1}{2}}$ and $D^{J=\frac{3}{2}}$, respectively. 
The relationship between $G_1$ and $G_2$ is the same as that between $A_1$ and $A_2$ as well as $T_1$ and $T_2$. 

For the moving system, the lattice symmetry groups are the subgroup of $O_h$: $C_{4v}$, $C_{2v}$ and $C_{3v}$ or their double cover for $\frac{\vec{P}L}{2\pi}=(0,0,1),\,(0,1,1)\,,(1,1,1)$, respectively. 
As the subgroup of $O_h$, the group elements are given by $C_{4v}= \{ g_{1},g_{14},g_{15},g_{24},g_{42},g_{43}, g_{46},g_{47}\}$, $C_{3v}= \{g_{1},g_2,g_3,g_{41},g_{43},g_{45}\}$ and $C_{2v}=\{g_1,g_{16},g_{41},g_{46}\}$. For other directions like $\frac{\vec{P}L}{2\pi}=(1,0,0)$ and $(1,1,0)$ we refer to Ref.~\cite{Wu:2021xvz}. 
We follow the Ref.~\cite{Morningstar:2013bda} for the conventions of the irreps of these groups.\footnote{Note that the irreps of different group may be denoted by the same notation.} 
Roughly speaking, for $C_{4v}$ there are five irreps $A_1,A_2,B_1,B_2,E$ and two additional ones $G_1,G_2$ for its double cover. 
The dimensions of these irreps are $1,1,1,1,2$ and $2,2$, respectively. 
For $C_{2v}$, there are four irreps $A_1,A_2,B_1,B_2$ and one additional one $G$ for its double cover. 
The dimensions are $1,1,1,1$ and $2$, respectively. 
(Note that the definitions of $B_1$ and $B_2$ here are reversed compared to Ref.~\cite{Dudek:2012gj,Dudek:2012xn}.)
For $C_{3v}$, there are three irreps $A_1,A_2,E$ and three additional ones $F_1,F_2,G$ for its double cover. 
The dimensions are $1,1,2$ and $1,1,2$, respectively. 
The correspondence of the notations of generators in Ref.~\cite{Morningstar:2013bda} and here are as follows: $(C_{4z},I_s C_{2y})\leftrightarrow (g_{15},g_{47})$ for $C_{4v}$, $(C_{3\delta},I_s C_{2b})\leftrightarrow(g_3,g_{43})$ for $C_{3v}$ and $(C_{2e},I_s C_{2f})\leftrightarrow (g_{16},g_{41})$ for $C_{2v}$, respectively. 
The specific matrix expressions of the generators for different irreps are provided in the Tab.(12-14) of Ref.~\cite{Morningstar:2013bda}. 
The matices for other elements can be obtained by appropriate multiplication. 
For convenience we present the multiplication as follows:\footnote{Since $D^\Gamma(\bar{g})=\pm D^\Gamma(g)$, for double cover group we only present half of the elements.}
\begin{itemize}
    \item For $C_{4v}$: $g_1=g_{15}^4,\,g_{14}=g_{15}^3,\,g_{24}=g_{15}^2,\,g_{42}=g_{47}g_{15},\,g_{43}=g_{47}g_{15}^3,\,g_{46}=g_{47}g_{15}^2$
    \item For $C_{3v}$: $g_1=g_3^3,\,g_2=g_3^2,\,g_{41}=g_{43}g_3,\, g_{45}=g_{43}g_3^2$
    \item For $C_{2v}$: $g_1=g_{16}^2,\,g_{46} = g_{41}g_{16}$
    \item For $\dc{C}_{4v}$: $g_1=g_{15}^8,\,g_{14}=g_{15}^7,\,g_{24}=g_{15}^6,\,g_{42}=g_{47}g_{15},\,g_{43}=g_{47}g_{15}^3,\, g_{46}=g_{47}g_{15}^2$
    \item For $\dc{C_{3v}}$: $g_1=g_3^6,\,g_2=g_3^5,\,g_{41}=g_{43}g_3,\, g_{45}=g_{43}g_3^2$ 
    \item For $\dc{C_{2v}}$: $g_1=g_{16}^4,\,g_{46}=g_{41}g_{16}$
\end{itemize}


\section{Polarization vector and helicity vector}\label{append:polarization and helicity}
In this appendix we present the polarization vector $\epsilon(\vec{p},\lambda)$ and helicty vector $\epsilon_H(\vec{p},\heli{\lambda})$ for a spin-$1$ particle with $\vec{p}=\left(p,\theta,\phi\right)$ and mass $m$. For polarization vector\cite{weinberg2005quantum},
\begin{align}
    \epsilon(\vec{p},\lambda=0) &= \begin{pmatrix}
         \frac{p}{m}\cos\theta \\
         \left(\frac{E}{m}-1\right) \cos\theta\sin\theta\cos\phi \\ \left(\frac{E}{m}-1\right) \cos\theta\sin\theta\sin\phi \\
         1 + \left(\frac{E}{m}-1\right) \cos^2\theta 
    \end{pmatrix} 
    \\
    \epsilon(\vec{p},\lambda=-1) &=\sqrt{\frac{1}{2}}
    \begin{pmatrix}
        \frac{p}{m} \sin\theta \,e^{-i\phi} \\
        1 + \left(\frac{E}{m} -1 \right)\sin^2\theta\cos\phi \,e^{-i\phi} \\
        -i + \left(\frac{E}{m} -1 \right)\sin^2\theta\sin\phi \,e^{-i\phi}
        \\
        \left(\frac{E}{m} -1 \right)\sin\theta\cos\theta \,e^{-i\phi}
    \end{pmatrix}
    \\
    \epsilon(\vec{p},\lambda=+1) &= - \epsilon^*(\vec{p},\lambda=-1)
\end{align}
For helicity vector, by definition, $\epsilon_H(\vec{p},\heli{\lambda}) = \sum\limits_{\lambda} e^{-i\lambda\phi}d^s_{\lambda \heli{\lambda}}(\theta) \epsilon(\vec{p},\lambda)$
\begin{align}
    \epsilon_H(\vec{p},\heli{\lambda}=0) &= \begin{pmatrix}
        \frac{p}{m} \\
        \frac{E}{m} \frac{\vec{p}}{\abs{\vec{p}}}
    \end{pmatrix} 
    \\
    \epsilon_H(\vec{p},\heli{\lambda}=+1) &= \sqrt{\frac{1}{2}} 
    \begin{pmatrix}
        0 \\
        -\cos\theta\cos\phi + i\sin\phi 
        \\
        -\cos\theta\sin\phi - i\cos\phi
        \\
        \sin\theta
    \end{pmatrix}
    \\
    \epsilon_H(\vec{p},\heli{\lambda}= -1) &= -  \epsilon_H^*(\vec{p},\heli{\lambda}=+1)
\end{align}

\section{Specific expressions of the pertinent L\"uscher function}\label{append:Luscher function}

In this appendix we present the specific expressions of the symmetric L\"uscher functions $\mathcal{M}_{l^\prime J^\prime,lJ}=\mathcal{M}_{lJ,l^\prime J^\prime}$ that are pertinent to this work. For the general definition of $\mathcal{M}$ we refer to Ref.~\cite{Woss:2018irj}. 
\begin{align}
    \mathcal{M}_{01,01} &= \mathcal{M}_{21,21} =  \frac{2 \mathcal{Z}_{00}(1;q^2)}{\sqrt{\pi} k_{\text{on}} L } 
    \\
    \mathcal{M}_{21,23} & = - \frac{2}{7\pi} \sqrt{\frac{6}{\pi}} \mathcal{Z}_{40}(1;q^2)
    \\
    \mathcal{M}_{23,23} & = \frac{2}{\pi}\left[ \frac{1}{q} \sqrt{\frac{1}{4\pi}} \mathcal{Z}_{00}(1;q^2) + \frac{1}{q^5} \frac{1}{7\sqrt{\pi}} \mathcal{Z}_{40}(1;q^2)  \right]
    \\
    \mathcal{M}_{01,43} &=  -\frac{2}{3\pi^{3/2}} \frac{1}{q^5} \mathcal{Z}_{40}(1;q^2) 
    \\
    \mathcal{M}_{21,43} &= \frac{20}{21\left(2\pi\right)^{3/2}} \frac{1}{q^5} \mathcal{Z}_{40}(1;q^2)
    \\
    \mathcal{M}_{23,43} & = -\frac{2}{\pi}\left[ \frac{1}{q^5} \frac{5}{77}\sqrt{\frac{3}{\pi}} \mathcal{Z}_{40}(1;q^2) + \frac{1}{q^7} \frac{25}{11\sqrt{39\pi}} \mathcal{Z}_{60}(1;q^2) \right] \\
    \mathcal{M}_{43,43} &= \frac{2}{\pi}\left[ \frac{1}{q}\sqrt{\frac{1}{4\pi}} \mathcal{Z}_{00}(1;q^2) + \frac{1}{q^5} \frac{27}{154\sqrt{\pi}}\mathcal{Z}_{40}(1;q^2) + \frac{1}{q^7}\frac{25}{66\sqrt{13\pi}} \mathcal{Z}_{60}(1;q^2) \right] \\
    \mathcal{M}_{01,44} &=  \frac{2\sqrt{2}}{\pi} \frac{1}{q^5} \sqrt{\frac{1}{5\pi}} \mathcal{Z}_{44}(1;q^2) \\
    \mathcal{M}_{21,44} &=  \frac{2\sqrt{2}}{\pi} \frac{1}{q^5}\sqrt{\frac{1}{10\pi}} \mathcal{Z}_{44}(1;q^2) \\ 
    \mathcal{M}_{23,44} &=  \frac{2\sqrt{2}}{\pi} \left[ -\frac{1}{q^5}\frac{3}{11}\sqrt{\frac{3}{5\pi}} \mathcal{Z}_{44}(1;q^2) + \frac{1}{q^7}\frac{5}{11}\sqrt{\frac{3}{13\pi}} \mathcal{Z}_{64}(1;q^2)  \right] \\
    \mathcal{M}_{43,44} & = \frac{2\sqrt{2}}{\pi}\left[ -\frac{1}{q^5}\frac{9}{22\sqrt{5\pi}} \mathcal{Z}_{44}(1;q^2) + \frac{1}{q^7}\frac{15}{22\sqrt{13\pi}} \mathcal{Z}_{64}(1;q^2) \right] \\
    \mathcal{M}_{44,44} &= \frac{2}{\pi} \left[ \frac{1}{q}\sqrt{\frac{1}{4\pi}} \mathcal{Z}_{00}(1;q^2) - \frac{1}{q^3}\frac{17}{22\sqrt{5\pi}}\mathcal{Z}_{20}(1;q^2) + \frac{1}{q^5} \frac{27}{286\sqrt{\pi}} \mathcal{Z}_{40}(1;q^2) +  \right.
    \\
    & \quad+\left. \frac{1}{q^7}\frac{1}{110\sqrt{13\pi}}\mathcal{Z}_{60}(1;q^2) + \frac{1}{q^9}\frac{896}{715\sqrt{17\pi}}\mathcal{Z}_{80}(1;q^2)  \right] 
\end{align}
where $q = \frac{k_{\text{on}}L}{2\pi}$ and L\"uscher Zeta function 
\[
\mathcal{Z}_{lm}(s;x^2) = \sum\limits_{\vec{n}\in\mathbb{Z}^3}\frac{\abs{\vec{n}}^l Y_{lm}(\vec{n})}{\left(\vec{n}^2 - x^2\right)^s}.
\]
The method for numerical evaluation of L\"uscher Zeta function can be found in Ref.~\cite{Gockeler:2012yj}.

\section{Ingredients for matrix elements}\label{append:matrix elements}
In subsection~\ref{subsec:matrix element} the expression of matrix element is provided. There are two parts, namely, kinematical part and dynamical part. The kinematical part is independent of the effective potential and entirely determined by the symmetry group and reference momentum. For convenience, we explicitly present the ingredients for kinematical part here. 

It should be noted that many results depend on the conventions of irreps and the choices of reference momentum, though the eigenvalues of Hamiltonian are invariant under different conventions and choices. Our conventions of irreps and reference momentum have been discussed in Appendix~\ref{append:convention} and Sec.~\ref{sec:projection}, respectively. For convenience we provide the reference momentum $\refvec{n}$ for $O_h$ with $|\refvec{n}|\leq 100$ as: $(0,0,0)$, $(0,0,1)$, $(0,1,1)$, $(1,1,1)$, $(0,0,2)$, $(0,1,2)$, $(1,1,2)$, $(0,2,2)$, $(0,0,3)$, $(2,2,1)$, $(0,1,3)$, $(1,1,3)$, $(2,2,2)$, $(0,2,3)$, $(1,2,3)$, $(0,0,4)$, $(0,1,4)$, $(2,2,3)$, $(1,1,4)$, $(0,3,3)$, $(3,3,1)$, $(0,2,4)$, $(1,2,4)$, $(3,3,2)$, $(2,2,4)$, $(0,0,5)$, $(0,3,4)$, $(0,1,5)$, $(1,3,4)$, $(1,1,5)$, $(3,3,3)$, $(0,2,5)$, $(2,3,4)$, $(1,2,5)$, $(0,4,4)$, $(2,2,5)$, $(4,4,1)$, $(0,3,5)$, $(3,3,4)$, $(1,3,5)$, $(0,0,6)$, $(4,4,2)$, $(0,1,6)$, $(1,1,6)$, $(2,3,5)$, $(0,2,6)$, $(1,2,6)$, $(0,4,5)$, $(4,4,3)$, $(1,4,5)$, $(3,3,5)$, $(2,2,6)$, $(0,3,6)$, $(2,4,5)$, $(1,3,6)$, $(4,4,4)$, $(0,0,7)$, $(2,3,6)$, $(0,1,7)$, $(0,5,5)$, $(3,4,5)$, $(1,1,7)$, $(5,5,1)$, $(0,4,6)$, $(0,2,7)$, $(1,4,6)$, $(1,2,7)$, $(3,3,6)$, $(5,5,2)$, $(2,4,6)$, $(2,2,7)$, $(4,4,5)$, $(0,3,7)$, $(1,3,7)$, $(5,5,3)$, $(0,5,6)$, $(3,4,6)$, $(2,3,7)$, $(1,5,6)$, $(0,0,8)$, $(0,1,8)$, $(0,4,7)$, $(2,5,6)$, $(1,1,8)$, $(1,4,7)$, $(5,5,4)$, $(3,3,7)$, $(0,2,8)$, $(4,4,6)$, $(1,2,8)$, $(2,4,7)$, $(3,5,6)$, $(2,2,8)$, $(0,6,6)$, $(0,3,8)$, $(6,6,1)$, $(1,3,8)$, $(0,5,7)$, $(3,4,7)$, $(1,5,7)$, $(5,5,5)$, $(6,6,2)$, $(2,3,8)$, $(4,5,6)$, $(2,5,7)$, $(0,4,8)$, $(0,0,9)$, $(1,4,8)$, $(4,4,7)$, $(6,6,3)$, $(0,1,9)$, $(3,3,8)$, $(1,1,9)$, $(3,5,7)$, $(2,4,8)$, $(0,2,9)$, $(0,6,7)$, $(1,2,9)$, $(1,6,7)$, $(5,5,6)$, $(6,6,4)$, $(2,2,9)$, $(0,5,8)$, $(3,4,8)$, $(2,6,7)$, $(0,3,9)$, $(1,5,8)$, $(4,5,7)$, $(1,3,9)$, $(2,5,8)$, $(2,3,9)$, $(3,6,7)$, $(4,4,8)$, $(0,4,9)$, $(6,6,5)$, $(1,4,9)$, $(3,5,8)$, $(0,7,7)$, $(3,3,9)$, $(7,7,1)$, $(5,5,7)$, $(0,0,10)$, $(0,6,8)$.

The elements of little group $\LG(\refvec{n})$ and the corresponding left coset are presented in Tab.~\ref{tab:LGandLC}.  For double cover group, the order of little group doubles while the left cosets remain unchanged. Specifically, if $g$ is an element of $\LG(\refvec{n})\subset G$, then both $g$ and $\bar{g}$ are elements of $\LG(\refvec{n})\subset \dc{G}$. 

The normalized eigenvectors with nonzero eigenvalues of $I$-matrix for $K\pi$, $N\pi$ and $\rho\pi$ system are presented in Tabs.~\ref{tab:eigenvec:2spinless}-\ref{tab:eigenvec:spin0spin1}. To make use of them correctly, several issues deserves to be mentioned:
\begin{enumerate}
    \item  Results for the patterns $(a,b,c),\,(a,b,*)_1,\,(*,a,b)_2,\,(a,b,c)_3$ are not shown since the their little group contains only the identity element and therefore $I^\Gamma=\mathbb{I}$. (Note that the helicity index needs to take all the values $\heli{\lambda}=-s,\cdots,s$ for these patterns.) 
    \item For $N\pi$ and $\rho\pi$ it is sufficient to present the results for non-negative helicity because, as shown in Tab.~\ref{tab:LGandLC}, $\LG^-(\refvec{n})$ are nonempty for all the patterns of  $\refvec{n}$ in Tabs.~\ref{tab:eigenvec:spin0spin0.5} and ~\ref{tab:eigenvec:spin0spin1}.
    \item For $N\pi$ and $\rho\pi$ with $\vec{P}\neq0$, the $I$-matrix may not entirely determined by the pattern of $\refvec{n}$. For example, both $\refvec{n}=(0,0,1)$ and $(0,0,-1)$ belong to $(0,0,*)_1$ but their $I$-matrix are not exactly the same. The reason is their little group contains non-trivial proper elements and the corresponding Wigner angle are different. Fortunately, the $I$-matrix for the system with $\vec{P}=0$ are entirely determined by the pattern of $\refvec{n}$ since either the pattern entirely determines the direction of $\refvec{n}$ or the only proper element in $\LG(\refvec{n})$ is identity element. 
\end{enumerate}

The Wigner rotation angle $\varphi_w(\refvec{n},g)$ for $g\in\LC(\refvec{n})$ is also required. With the help of Mathematica or other scientific languages, they can be computed straightforwardly from their definition. For brevity, explicit results are not shown here. It is worth emphasizing that, for bosonic systems, $\varphi_w$ can be calculated using either bosonic or fermionic Wigner-D matrices. However, for fermionic systems, the fermionic Wigner-D matrices must be used or $\varphi_w$ may differ by $2\pi$, which makes difference.

\begin{table}[]
    \centering
    \begin{tabular}{c|c|c|c}
      & pattern  & $\LG(\refvec{n})=\left\{g_i\right\}$ & $\LC(\refvec{n})=\left\{g_i\right\}$  \\ \hline
      \multirow{7}{*}{$O_h$} & $(0,0,0)$  &  $i=1,\cdots,48$ & $i=1$ \\ \cline{2-4}
      & $(0,0,a)$ & $i=1,14,15,24,42,43,46,47$ & $i=1,2,3,4,6,18$  \\ \cline{2-4}
      & $(0,a,a)$ & $i=1,16,41,46$ & $i=1,\cdots,10,11,17$ \\ \cline{2-4}
      & $(a,a,a)$ & $i=1,2,3,41,43,45$ & $i=1,4,5,6,10,11,12,17$ \\ \cline{2-4}
      & $(0,a,b)$ & $i=1,46$ & $i=1\cdots,24$ \\ \cline{2-4}
      & $(a,a,b)$ or $(b,b,a)$ & $i=1,43$ & $i=1,\cdots,24$  \\ \cline{2-4} 
      & $(a,b,c)$ & $i=1$ & $i=1,\cdots,48$ \\ \hline
      \multirow{4}{*}{$C_{4v}$} & $(0,0,*)_1$ &   $i={1},{14},{15},{24},{42},{43}, {46}, {47}$ & $i=1$ \\ \cline{2-4}
      & $(0,a,*)_1$ & $i=1,46$ & $i=1,14,15,24$ \\ \cline{2-4}
      & $(a,a,*)_1$ & $i=1,43$ & $i=1,14,15,24$ \\ \cline{2-4}
      & $(a,b,*)_1$ & $i=1$ & $i={1},{14},{15},{24},{42},{43}, {46}, {47}$ 
      \\ \hline
       \multirow{4}{*}{$C_{2v}$} & $(0,a,a)_1$ & $i=1,16,41,46$ & $i=1$
       \\ \cline{2-4}
       & $(0,a,b)_2$ &  $i=1,46$ & $i=1,16$ \\ \cline{2-4}
       & $(*,a,a)_2$ & $i=1,41$ & $i=1,16$  \\ \cline{2-4}
       & $(*,a,b)_2$ & $i=1$ & $i=1,16,41,46$
       \\ \hline 
       \multirow{3}{*}{$C_{3v}$} & $(a,a,a)_3$ & $i=1,2,3,41,43,45$ & $i=1$
       \\ \cline{2-4}
       & $(a,a,b)_3$ & $i=1,43$ & $i=1,2,3$
       \\ \cline{2-4}
       & $(a,b,c)_3$ & $i=1$ & $i=1,2,3,41,43,45$
       \\ \hline 
    \end{tabular}
    \caption{The little group $\LG(\refvec{n})$ and  the chosen left coset element $\LC(\refvec{n})$ for $O_h$, $C_{4v}$, $C_{2v}$ and $C_{3v}$. The conventions of elements are discussed in Appendix~\ref{append:convention}. }
    \label{tab:LGandLC}
\end{table}

\begin{table}[tbp]
    \centering
    \begin{tabular}{c|c|l}
    $G$ & pattern & $ \text{[eigenvectors]}^\Gamma$ \\ \hline
    \multirow{7}{*}{$O_h$} &  $(0,0,0)$  &  $\left[ (1) \right]^{A_1^+} $  \\ \cline{2-3} 
    & $(0,0,a)$  &  $\left[  (1) \right]^{A_1^+ }$,   $\left[ (1,0)\right]^{E^+}$, $\left[ (0,0,1)\right]^{T_1^-}$  \\ \cline{2-3}
    & $(0,a,a)$ & $\left[ (1) \right]^{A_1^+}$,  $\left[\frac{1}{2}\left(-1,\sqrt{3}\right)\right]^{E^+}$, $\left[\frac{1}{\sqrt{2}}\left(0,1,1\right)\right]^{T_1^-}$, $\left[(1,0,0)\right]^{T_2^+}$,  $\left[\frac{1}{\sqrt{2}}\left(0,-1,1\right)\right]^{T_2^-}$ \\ \cline{2-3}
    & $(a,a,a)$  & $\left[(1)\right]^{A_1^+,\,A_2^-}$, $\left[\frac{1}{\sqrt{3}}\left(1,1,1\right)\right]^{T_1^-,\,T_2^+}$  \\ \cline{2-3}
    & $(0,a,b)$ & $\left[(1)\right]^{A_1^+,\,A_2^+}$,  $\left[(0,1),\,(1,0)\right]^{E^+}$, $\left[(1,0,0)\right]^{T_1^+,\,T_2^+}$, $\left[\left(0,0,1\right),\left(0,1,0\right)\right]^{T_1^-,\,T_2^-}$ \\ \cline{2-3}
    & $(a,a,b)$  & $\left[(1)\right]^{A_1^+,\,A_2^-}$,  $\left[(1,0)\right]^{E^+}$, $\left[(0,1)\right]^{E^-}$, $\left[\frac{1}{\sqrt{2}}\left(-1,1,0\right)\right]^{T_1^+,\,T_2^-}$ \\
    & or $(b,b,a)$ & $\left[\left(0,0,1\right),\frac{1}{\sqrt{2}}\left(1,1,0\right) \right]^{T_1^-,\,T_2^+}$ \\
    \hline
    \multirow{4}{*}{$C_{4v}$} & $(0,0,*)_1$ & $\left[(1)\right]^{A_1}$
    \\ \cline{2-3}
    & $(0,a,*)_1$ &  $\left[(1)\right]^{A_1,B_1} ,\left[(0,1)\right]^{E}$ 
    \\ \cline{2-3} 
    & $(a,a,*)_1$ & $\left[(1)\right]^{A_1,\,B_2},\left[\frac{1}{\sqrt{2}}\left(1,1\right)\right]^{E}$ 
    \\ \hline
    \multirow{3}{*}{$C_{2v}$} & $(0,a,a)_2$ & $\left[(1)\right]^{A_1}$
    \\ \cline{2-3}
    & $(0,a,b)_2$ & $\left[(1)\right]^{A_1,\,B_2}$ 
    \\ \cline{2-3} 
    & $(*,a,a)_2$ & $\left[(1)\right]^{A_1,\,B_1}$ 
    \\ \hline 
    \multirow{2}{*}{$C_{3v}$} & $(a,a,a)_3$ & $[(1)]^{A_1}$ 
    \\ \cline{2-3} 
    & $(a,a,b)_3$ & $[(1)]^{A_1},\,[(0,1)]^{E}$ 
    \\ \hline
    \end{tabular}
    \caption{The normalized eigenvectors with nonzero eigenvalue of $I$-matrix for $K\pi$-system. The intrinsic parity is assumed to be positive.}
    \label{tab:eigenvec:2spinless}
\end{table}

\begin{table}[htbp]
    \centering
    \begin{tabular}{c|c|l}
      $G$ & pattern &  $ \text{[eigenvectors]}^\Gamma_{\heli{\lambda}=+\frac{1}{2}}$ \\ \hline
     \multirow{7}{*}{$\dc{O_h}$} &
     $(0,0,a)$ &  $\left[(1,0)\right]^{G_1^\pm}$, $\left[(0,1,0,0)\right]^{H^\pm}$
     \\ \cline{2-3}
     &  \multirow{2}{*}{$(0,a,a)$} & $\left[\frac{\left((\sqrt{2}+1)i,\,1\right)}{\sqrt{4+2\sqrt{2}}}\right]^{G_1^\pm}$, $\left[\frac{\left((-\sqrt{2}+1)i,\,1\right)}{\sqrt{4-2\sqrt{2}}}\right]^{G_2^\pm}$,
     \\
     & & $\left[\frac{\left(i(2-\sqrt{2}),\,\sqrt{3}-\sqrt{6},\,0,\,1\right)}{\sqrt{16-10\sqrt{2}}},\, \frac{\left(-\sqrt{3}+\sqrt{6},\,i(2-\sqrt{2}),\,1,\,0\right)}{\sqrt{16-10\sqrt{2}}} \right]^{H^\pm}$  \\
     \cline{2-3} 
     & $(a,a,a)$ & $\left[\frac{\left((1+i)(1+\sqrt{3}),\,2\right)}{2\sqrt{3+\sqrt{3}}}\right]^{G_1^\pm, G_2^\pm}$, 
       $\left[\frac{\left((1-i)(1+\sqrt{3}),\,2i,\,(1+i)(1+\sqrt{3}),\,2\right)}{2\sqrt{2(3+\sqrt{3})}}\right]^{H^\pm}$ \\
     \cline{2-3} 
     & $(0,a,b)$ & $\left[\left(0,1\right),\left(1,0\right)\right]^{G_1^\pm,G_2^\pm}$, $\left[\left(1,0,0,0\right),\left(0,1,0,0\right),\left(0,0,1,0\right),\left(0,0,0,1\right)\right]^{H^\pm}$
     \\
     \cline{2-3}
    & $(a,a,b)$ & \multirow{2}{*}{$\left[\left(0,1\right),\left(1,0\right)\right]^{G_1^\pm,G_2^\pm}$, $\left[\left(1,0,0,0\right),\left(0,1,0,0\right),\left(0,0,1,0\right),\left(0,0,0,1\right)\right]^{H^\pm}$} 
    \\ 
    & or $(b,b,a)$ &   
    \\ \hline
     \multirow{3}{*}{$\dc{C_{4v}}$} & $(0,0,*)_1$  & $\left[(1,0)\right]^{G_1}(\text{if}\,\,*>0),\,\left[(0,1)\right]^{G_1}(\text{if}\,\,*<0)$ 
     \\ \cline{2-3}
     & $(0,a,*)_1$ &  $\left[\left(1,0\right),\left(0,1\right)\right]^{G_1,G_2}$
     \\ \cline{2-3} 
     & $(a,a,*)_1$ & $\left[\left(1,0\right),\left(0,1\right)\right]^{G_1,G_2}$
     \\ \hline
     \multirow{3}{*}{$\dc{C_{2v}}$ } &   $(0,a,a)_2$ & $\left[\frac{1}{\sqrt{2}}\left(i,1\right)\right]^{G}$ 
     \\ \cline{2-3} 
     & $(0,a,b)_2$ & $\left[\left(0,1\right),\left(1,0\right)\right]^G$ 
     \\ \cline{2-3} 
    & $(*,a,a)_2$ & $\left[\left(0,1\right),\left(1,0\right)\right]^G$ 
    \\ \hline
    \multirow{2}{*}{$\dc{C_{3v}}$} & $(a,a,a)_3$  &  
    $\left[\frac{\left((1+i)(\sqrt{3}+1),2\right)}{2\sqrt{3+\sqrt{3}}}\right]^{G}$
    \\ \cline{2-3}
    & $(a,a,b)_3$ & $\left[(1)\right]^{F_1,\,F_2}$, $\left[(1,0),(0,1)\right]^{G}$
    \\ \hline
    \end{tabular}
    \caption{The normalized eigenvectors with nonzero eigenvalue of $I$-matrix for $N\pi$-system. The intrinsic parity is assumed to be positive.}
    \label{tab:eigenvec:spin0spin0.5}
\end{table}

\begin{table}[]
    \centering
    \begin{tabular}{c|c|l}
      $G$ & pattern &  $ \text{[ eigenvectors ]}^\Gamma_{\heli{\lambda}\geq0}$ \\ \hline
      \multirow{9}{*}{$O_h$} & $(0,0,a)$ & $\left[(1)\right]^{A_1^-}_0$ $\left[(1,0)\right]^{E^-}_{0},\left[(0,0,1)\right]^{T_1^+}_{0}$, $\left[\frac{1}{\sqrt{2}}\left(i,1,0\right)\right]^{T_1^\pm}_1$, $\left[\frac{1}{\sqrt{2}}(-i,1,0)\right]_1^{T_2^\pm}$
      \\ \cline{2-3}
      & \multirow{2}{*}{$(0,a,a)$} & $\left[(1)\right]^{A_1^-}_0,\, \left[\frac{1}{2}(-1,\sqrt{3})\right]_0^{E^-},\,\left[\frac{1}{\sqrt{2}}(0,1,1)\right]_0^{T_1^+},\,\left[\frac{1}{\sqrt{2}}(0,-1,1)\right]_0^{T_2^+},\,\left[(1,0,0)\right]_0^{T_2^-}$,
      \\ 
      & & $\left[(1)\right]_1^{A_2^\pm},\,\left[\frac{1}{2}(\sqrt{3},1)\right]_1^{E^\pm},\,\left[(1,0,0),\frac{1}{\sqrt{2}}(0,-1,1)\right]_1^{T_1^\pm},\left[\frac{1}{\sqrt{2}}(0,1,1)\right]_1^{T_2^\pm}$
      \\  \cline{2-3}
      & $(a,a,a)$ & $ \left[(1)\right]_0^{A_1^-,\,A_2^+},\,\left[\frac{1}{\sqrt{3}}(1,1,1)\right]_0^{T_1^+,\,T_2^-},\,\left[\frac{1}{\sqrt{2}}(i,1)\right]_1^{E^\pm},\,\left[\frac{1}{\sqrt{3}}(e^{-\frac{2\pi i}{3}}, e^{\frac{2\pi i}{3}},1)\right]_1^{T_1^\pm,T_2^\pm}$ 
      \\ \cline{2-3}
      & \multirow{2}{*}{$(0,a,b)$} & $\left[(1)\right]_0^{A_1^-,\,A_2^-}$, $\left[(0,1),(1,0)\right]_0^{E^-}$, $\left[(0,1,0),(0,0,1)\right]_0^{T_1^+,T_2^+}$, $\left[(1,0,0)\right]_0^{T_1^-,T_2^-}$, 
      \\
      & & $\left[(1)\right]_1^{A_1^\pm,\,A_2^\pm}$, $\left[(0,1),(1,0)\right]_1^{E^\pm}$, $\left[(1,0,0),(0,1,0),(0,0,1)\right]_1^{T_1^\pm,T_2^\pm}$
      \\ \cline{2-3}
      & \multirow{2}{*}{$(a,a,b)$} & $\left[(1)\right]_0^{A_1^-,\,A_2^+},\,\left[(0,1)\right]_0^{E^+},\,\left[(1,0)\right]_0^{E^-},\,\left[(-\frac{1}{\sqrt{2}},\frac{1}{\sqrt{2}},0)\right]_0^{T_1^-,\,T_2^+}$
      \\
     & or $(b,b,a)$ & $\left[(0,0,1),(\frac{1}{\sqrt{2}},\frac{1}{\sqrt{2}},0)\right]_0^{T_1^+,\,T_2^-}$, 
     \\
     & & $\left[(1)\right]_1^{A_1^\pm,\,A_2^\pm},\,\left[(0,1),(1,0)\right]_1^{E^\pm},\,\left[(1,0,0),(0,1,0),(0,0,1)\right]_1^{T_1^\pm,\,T_2^\pm}$
     \\ \hline
     \multirow{3}{*}{$C_{4v}$} & $(0,0,*)_1$ & $\left[(1)\right]^{A_2}_0,\,\left[\frac{1}{\sqrt{2}}(i,1)\right]^{E}_1(\text{if}\,\, *>0),\,\left[\frac{1}{\sqrt{2}}(-i,1)\right]^{E}_1(\text{if}\,\, *<0)$
     \\ \cline{2-3}
     & $(0,a,*)_1$ & $[(1)]_0^{A_2,\,B_2},\left[(1,0)\right]_0^E$, $\left[(1)\right]_1^{A_1,\,A_2,\,B_1,\,B_2},\,\left[(1,0),(0,1)\right]_1^{E}$
     \\ \cline{2-3}
     & $(a,a,*)_1$ & $\left[(1)\right]_0^{A_2,B_1},\,\left[\frac{1}{\sqrt{2}}(-1,1)\right]_0^{E}$, $\left[(1)\right]_1^{A_1,\,A_2,\,B_1,\,B_2},\,\left[(1,0),(0,1)\right]_1^{E}$ 
     \\ \hline
    \multirow{3}{*}{$C_{2v}$} & $(0,a,a)_2$ & $\left[(1)\right]_0^{A_2}$, $\left[(1)\right]_1^{B_1,\,B_2}$ 
    \\ \cline{2-3}
    & $(0,a,b)_2$ & $\left[(1)\right]_0^{A_2,\,B_1}$, $\left[(1)\right]_1^{A_1,\,A_2,\,B_1,\,B_2}$
    \\ \cline{2-3}
    & $(*,a,a)_2$ & $\left[(1)\right]_0^{A_2,\,B_2}$, $\left[(1)\right]_1^{A_1,\,A_2,\,B_1,\,B_2}$
    \\ \hline
    \multirow{2}{*}{$C_{3v}$} & $(a,a,a)_3$ & $\left[(1)\right]_0^{A_2}$, $\left[\frac{1}{\sqrt{2}}(-i,1)\right]^E_1(\text{if}\,\,a>0),\,\left[\frac{1}{\sqrt{2}}(i,1)\right]^E_1(\text{if}\,\,a<0)$
    \\ \cline{2-3}
    & $(a,a,b)_3$ & $\left[(1)\right]_0^{A_2},\left[(1,0)\right]_0^{E}$, $\left[(1)\right]_1^{A_1,A_2}$, $\left[(1,0),(0,1)\right]_1^{E}$
    \\ \hline
    \end{tabular}
    \caption{The normalized eigenvectors with nonzero eigenvalue of $I$-matrix for $\rho\pi$-system. The intrinsic parity is assumed to be positive.}
    \label{tab:eigenvec:spin0spin1}
\end{table}

\clearpage
\bibliography{ref}

\end{document}